\documentclass[twocolumn,seceq]{jpsj2} 

\newcommand{\bigcite}[1]{\cite{#1}}
\newcommand{\Ham}{{\cal H}}

\newcommand{\vect}[1]{{\mbox{\boldmath $#1$}}}

\newcommand{\Prob}[1]{{\rm Prob}(#1)}
\newcommand{\Tr}[1]{{\rm Tr}\left(#1\right)}
\newcommand{\eqlabel}[1]{\label{eq:#1}}
\newcommand{\figlabel}[1]{\label{fig:#1}}
\newcommand{\tablabel}[1]{\label{tab:#1}}
\newcommand{\seclabel}[1]{\label{sec:#1}}
\newcommand{\ssclabel}[1]{\label{ssc:#1}}
\newcommand{\ssslabel}[1]{\label{sss:#1}}
\newcommand{\Equation}[1]{Equation (\ref{eq:#1})}
\newcommand{\Eq}[1]{(\ref{eq:#1})}
\newcommand{\Fig}[1]{Fig.\ \ref{fig:#1}}
\newcommand{\Figure}[1]{Figure \ref{fig:#1}}
\newcommand{\Tab}[1]{Table \ref{tab:#1}}

\newcommand{\msec}[1]{\S\,\ref{sec:#1}}
\newcommand{\mssc}[1]{\S\,\ref{ssc:#1}}
\newcommand{\msss}[1]{\S\,\ref{sss:#1}}
\newcommand{\Sec}[1]{\S\,\ref{sec:#1}}

\newcommand{\dudu}{\begin{array}{cc} 0 & 1 \\ 0 & 1 \end{array}}
\newcommand{\udud}{\begin{array}{cc} 1 & 0 \\ 1 & 0 \end{array}}

\newcommand{\MCave}[1]{\left\langle #1 \right\rangle_{\rm MC}}
\newcommand{\sgn}[1]{\,{\rm sgn}\left(#1\right)}
\newcommand{\diag}[1]{{\rm diag}\left(#1\right)}
\newcommand{\Step}[1]{{Step \ref{st:#1}}}

\newcommand{\deffig}[4]{
\begin{figure}[tb]
  \begin{center}
  \includegraphics[width=#3 \textwidth]{FIG/#2}
  \end{center}
  \caption{ \figlabel{#1} #4}
\end{figure}
}

\newenvironment{liststep}{%
\begin{list}{{\bf Step \arabic{enumi}}}{\usecounter{enumi}%
\leftmargin=0.4cm\labelwidth=0.2cm\labelsep=0.2cm%
\listparindent=0cm\itemindent=0cm
}}{\end{list}}

\newcommand{\mattwo}[4]{
  ${\scriptsize 
  \left( \begin{array}{ll} #1 & #2 \\ #3 & #4 \end{array} \right)
  }$
}
\newcommand{\figbox}[1]{
  \parbox{14mm}{\includegraphics[width=14mm]{FIG/#1}}
}

\title{Recent Developments of World-Line Monte Carlo Methods}
\author{
  Naoki \textsc{KAWASHIMA}$^{1}$
  \thanks{E-mail address: nao@phys.metro-u.ac.jp} and
  Kenji \textsc{HARADA}$^{2}$
  \thanks{E-mail address: harada@acs.i.kyoto-u.ac.jp}
}
\inst{
  $^{1}$Department of Physics, Tokyo Metropolitan University, Tokyo 192-0397\\
  $^{2}$Department of Applied Analysis and Complex Dynamical Systems, Kyoto University, Kyoto 606-8501
}
\abst{
World-line quantum Monte Carlo methods are reviewed
with an emphasis on breakthroughs made in recent years.
In particular, three algorithms --- 
the loop algorithm, the worm algorithm, and the
directed-loop algorithm --- for updating world-line configurations 
are presented in a unified perspective.
Detailed descriptions of the algorithms in specific cases are also given.
}
\kword{quantum spin system, Heisenberg model, quantum Monte Carlo,
XY model, XXZ model, cluster algorithm, loop algorithm,
worm algorithm, directed-loop algorithm, world-line method}
\begin{document}
\maketitle


\section{Introduction} 
The Monte Carlo method based on a Markov process has been quite a powerful
tool of the model analysis in many-body physics such as
condensed matter physics, statistical physics and field theory.
In the present review, we focus on a branch of Markov-chain Monte Carlo
methods that have been developed remarkably during 
the past decade, i.e., the quantum Monte Carlo\cite{Suzuki1976} that samples from
an ensemble of {\it world-line} configurations in the path-integral
representation of the partition function.
The methodological advancement is largely due to the global update
of the world-line configurations.
The breakthrough was made 
by Evertz, Lana and Marcu\cite{EvertzLM1993},
who proposed a new algorithm, called a {\it loop algorithm},
for the $s=1/2$ $XXZ$ model, and later also by 
Prokof'ev, Svistunov and Tupitsyn,\cite{ProkofevST1998}
whose approach, called the worm algorithm, seemed quite different at first sight.
In a loop algorithm, the world-line configuration is updated
in the unit of loops in the space-time formed by a stochastic procedure.
It turned out that the loop update does not only reduce the critical
slowing-down, but it also removes several other drawbacks of 
the conventional quantum Monte Carlo.
In the worm algorithm\cite{ProkofevST1998}, on the other hand, 
the world-line configuration is updated by the movements of
a {\it worm}, i.e., a pair of artificial singular points
at which world-lines are discontinuous.
A framework was proposed\cite{SyljuasenS2002} recently that
unifies these two ways of updating and enjoys the virtues of both.
In the present article, therefore,
we focus on three important algorithms
(or, to be more precise, three frameworks for algorithms);
the loop algorithm\cite{EvertzLM1993},
the worm algorithm\cite{ProkofevST1998}, 
and the directed-loop algorithm\cite{SyljuasenS2002}.  
In some special cases two of them are identical, e.g., the
directed-loop algorithm applied to the $s=1/2$ antiferromagnetic
Heisenberg model is nothing but a single-cluster version of the loop
algorithm.

Before the proposal of these new algorithms,
simulations had been done with local updating rule 
on the discretized imaginary time.
The local updating rule is analogous to the single-spin-flip Metropolis 
algorithm of the Ising model.
While it provided the first systematic means of numerical
study of systems at finite temperatures, 
it had a number of drawbacks;
(i) the critical slowing-down, 
(ii) the fine-mesh slowing-down 
    (i.e., the slowing-down when the discretization step of the imaginary time 
    is decreased),
(iii) non-ergodicity (the temporal and the spatial winding numbers are conserved),
(iv) the discretization error, and
(v) difficulty in measuring the off-diagonal quantities,
(vi) the negative-sign problem.
These drawbacks have been removed (or at least reduced) by the recent 
development of the quantum Monte Carlo method mentioned above.
The critical slowing-down and the fine-mesh slowing-down have been reduced
to the negligible level in most applications.
\cite{EvertzLM1993,KawashimaG1994,ProkofevST1998,Sandvik1999,SyljuasenS2002}
The non-ergodicity and the discretization error have been completely removed.
\cite{BeardW1996}
In addition, most of the off-diagonal quantities of interest can be measured.
\cite{BrowerCW1998,ProkofevST1998,KashurnikovPST1999}
The negative-sign problem is the toughest and only very limited solution
is available. 
However, there is at least a few cases where this difficulty 
can be overcome by the loop algorithm.
\cite{ChandrasekharanW1999,ChandrasekharanCOW2002}

There are a number of articles already published on the quantum Monte Carlo.
We here only refer the reader to a review article\cite{HatanoS1994} for
the achievement made before the loop algorithm was proposed.
For the loop algorithm and related algorithms, 
an excellent overview\cite{Evertz2003} has been written 
on the loop algorithm by one of the founders of the algorithm.
Still there remains a lot of technical difficulties
for an unfamiliar reader to start simulations from scratch.
Therefore, we feel it useful to put various technical details 
together with the background mathematics.
The purpose of the present article is, therefore,
to present various ingredients in a single article 
in a form comprehensible to non-specialists
and ready to use for practitioners.
In what follows we describe how we perform
the simulation in detail and take a particular care in
making the article practical.
On the other hand, 
we do not intend to make the present article to be comprehensive;
we mention applications only when it is necessary for
illustrating a new idea and show how effective it is.
As a result, only a few applications are discussed in the rest of the article.
We refer the readers who are interested in various applications,
as well as other things that we omit in the present article,
to the review articles mentioned above.

To make this article usable, we separate {\it how} from {\it why}, 
i.e., the description of the algorithms from their mathematical derivations.
Therefore, those who need a quick start,
rather than knowing why an algorithm gives correct results,
may skip the theoretical part (\msec{Theory}) and immediately go to
\msec{Recipe} entitled ``Numerical Recipe''.
This section is almost self-contained and only the minimal references
are made to the other parts of the article.


\section{Theory}
\seclabel{Theory}
In this section, we present a few algorithms, 
roughly in chronological order.
Descriptions will constitute a mathematical justification 
of the algorithms' validity, though they do not follow
the conventional theory-and-proof format.
Some examples are also presented for illustrating relevant ideas
and the efficiency of the resulting algorithms.
While this style would make it easier to follow the logic that 
establishes the validity of the algorithms,
it may make it hard to find the precise and detailed definitions
of the procedures.
Therefore, we add another section following the present one,
in which we concentrate on describing the procedures precisely.

\subsection{Cluster Update}
\ssclabel{ClusterUpdate}
The improvements accomplished on the quantum Monte Carlo simulation
during the last decade was largely due to the global update,
in which configurations are updated in units of some non-local clusters.
Such a method of updating is inspired by the Swendsen-Wang (SW) algorithm
\cite{SwendsenW1987} for the Ising model.
In fact, it is not merely inspired but has the same mathematical
back-ground as the SW algorithm.
This is manifested by the fact that the loop algorithm proposed by
Evertz {\it et al.} for the $s=1/2$ $XXZ$ quantum spin model
depends continuously on the anisotropy and in the 
limit of a large uni-axial anisotropy (i.e., the Ising limit),
the algorithm converges to something equivalent to the SW algorithm
\cite{KawashimaG1995B}.
In this sense, the loop algorithm for the quantum spin systems
can be considered as a generalization of the SW algorithm.
The same is true for the single-cluster variant of the cluster algorithm
by Wolff;\cite{Wolff1989}
the single-cluster variant of the loop algorithm for the quantum
Monte Carlo can be derived from the Wolff algorithm in 
exactly the same way as we can derive its multiple-cluster variant
from the SW algorithm.
In what follows, we consider the multiple-cluster variant,
when we have to choose one, while the generalization to the
single-cluster variant is straightforward in many cases.

We start with describing the SW algorithm
to clarify the mathematical basis underlying almost all the 
algorithms discussed in the present article.
Simply stated, the SW algorithm and other algorithms presented below
are special cases of the dual Monte Carlo algorithm
\cite{KandelD1991,KawashimaG1995}.
In a dual Monte Carlo algorithm, the Markov process alternates between
two configuration spaces; the space of the original configurations that 
naturally arise from the model (such as the spin configurations 
in the Ising model and the world-line configurations 
in the quantum lattice models) and the space of 
the configurations of auxiliary variables.
It is up to us to define the auxiliary variables and
the resulting algorithm depends on the definition.
In what follows, we denote the original configuration by $S$ and 
the auxiliary one by $G$.
(We denote the size of spins by $s$ instead of $S$, 
to avoid confusion.)
Once the auxiliary variables are defined, 
a stochastic process is characterized by the transition 
probabilities $T(G|S)$ and $T(S|G)$, the former being
of generating $G$ with $S$ given and the latter 
of generating $S$ with $G$ given.
The stochastic process as depicted in \Fig{DualMonteCarlo}
yields the limiting distribution
$$
  \lim_{n\to\infty} P_n(S) \equiv 
  \lim_{n\to\infty} \Prob{S(n) = S}
  \propto W(S)
$$
provided that we define the transition probabilities
so that the ergodicity and the extended detailed balance
\begin{equation}
  T(G|S)W(S) = T(S|G)W(G)
  \eqlabel{GeneralizedDetailedBalance}
\end{equation}
may hold.
Here $W(S)$ ($W(G)$) is an arbitrary positive function of $S$ ($G$).
It is specified for each individual case as we see below.

\deffig{DualMonteCarlo}{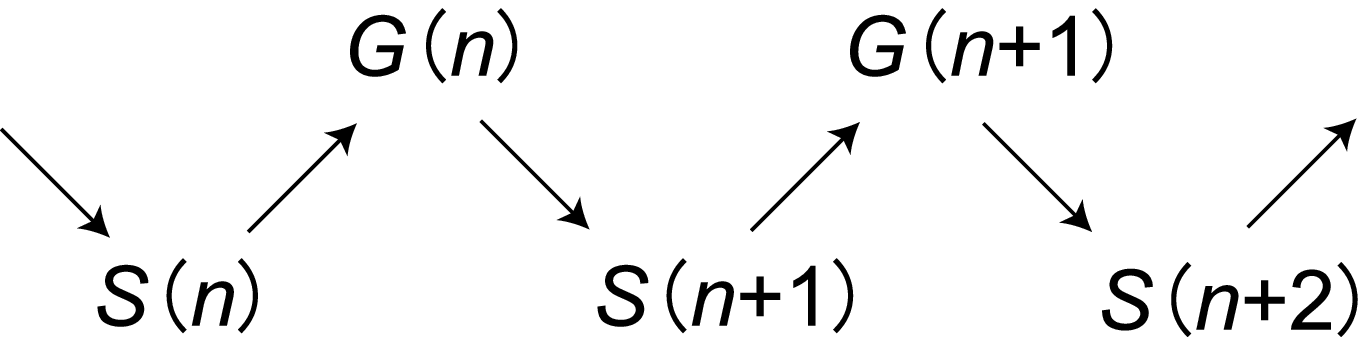}{0.4}{
A schematic process of a dual Monte Carlo.
The arrow indicates dependencies.
It should be noted that the new state $S(n+1)$ depends on
the previous one $S(n)$ only through $G(n)$.
The same is true for $G(n+1)$.
}

Swendsen and Wang chose the auxiliary variable $G_u$
to be a one-bit (i.e., 0-or-1) variable defined on each 
pair $u$ of interacting spins.
The auxiliary configuration $G$ is defined as the set of all
such variables: $G \equiv (G_1, G_2, \cdots, G_{N_B})$, where
$N_B$ is the total number of the nearest-neighbor pairs.
It is very cumbersome to describe the procedure in terms
only of variables, with no picture, though in the end such 
a description is needed for coding.
We, therefore, resort to visual means
whenever a suitable visualization is available.
The local unit $u$ on which we define an auxiliary 
variable $G_u$ is not necessarily a pair of sites.
In addition, $G_u$ is not necessarily a 0/1 variable either.
Therefore, the visualization varies depending on the problem.
In the case of the Ising model, however, the visualization is 
done most naturally by representing an up-spin and a down-spin 
by an open and a solid circle, respectively,
and an auxiliary variable by the presence
(corresponding to $G_u = 1$) or the absence 
(corresponding to $G_u = 0$) of a solid line connecting 
two neighboring circles. (See \Fig{Ising}.)

Swendsen and Wang\cite{SwendsenW1987} proposed the following procedure
of updating the spin configuration. For a given configuration,
(i) assign $G_u=0$ or $1$ probabilistically to each $u$,
(ii) identify connected spins to form clusters, and for each cluster
(iii) assign a common value $\pm 1$ to all the spin variables
    on it.
In the graphical terms, this yields the following
(i) connect nearest-neighbor circles with some probability, 
(ii) recognize the connected sets of circles, and for each connected set,
(iii) change the color of all circles simultaneously 
with a certain probability.
The step (iii) is often called a `cluster-flip'.

In the following, we show
that this stochastic process produces the distribution 
of spin configuration proportional to the Boltzmann weight 
\begin{equation}
  W(S) = {\rm exp}\left(K\sum_{(ij)}S_iS_j \right)
       = \prod_{u} w(S_u),
  \eqlabel{BoltzmannWeight}
\end{equation}
where $u \equiv (i,j)$, $S_u = (S_i, S_j)$, and
$w(S_u) \equiv {\rm exp}(KS_iS_j)$.
Following Fortuin and Kasteleyn\cite{KasteleynF1969, FortuinK1972},
we decompose the local Boltzmann weight as
\begin{equation}
  w(S_u) = \sum_{G_u} w( S_u, G_u ).
  \eqlabel{LocalFK}
\end{equation}
The function $w(S_u, G_u)$ is defined as
\begin{equation}
  w(S_u, G_u) = 
  \left\{ \begin{array}{ll}
  e^{-K}       & \mbox{($G_u = 0$)} \\
  e^K - e^{-K} & \mbox{($S_i = S_j$ and $G_u = 1$)} \\
  0            & \mbox{($S_i \ne S_j$ and $G_u = 1$)}
  \end{array} \right..
  \eqlabel{LocalExtendedWeight}
\end{equation}
It is easy to verify that this definition
satisfies \Eq{LocalFK}.
Using \Eq{LocalFK},
eq.\ \Eq{BoltzmannWeight} can be formally rewritten as
\begin{equation}
  W(S) = \sum_G W(S,G) \eqlabel{GlobalFK},
\end{equation}
where
\begin{equation}
  W(S,G) \equiv \prod_u w(S_u,G_u).
  \eqlabel{ExtendedWeight}
\end{equation}
Once the target weight $W(S)$ is written in
the form of \Eq{GlobalFK} with \Eq{ExtendedWeight}
and \Eq{LocalFK}, we can in general satisfy 
the detailed balance \Eq{GeneralizedDetailedBalance} 
by defining the transition probabilities as
\begin{equation}
  T(G|S) \equiv \frac{W(S,G)}{W(S)}, \qquad
  T(S|G) \equiv \frac{W(S,G)}{W(G)},
  \eqlabel{TransitionProbabilities}
\end{equation}
where $W(G) \equiv \sum_S W(S,G)$.
As stated above, 
a Markov process with these transition probabilities
yields the target distribution $W(S)$.

For the graph-assignment probability $T(G|S)$,
we can rewrite \Eq{TransitionProbabilities} using
\Eq{BoltzmannWeight} and \Eq{LocalFK} as
$T(G|S) = \prod_u t(G_u | S_u)$ with $t(G_u | S_u)$ being
\begin{equation}
  t(G_u|S_u) \equiv w(S_u,G_u)/w(S_u).
  \eqlabel{GraphAssignmentProbability}
\end{equation}
Since $T(G|S)$ is factorized into the local factors $t(G_u|S_u)$,
the graph assignment can be done locally;
we can assign a graph element to
each local unit independently 
with the probability $t(G_u | S_u)$.
For the Ising model, in particular, this 
transition probability is realized by the well-known
Swendsen-Wang procedure, i.e.,
connecting each nearest-neighbor pair of parallel
spins with the probability $1-e^{-2K}$ and 
leaving them unconnected otherwise.

For the spin-updating probability $T(S|G)$, we similarly obtain
\begin{equation}
  T(S|G) = 2^{-N_{\rm C}(G)} \Delta(S,G),
  \eqlabel{ClusterFlippingProbability}
\end{equation}
where $N_{\rm C}(G)$ is the number of connected clusters in $G$
and $\Delta(S,G)$ is the function that takes a value 1
if and only if all spins in each cluster are aligned in 
the same direction in $S$.
Therefore, this transition probability is realized by
the step (iii) in Swendsen and Wang's procedure.
Thus the validity of the procedure has been proved.

\deffig{Ising}{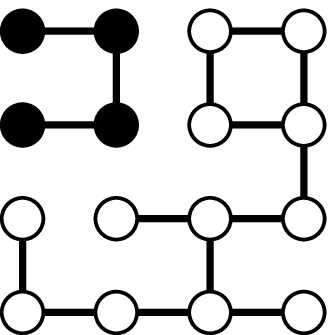}{0.2}{
A spin configuration $S$ and 
a matching graph $G$ of the Ising model.
The spin configuration is represented by the open and solid circles
whereas the graph consists of the lines connecting circles.
}

\subsection{Path Integral and Quantum Monte Carlo}
\ssclabel{PathIntegral}
The description of the SW algorithm given in the previous subsection
is quite general; the only model-specific part is
the first equality in \Eq{BoltzmannWeight}
and \Eq{LocalExtendedWeight}.
In fact, the loop algorithm for the quantum Monte Carlo
can be regarded as a special case of this framework.
As long as the target weight $W(S)$ can be expressed as
\Eq{GlobalFK} with some $W(S,G)$,
the transition probabilities \Eq{TransitionProbabilities}
constitute a valid algorithm (provided, of course, that
the ergodicity holds), regardless of the model we consider.
Therefore, the only that we have to do is to
specify ingredients such as $S$, $G$, and $W(S,G)$,
which we do in this subsection.

There are two ways of introducing $S$;
one by the path integral\cite{Suzuki1976}
and the other by the high-temperature series expansion\cite{Handscomb1962}.
While the latter leads to a discrete-time algorithm
with an exponentially small systematic error,
the former is simpler to describe.
In addition, both the representations reduce to the same 
algorithm in the continuous-time limit.
Therefore, we describe the framework starting from the path-integral
representation.
The formulation based on the high-temperature series expansion
is discussed in \mssc{SSE}.

For the derivation of the algorithm presented below,
it is often useful to consider the path integral in the
discretized imaginary time (though 
the discretization is not needed in the final algorithm).
Such an expression can be obtained as follows.
First we consider the identity,
\begin{equation}
  Z = \sum_{\psi} \left\langle \psi \left| 
      \lim_{L\to\infty} \prod_{k=1}^L
      \left(1-\frac{\beta}{L} \Ham \right)
      \right| \psi \right\rangle.
  \eqlabel{ProductRepresentation}
\end{equation}
In particular, when the Hamiltonian is a sum of $M$ terms,
\begin{equation}
  \Ham = \sum_{b=1}^M \Ham_b,
  \eqlabel{HamiltonianDecomposition}
\end{equation}
then, \Eq{ProductRepresentation} leads to 
$$
  Z = 
      \sum_{\psi} \left\langle \psi \left| 
      \lim_{L\to\infty} \prod_{k=1}^L \prod_{b=1}^M
      \left(1-\frac{\beta}{L} \Ham_b \right)
      \right| \psi \right\rangle.
$$
Here, the summation is taken over some complete orthonormal basis $\{\psi\}$
of the Hilbert space.
Inserting $1 = \sum_{\psi} |\psi\rangle\langle\psi|$
between two adjacent factors, we obtain
\begin{equation}
  Z = \lim_{L\to\infty} \sum_{\{\psi_b(k)\}} 
  \prod_{k=1}^L\prod_{b=1}^M
  \left\langle 
  \psi_{b+1}(k)\left|\left(1-(\Delta\tau) \Ham_b \right)\right| \psi_b(k)
  \right\rangle,
  \eqlabel{Quantum}
\end{equation}
where $\Delta\tau \equiv \beta/L$.
For simplifying the notation, we denote
$(b,k)$ as $u$, $\psi_b(k)$ as $\psi_u$, 
$\psi_{b+1}(k)$ as $\psi'_u$, $\Ham_b$ as $\Ham_u$, and
$\{\psi_b(k)\}$ as $S$.
It follows that
\begin{eqnarray}
  Z & = & \lim_{L\to\infty} \sum_{S} W_L(S),
  \eqlabel{PathIntegralPartitionFunction}\\
  W_L(S) 
  & \equiv & \prod_u w(S_u) ,\nonumber \\
  w(S_u) & \equiv &
  \left\langle \psi'_u \left|\left(1- (\Delta\tau) \Ham_u \right)\right| \psi_u \right\rangle.
  \eqlabel{BoltzmannWeightQ}
\end{eqnarray}
Thus the partition function is expressed as the sum of a weight
that is a function of a space-time configuration.
In the case of the particle systems, 
with the basis that diagonalizes the local number operators,
the space-time configuration is called a world-line configuration,
since the configuration is visualized by trajectories of particles
in the space-time.
The configuration for spin models is also called the world-line configuration
by regarding up-spins and down-spins as particles and holes respectively.
An example of the world-line configuration 
is shown in \Fig{DiscreteTimeWorldLine}.
The whole system consists of $L$ layers whereas each layer
contains $M$ sub-layers. 
(The number $L$ is called the Trotter number.)
The ``height'' of each layer is $\Delta\tau$
and the height of the whole system is always $\beta$
regardless of $L$.
Every $\Ham_b$ has its representative
in each layer, i.e., a unit $u$ called a {\it plaquette}.
Each of $M$ sub-layers contains a plaquette.

\deffig{DiscreteTimeWorldLine}{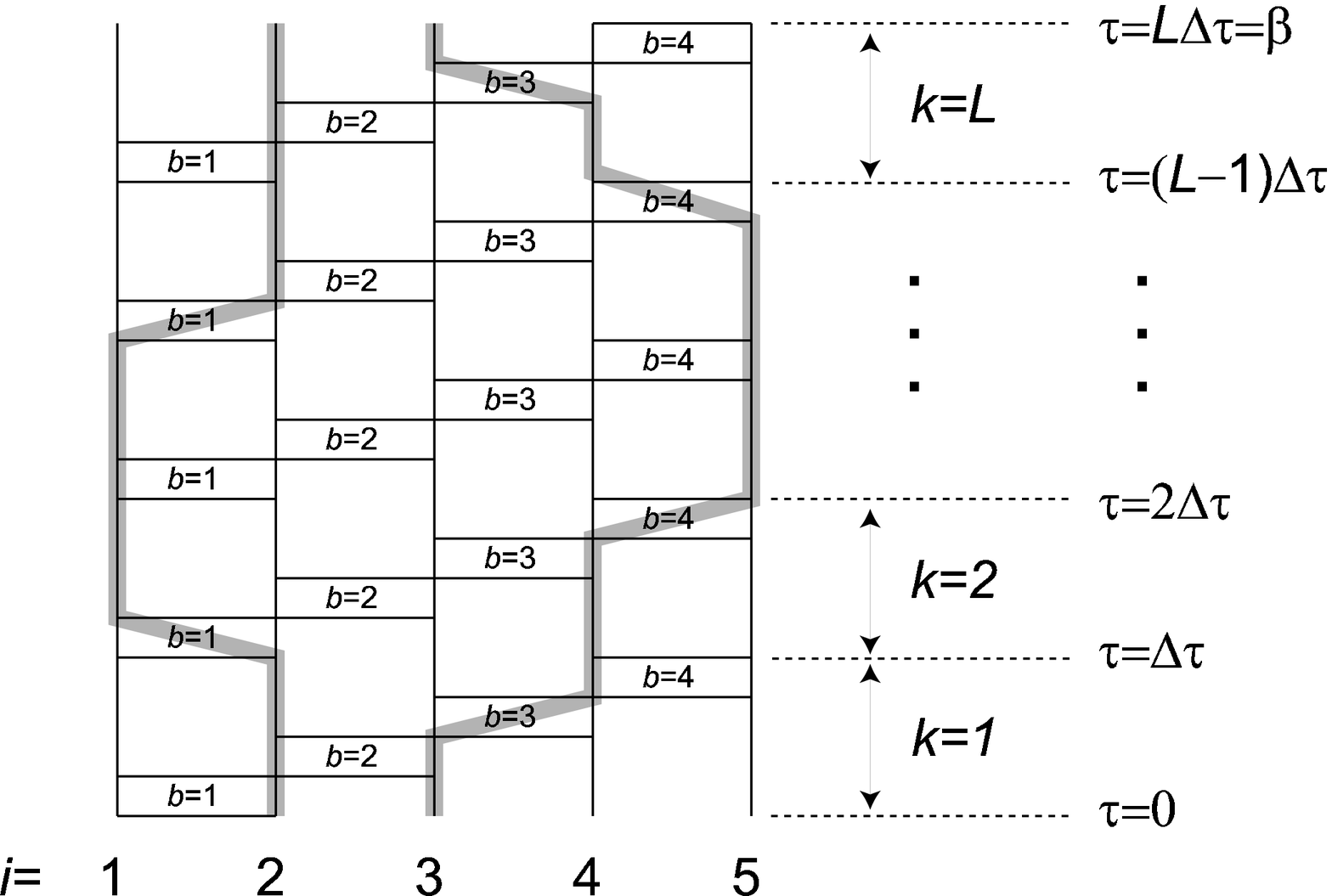}{0.4}{
A visualization of the path integral of a five-site system
with discrete imaginary time.
The world-lines are represented by two thick gray lines.
}

In early world-line Monte Carlo algorithms, updates of
a configuration were done in many steps,
\cite{Suzuki1976,SuzukiMK1977,Marcu1986}
each being a local update that modifies
only a small part of the system.
Before the loop algorithm, the unit of the local update
was a square whose spatial dimension equals 
the lattice spacing and the temporal dimension 
the discretization unit of time.
The square is shown in \Fig{LocalUpdate} together with
the world-line configurations before and after the update.
Because of the local nature of the updating unit,
the algorithm exhibits a severe slowing-down.
It happens when we approach a critical point or zero temperature.
This can be intuitively understood as the discrepancy of 
the physical correlation length and the spatial scale of the updating unit.
Another slowing-down, pointed out by Wiesler\cite{Wiesler1982},
when the temporal scale of the system
(i.e., the inverse evergy gap) largely differs from the temporal scale of the
updating unit.
The situation occurs when one decreases the discretization unit of
the imaginary time in order to reduce the systematic error due to the
discretization.
It was proposed\cite{NakamuraI2003} that 
this slowing-down can be removed by applying the loop
method only to the temperal direction .
The algorithms discussed below solve both types of slowing-down
in many cases of interest.

\deffig{LocalUpdate}{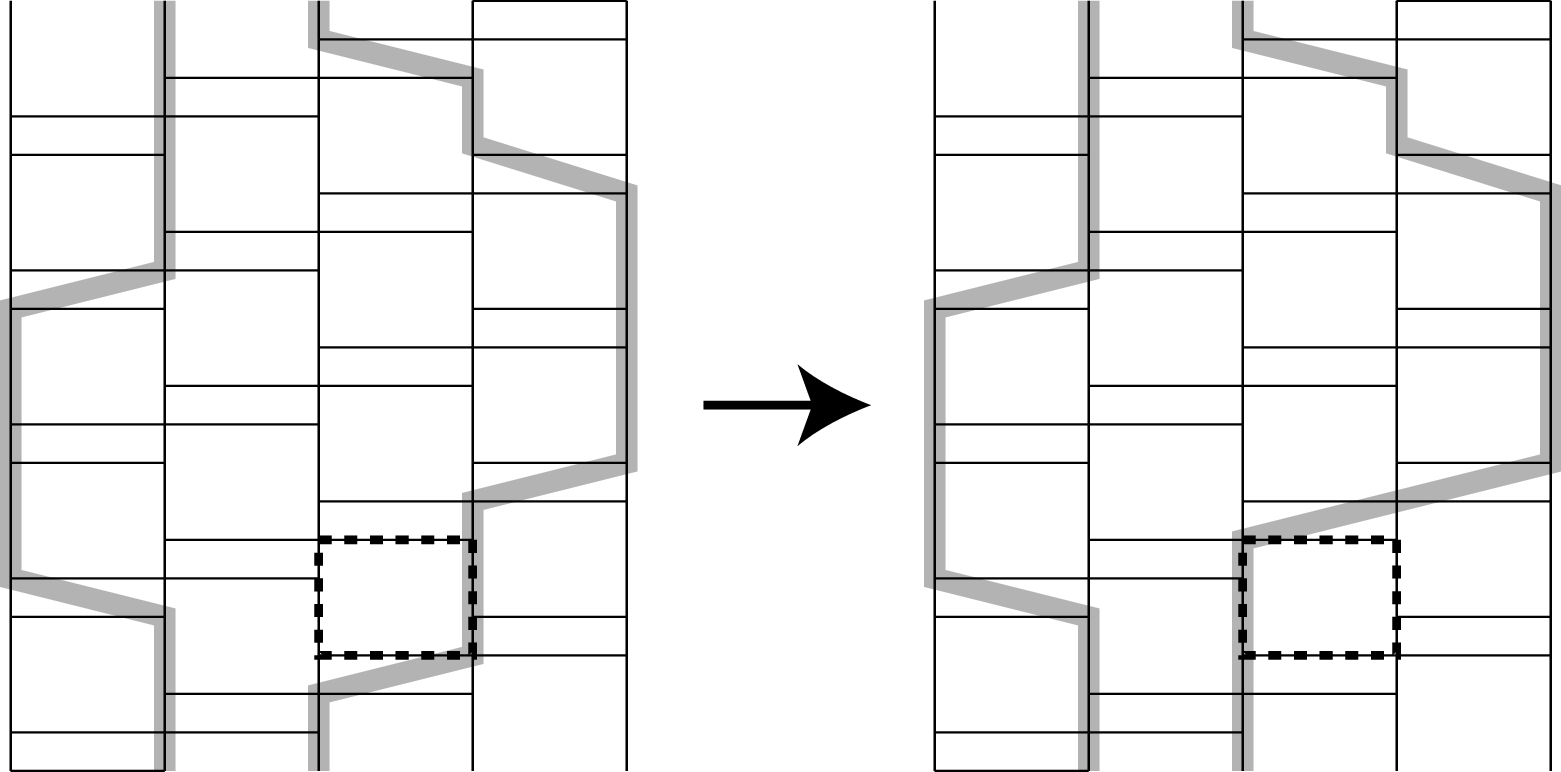}{0.4}{
One step in the local update algorithm.
The squares such as the one drawn with dashed lines 
are the units of the update.
At every step, one of the squares is chosen at random.
The flip of the chosen square is accepted with a probability
that depends on the weights of the configurations before and
after the flip.
}

\subsection{Loop Update}
\ssclabel{LoopUpdate}
A loop algorithm for a quantum system can be constructed 
in a similar fashion as the Swendsen-Wang algorithm
mentioned in \mssc{ClusterUpdate}.
That is, by introducing additional variables $G_u = G(b,k)=0,1$, 
we can rewrite $W_L(S)$ in \Eq{BoltzmannWeightQ} as
\begin{eqnarray}
  & & W_L(S)
    \equiv
    \prod_u \sum_{G_u=0,1}
    \left\langle \psi'_u \left|(-(\Delta\tau) \Ham_u)^{G_u} \right| \psi_u \right\rangle 
    \nonumber
    \\
  & & \quad =
    \sum_G \left(\Delta\tau\right)^{n(G)}
    \prod_u
    \left\langle \psi'_u \left|(- \Ham_u)^{G_u} \right| \psi_u \right\rangle 
    \eqlabel{TheWeight}
    \\
  & & \quad =
    \sum_G W_L(S,G),
    \eqlabel{GlobalFKQ}
\end{eqnarray}
where $G \equiv \{ G_u \}$, $n(G) \equiv \sum_u G_u$, and 
\begin{eqnarray}
  W_L(S,G) & \equiv & \prod_u w(S_u,G_u), \eqlabel{GEW} \\
  w(S_u,G_u) & \equiv & 
      \left\langle 
      \psi'_u \left|(-(\Delta\tau) \Ham_u)^{G_u}\right| \psi_u
      \right\rangle.
  \eqlabel{ExtendedWeightQ}
\end{eqnarray}
Since these expressions \Eq{GlobalFKQ} and \Eq{GEW}
have the same form as \Eq{GlobalFK} and \Eq{ExtendedWeight},
we can apply the prescription presented in \mssc{ClusterUpdate}
for defining the transition probabilities 
$T(G|S)$ and $T(S|G)$ through \Eq{TransitionProbabilities},
thereby constructing an algorithm 
of simulating a target distribution $W_L(S)$.
Since $W_L(S)$ is an approximation to $W(S)$,
such an algorithm can be used as an `approximate' algorithm
of simulating the distribution $W(S)$.
(In \mssc{ContinuousImaginaryTimeLimit}, 
we see that this `approximation' can be made exact.)

In order to complete the definition of the algorithm,
we have to specify the Hamiltonian, 
what orthonormal set we use, and
how we decompose it in \Eq{HamiltonianDecomposition}.
In what follows, we do these and examine what procedure
corresponds to the resulting transition probabilities.

First of all, we specify the meaning of the
decomposition \Eq{HamiltonianDecomposition} of the Hamiltonian.
We start with the graphical decomposition
\cite{KandelD1991,EvertzM1994,AizenmanN1994,KawashimaG1994,Kawashima1996,BrowerCW1998,HaradaK2001}
of the pair Hamiltonian $\Ham_{ij}$:
\begin{equation}
  \Ham_{ij} = \sum_{g} \Ham_{ij}(g), \quad
  \Ham_{ij}(g) = - a(g)\, \hat\Delta_{ij}(g),
  \eqlabel{GraphDecomposition}
\end{equation}
where $g$ specifies a type of a graph element,
$a(g)$ is some positive constant,
and $\hat\Delta_{ij}(g)$ is an operator whose matrix 
elements are 0 or 1.
As shown in \Tab{GraphElements}, 
a $\hat\Delta$-operator corresponds to 
a graph element with two types of lines,  each representing
a condition for making the matrix element 1.
A solid line connecting two spins represents the condition
that the two must be parallel whereas a dashed line 
requires that the two be anti-parallel.

In the case of the  $s=1/2$ anti-ferromagnetic Heisenberg model,
$$
  \Ham_{ij} = -J ( S^x_iS^x_j + S^y_iS^y_j - (S^z_iS^z_j-1/4) ),
$$
which we use in the following as an example,
the summation in eq.\ \Eq{GraphDecomposition} contains only one term:
\begin{equation}
  \Ham_{ij} = - \frac{J}{2} \hat\Delta_{ij}(g_{\rm H}),
  \eqlabel{GraphDecompositionAFH}
\end{equation}
where $g_{\rm H}$ is the graph element shown in the third graph
from the top in \Tab{GraphElements}.
Here, for the orthonormal complete set ($\{\psi\}$ in
\Eq{ProductRepresentation}), we have chosen
the set of the simultaneous eigenstates of 
the $z$-components of all spin operators,
as we do in most of the present article.
The operator $\hat\Delta_{ij}(g_{\rm H})$ 
is explicitly defined in terms of the matrix elements as
\begin{equation}
  \left\langle \sigma'_i \sigma'_j \left|
  \hat\Delta_{ij}(g_{\rm H})
  \right| \sigma_i \sigma_j \right\rangle
  =
  \delta_{\sigma_i,-\sigma_j}
  \delta_{\sigma'_i,-\sigma'_j}
  \eqlabel{MatrixElementOfHorizontalGraph}.
\end{equation}

\Equation{GraphDecomposition} results in the decomposition
of the total Hamiltonian as
$\Ham = \sum_{(ij)}\sum_{g} \Ham_{ij}(g)$.
Thus the Hamiltonian is decomposed in the form of 
\Eq{HamiltonianDecomposition} by
identifying $b$ and $((i,j),g)$, i.e., $u$ and $((i,j),g,k)$.
Then, the algorithm follows from 
the prescription given in \mssc{ClusterUpdate}.

For example, the graph assignment probability
$t(G|S)$ in \Eq{GraphAssignmentProbability} becomes
\begin{equation}
  t(1|S_u) = 
    (\Delta\tau) \times
    a(g) \left\langle\psi'_u \left| \hat\Delta_{ij}(g) \right| \psi_u \right\rangle
  \eqlabel{GAP}
\end{equation}
for $S_u$ being a non-kink, i.e., $\psi'_u = \psi_u$.
On the other hand, if $S_u$ is a kink, or $\psi'_u \ne \psi_u$, 
then $t(1|S_u) = 1$.
Thus, for the case of $s=1/2$ antiferromagnetic 
Heisenberg model, the probability is
\begin{equation}
  t(1|S_u) = 
  \left\{ \begin{array}{ll}
  1 & \mbox{($S_u$ is a kink.)} \\
  \frac{J}{2} \, (\Delta\tau) & \left(\mbox{
    $S_u = 
    {\tiny \left(\dudu\right)}$ 
    or 
    ${\tiny \left(\udud\right)}$ 
  }\right)\\
  0 & \mbox{(otherwise)}
  \end{array}\right..
  \eqlabel{GAPforAFH}
\end{equation}

Choosing the value 1 for $G_u$ means that we place a
graph of the type $g$ on the plaquette $u$.
For all the plaquettes with the value 0, we assign the `identity',
or `trivial' graph (the top row in \Tab{GraphElements}) representing 
the identity operator.
In what follows, we call a plaquette on which a non-trivial graph-element
is assigned a {\it vertex}.

The procedure that realizes the probability $T(S|G)$ is 
the same as that in the SW algorithm;
first identify the points connected by the lines
and then flip each cluster with probability 1/2.
(When applied to the isotropic Heisenberg model, 
the graph elements $g_{\rm CB}$ or $g_{\rm HB}$ 
in \Tab{GraphElements} do not appear,
and the resulting clusters are simple loops.
The name of the algorithm follows from this fact.
In the present paper, we use the name even for the cases
where clusters are not simple loops.)
A `space-time' point $(i,k)$ plays 
the same role as a site $i$ in the SW algorithm.
An example of one step in the loop algorithm is depicted
in \Fig{LoopUpdate} for the $s=1/2$ antiferromagnetic
Heisenberg model in one dimension.

\begin{table}[t]
\caption{
The graph elements, 
and the matrix elements of the delta operators $\hat\Delta_{ij}(g)$
corresponding to each element.
The base vectors of the two-spin Hilbert space are 
$|\frac12,\frac12\rangle$,
$|\frac12,-\frac12\rangle$,
$|-\frac12,\frac12\rangle$, and
$|-\frac12,-\frac12\rangle$.
To emphasize the difference in the constraints indicated by
the lines, dashed lines are used when the connected spins must
be anti-parallel to each other
whereas solid lines connect parallel spins.
\tablabel{GraphElements}
} 
\ \\[-3mm]
\begin{tabular}{c|c|c}
\hline
\rule{0mm}{8mm} Symbol & 
Graph & 
$\left\langle \sigma'_i \sigma'_j \left| \hat\Delta_{ij}(g) 
\right| \sigma_i \sigma_j \right\rangle $ \\[5mm]
\hline
\hline
\rule{0mm}{10mm}
$g_{\rm I}$ & 
\rule{3mm}{0mm}\parbox{12mm}{\includegraphics[width=12mm]{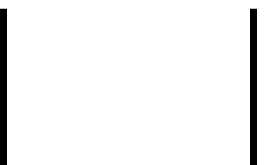}}\rule{3mm}{0mm} &
${\displaystyle 
\left(\begin{array}{cccc}
 1 & 0 & 0 & 0 \\
 0 & 1 & 0 & 0 \\
 0 & 0 & 1 & 0 \\
 0 & 0 & 0 & 1 
\end{array}\right)
}$ 
\\[6mm]
\hline
\rule{0mm}{10mm}
$g_{\rm C}$ & 
\rule{3mm}{0mm}\parbox{12mm}{\includegraphics[width=12mm]{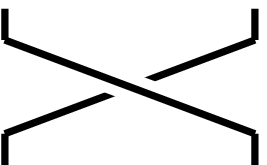}}\rule{3mm}{0mm} &
${\displaystyle 
\left(\begin{array}{cccc}
 1 & 0 & 0 & 0 \\
 0 & 0 & 1 & 0 \\
 0 & 1 & 0 & 0 \\
 0 & 0 & 0 & 1 
\end{array}\right)
}$ 
\\[6mm]
\hline
\rule{0mm}{10mm}
$g_{\rm H}$ & 
\rule{3mm}{0mm}\parbox{12mm}{\includegraphics[width=12mm]{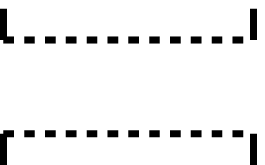}}\rule{3mm}{0mm} &
${\displaystyle 
\left(\begin{array}{cccc}
 0 & 0 & 0 & 0 \\
 0 & 1 & 1 & 0 \\
 0 & 1 & 1 & 0 \\
 0 & 0 & 0 & 0 
\end{array}\right)
}$ 
\\[6mm]
\hline
\rule{0mm}{10mm}
$g_{\rm CB}$ & 
\rule{3mm}{0mm}\parbox{12mm}{\includegraphics[width=12mm]{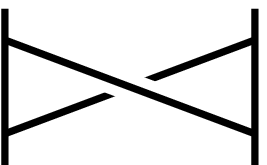}}\rule{3mm}{0mm} &
${\displaystyle 
\left(\begin{array}{cccc}
 1 & 0 & 0 & 0 \\
 0 & 0 & 0 & 0 \\
 0 & 0 & 0 & 0 \\
 0 & 0 & 0 & 1 
\end{array}\right)
}$ 
\\[6mm]
\hline
\rule{0mm}{10mm}
$g_{\rm HB}$ & 
\rule{3mm}{0mm}\parbox{12mm}{\includegraphics[width=12mm]{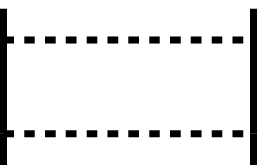}}\rule{3mm}{0mm} &
${\displaystyle 
\left(\begin{array}{cccc}
 0 & 0 & 0 & 0 \\
 0 & 1 & 0 & 0 \\
 0 & 0 & 1 & 0 \\
 0 & 0 & 0 & 0 
\end{array}\right)
}$ 
\\[6mm]
\hline
\end{tabular}
\end{table}

\deffig{LoopUpdate}{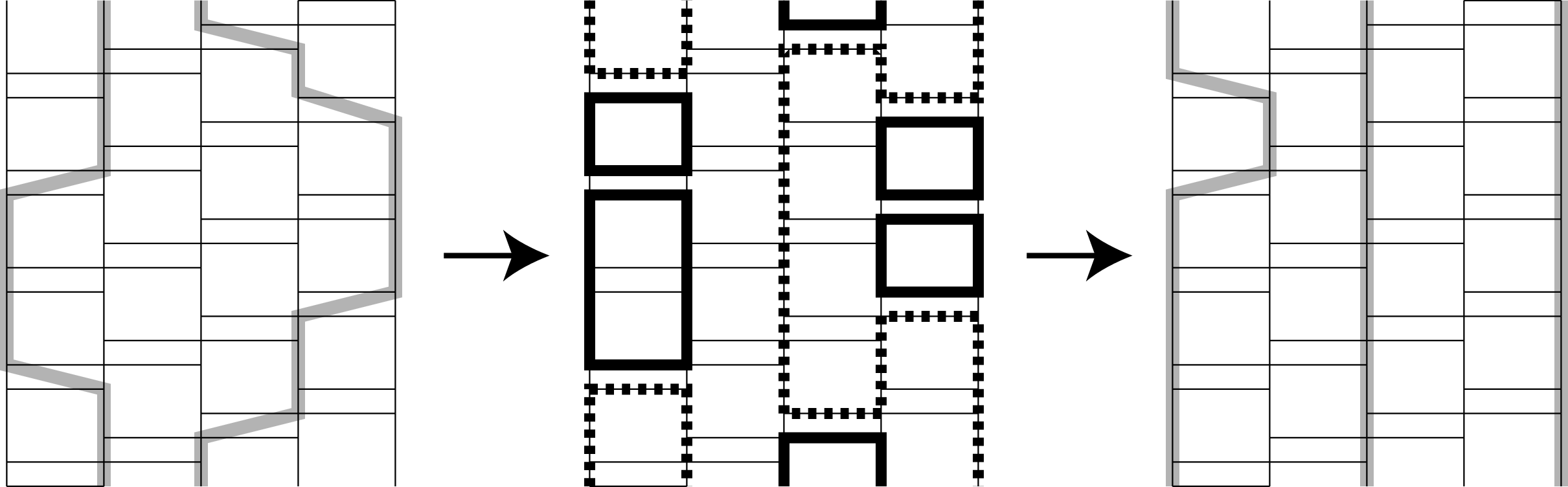}{0.45}{
A loop update.
The decomposition into loops (from the left to the middle)
and the flipping of the loops (from the middle to the right) 
are shown for the $s=1/2$ antiferromagnetic Heisenberg model 
in one dimension.
The only non-trivial graph elements are the `horizontal' ones,
$g_{\rm H}$, in \Tab{GraphElements}.
The loops flipped in the transition from the middle diagram to
the right are indicated by dashed lines in the middle diagram 
whereas other loops are drawn with solid lines.
}

It is useful to see what kind of loops and clusters are formed
\cite{EvertzLM1993,EvertzM1994,Kawashima1996} in various other cases.
We consider the $XYZ$ model
described by the Hamiltonian
\begin{eqnarray}
  \Ham & = & \sum_{(ij)} {\Ham}_{{ij}}, \nonumber \\
  \Ham_{ij} & = & c - J_x S^x_i S^x_j - J_y S^y_i S^y_j - J_z S^z_i S^z_j,
  \eqlabel{XYZHamiltonian}
\end{eqnarray}
where $c$ is a constant.
As for the orthonormal complete set,
we take the set of the simultaneous eigenstates of 
the $z$-components of all spin operators as above.
Therefore, a basis vector $\psi$ can be uniquely specified by
the eigenvalues of the $N$ operators, $S^z_1, S^z_2, \cdots, S^z_N$.
In this representation, when $J_x$ and/or $J_y$ are negative,
the off-diagonal matrix elements of $-\Ham$ may be negative.
For the bipartite lattices, however,
the number of the negative matrix elements in the whole configuration
is even, which makes the weight $W(S)$ always positive.
Another way of seeing this\cite{MakivicD1991} is to divide the whole lattice into
two sub-lattices, A and B, so that a site on the sub-lattice A is 
surrounded by sites on the sub-lattice B, and rotate the spins on the sub-lattice B.
For example, when $J_x = J_y < 0$, we rotate spins on the sub-lattice B
around the $z$-axis, so that $S^x_i \to -S^x_i$ and $S^y_i \to -S^y_i$.
This rotation makes all the off-diagonal elements positive.
In what follows, therefore, we consider the cases with no negative-sign 
problem and assume that all the off-diagonal matrix elements
of $-\Ham_{ij}$ are non-negative.

Then, the pair Hamiltonian $\Ham_{ij}$, eq.\ \Eq{XYZHamiltonian}, with
$J_x = J_y > 0$ can be expressed with two graph elements.
We can see this in the following graphical decomposition analogous
to \Eq{GraphDecompositionAFH},
$$
  -\Ham_{ij} = 
  \left\{ \begin{array}{lc}
    \frac{J_x}{2}    \hat\Delta_{ij}(g_{\rm C})
  + \frac{J_z-J_x}{2} \hat\Delta_{ij}(g_{\rm CB})
  & \mbox{(I)}
  \\[5mm]
    \frac{J_x+J_z}{4} \hat\Delta_{ij}(g_{\rm C})
  + \frac{J_x-J_z}{4} \hat\Delta_{ij}(g_{\rm H})
  & \mbox{(II)}
  \\[5mm]
    \frac{J_x}{2} \hat\Delta_{ij}(g_{\rm H})
  + \frac{-J_z-J_x}{2} \hat\Delta_{ij}(g_{\rm HB})
  & \mbox{(III)}
  \end{array} \right.,
$$
where the constant $c$ in \Eq{XYZHamiltonian} has been chosen 
so that the matrix elements are positive in each form.
Since we need an expression of the form
\Eq{GraphDecomposition} with positive $a(g)$,
only one of these three expressions can be used 
for a particular set of the values of $J_x$ and $J_z$.
The form (I) can be used for
the easy-axis ferromagnetic model ($0 < J_x = J_y \le J_z$),
the form (II) for the easy-plane model ($0 \le |J_z| < J_x = J_y$), and
the form (III) for the easy-axis antiferromagnetic model 
($0 < J_x = J_y \le -J_z$).
The five types of graph elements shown in \Tab{GraphElements}
are sufficient for expressing the pair Hamiltonian in all the three cases.

The second and the fourth elements in \Tab{GraphElements}
are required for expressing 
the pair Hamiltonian of the easy-axis ferromagnetic model (the case (I)).
The fourth graph-element binds all the four spins 
$\sigma_i$, $\sigma_j$, $\sigma'_i$, and $\sigma'_j$.
Therefore, in this case,
a resulting cluster of spins that is to be flipped simultaneously
is not generally a single loop but a number of loops bound together.
On the other hand, the second and the third graph elements are
sufficient for expressing the pair Hamiltonian of the
easy-plane model (the case (II)), such as the $XY$ model.
In either graph element, the four spins are bound only pair-wise.
Therefore, a graph $G$ consists of loops in this case.
For the easy-axis antiferromagnetic model (the case (III)),
the graph elements required are the third and the last.
Therefore, in this case, a cluster is not a single loop,
in general, similar to the case (I).
For more details of the algorithm and the case with 
a lower symmetry (i.e., the $XYZ$ model), see \msec{Recipe}.
(A sample program may be found at a web-site.\cite{WebSite})

Many applications of the loop updating method have been done.
Here, we only show the result for the quantum $s=1/2$ $XY$
model in two dimensions\cite{HaradaK1997, HaradaK1998}, 
which clearly demonstrates the utility of the loop algorithm.

It is well-known that the helicity modulus exhibits the 
universal jump at the Kousterlitz-Thouless type phase transition.
\cite{NelsonK1977}
The system-size dependence of the quantity near the critical point
is also predicted theoretically.
In the quantum spin model, such as the $s=1/2$ $XY$ model,
the helicity modulus is related
to the fluctuation in the total winding number of 
the world-lines by 
$
  \Upsilon = (T/2) \langle {\bf W}^2 \rangle,
$
\cite{PollockC1987}
where ${\bf W} \equiv (W_x,W_y)$ with $W_x$ ($W_y$) being the total
winding number in the $x$ ($y$) direction.
Therefore, we can estimate the critical temperature accurately
by measuring the winding number.
In \Fig{Helicity}, the raw data of the helicity modulus is shown.
We can see the universal jump even in the raw data and
obtain a rough estimate of the transition temperature 
$T_{\rm KT} \sim 0.35$.
We can obtain a much more precise estimate for the critical temperature
by fitting the data to the theoretically predicted
form of the size dependence.
The best estimate of the transition temperature has been obtained
in this way.
Note that it is difficult to estimate the transition temperature
by means of a conventional world-line quantum Monte Carlo method,
such as the one shown in \Fig{LocalUpdate}.
This is because the auto-correlation time becomes too long
as we approach the critical temperature from above.
Unlike the ordinary phase transition, it is increasingly more
difficult to equilibrate the system even after passing the
transition temperature since the system remains critical
in the whole low-temperature region.
Therefore, with conventional methods, we can obtain
reliable estimates of various quantities only in the high-temperature region.
Another reason that makes difficult the estimation by 
the local update is that it does not yield an ergodic algorithm;
the winding number of world-lines is not allowed to vary. 
Therefore, one must observe other quantities,
for which the size dependence is known less precisely, or 
introduce some additional global updates for making 
the winding number vary.
The latter was done in an early simulations,\cite{MarcuW1985}
and later in a simulation of a bosonic system\cite{Hatano1995}.
However, these additional global flips tend to form 
bottlenecks in the configuration space, slowing down the
whole simulation.

\deffig{Helicity}{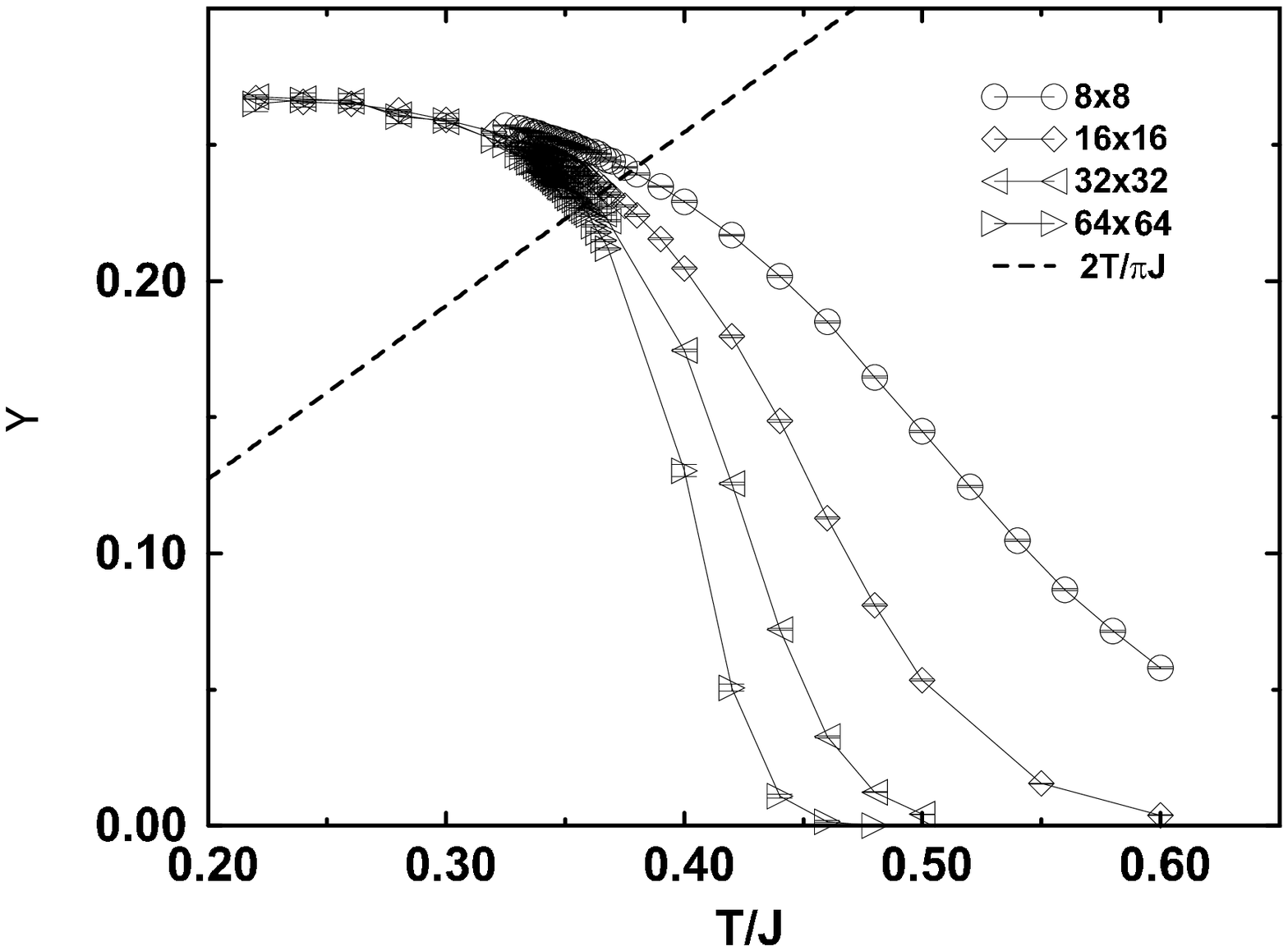}{0.45} {
  The helicity modulus (or the superfluid
  density) $\Upsilon = (T/2) \langle {\bf W}^2 \rangle$ as a function
  of the temperature.  The universal jump is expected at the point where
  $\Upsilon = 2T/\pi J$.  The error bars are drawn, but most of them are
  too small to be recognized.
  (Adopted from Harada and Kawashima\cite{HaradaK1997}.)
}

\subsection{Formulation Based on the Series Expansion}
\ssclabel{SSE}
So far we have been using the approximation of the imaginary-time
discretization.
While we can use the finite-$L$ expressions for constructing an approximate
algorithm in order to obtain all the results that we need,
the results would come with a systematic error due to the imaginary-time 
discretization.
Therefore, we would have to do an extrapolation to get the final result free 
from this systematic error.
However, there are two ways to get rid of the discretization error.
In the first method, 
which we discuss in \mssc{ContinuousImaginaryTimeLimit},
we perform the extrapolation to the continuous
imaginary time in the algorithm, not in the numerical results.
In other words, there exists a computational procedure that operates 
directly on continuous degrees of freedom 
(on the floating-point variables, to be precise).
In this method there is no discretization error.
In the present section, on the other hand, we present a method with 
discretized degrees of freedom that yields an algorithm with 
a much smaller error (negligible for most purposes) than the naive
discretized-time method.

The formulation is based on the high-temperature series expansion,
and is originated in Handscomb's method\cite{Handscomb1962}.
It was later elaborated by Sandvik and coworkers.
\cite{SandvikK1991,Sandvik1992,Sandvik1999}
It starts from the expansion of the partition function,
$$
  Z = \lim_{L\to\infty} Z_L,
$$
where
$$
  Z_L \equiv \sum_{n=0}^{L} \frac{\beta^n}{n!} {\rm Tr} (-\Ham)^n.
$$
Then, we visualize each term by considering $L$ ``boxes'' and 
put $n$ ``marbles'', each corresponds to $-\Ham$, into these boxes.
Each box can contain one marble at most.
Therefore, there are
$\left(\begin{array}{c} L \\ n \end{array}\right)$ distinct
pictures corresponding to the same term.
Thus we have
$$
  Z_L = \sum_{\{\gamma_k\}} \beta^{(\sum_{k}\gamma_k)}\  \frac{(L-n)!}{L!}
  \Tr{\prod_{k=1}^L (-\Ham)^{\gamma_k}},
$$
where $\gamma_k = 0,1$ represents a filled or an empty box, respectively.
When the Hamiltonian is decomposed into a product of local factors
as in \Eq{HamiltonianDecomposition}, we can rewrite the above as
$$
  Z_L = \sum_G \beta^{n(G)} \frac{(L-n(G))!}{L!} \Tr{\prod_u (-\Ham_u)^{G_u}},
$$
where $u \equiv (b,k)$, $\Ham_u \equiv \Ham_b$, 
$G_u = 0,1$, $G \equiv \{ G_u \}$ and $n(G)\equiv\sum_u G_u$.
The summation is restricted to the graphs $G$ such that
$\sum_{b} G_{(b,k)} = \gamma_k \le 1$.
As we did in the path-integral formulation to obtain \Eq{Quantum},
we can insert the identity operators expanded in the 
orthonormal complete set to obtain,
$$
  Z_L     =  \sum_S W_L(S),
$$
\begin{equation}
  W_L(S) = \sum_{G} \beta^{n(G)} \frac{(L-n(G))!}{L!} 
  \prod_u \langle \psi'_u | (-\Ham_u)^{G_u} | \psi_u \rangle.
  \eqlabel{SE}
\end{equation}

Apart from the factor $(L-n(G))!/L!$ and the restriction on
the summation, eq.\ \Eq{SE} looks similar to \Eq{TheWeight}.
In fact, when $L \gg 1$, the difference in the factor is small because
$\beta^n (L-n)!/L!$ is approximately equal to $(\beta/L)^n$.
In addition, for large $L$,
the difference due to the restriction produces
only a small difference, because 
the typical value of $n(G)$ in an actual simulation should not depend on 
$L$ (as long as $L$ is large enough) and therefore
having more than one vertices in the same layer is 
an increasingly rare event for large $L$
even if such an event is allowed.
Therefore, eq.\ \Eq{SE} derived from the
series expansion is approximately equal to \Eq{TheWeight} 
derived from the path-integral formulation for the same $L$,
and they become identical in the large-$L$ limit.
This means that the algorithms that follow are similar for a finite $L$
and identical for the infinite $L$.
The important difference, however, is that the discretization error
for a finite $L$ is exponentially small for \Eq{SE} whereas
that for \Eq{TheWeight} vanishes only algebraically.
Therefore, in the path-integral formulation, we need to take the
infinite-$L$ limit whereas in the series-expansion formulation,
it is not necessary as long as $L$ is large enough.

All the algorithms, such as the loop algorithm and 
the directed-loop algorithm discussed below, can be derived from
the series-expansion formulation as well as the path-integral
formulation, and in most cases
the mapping from an algorithm derived from the latter
to the one derived from the former is straightforward.
While a ``discretized-time'' algorithm based 
on the series-expansion representation can be advantageous 
for efficient implementation
(because only integral variables are needed there),
we discuss in what follows a number of algorithms 
using the path-integral formulation to avoid 
the factor $(L-n)!/L!$ appearing in the expressions.

\subsection{Continuous-Imaginary-Time Limit}
\ssclabel{ContinuousImaginaryTimeLimit}
The loop algorithm is useful not only 
in speeding up the simulation but also in taking the $L\to \infty$ limit
in the algorithm\cite{BeardW1996}, which makes the algorithm free from the 
systematic error due to the discretization of the imaginary time.

For a large $L$, the target distribution $W_L(S)$ of the spin
configuration mimics the distribution 
$W(S) \equiv \lim_{L\to\infty} W_L(S)$. 
It means that if we look at
a configuration with a poor resolution in the imaginary time,
we cannot tell whether the configuration is generated with
the weight $W(S)$ or $W_L(S)$.
Therefore, when we have a finite correlation time $\xi_{\tau}$
in the target distribution $W(S)$,
we do not have a kink at which a state changes, i.e.,
$\psi_{b+1}(k) \ne \psi_b(k)$
in (typically) $\xi_{\tau}/\Delta\tau$ consecutive layers.
Let us consider an imaginary-time interval of the length $I$
that includes many layers with no kink.
As can be seen in \Eq{GAP},
we assign a graph element of type $g$ with probability
$(\Delta\tau) a(g)$ to a unit $u$
when $S_u$ makes the matrix element of $\hat\Delta(g)$ unity.
Since there are $I/\Delta\tau$ layers in this interval,
the probability of assigning $n$ graph elements of type $g$
to this interval is given by
$$
  \left(\begin{array}{c} I/(\Delta\tau) \\ n \end{array}\right)
  \left( (\Delta\tau) a(g) \right)^n
  \left( 1- (\Delta\tau) a(g) \right)^{I/(\Delta\tau) - n}.
$$
In the continuous-time limit $L\to\infty$ this reduces to
$$
  \frac{1}{n!}\left( I a(g) \right)^n e^{-I a(g)}.
$$
This is nothing but the Poisson distribution with mean $I a(g)$.
Therefore, instead of repeating the graph assignment 
procedure for all the plaquettes,
we can generate a number $n$
with the Poisson distribution with mean $I a(g)$,
and choose $n$ points from $I$ uniform-randomly.
The result would be statistically the same as what we would 
obtain from the discrete-time procedure described in 
\mssc{LoopUpdate} (with extremely large $L$).

\deffig{ContinuousTime}{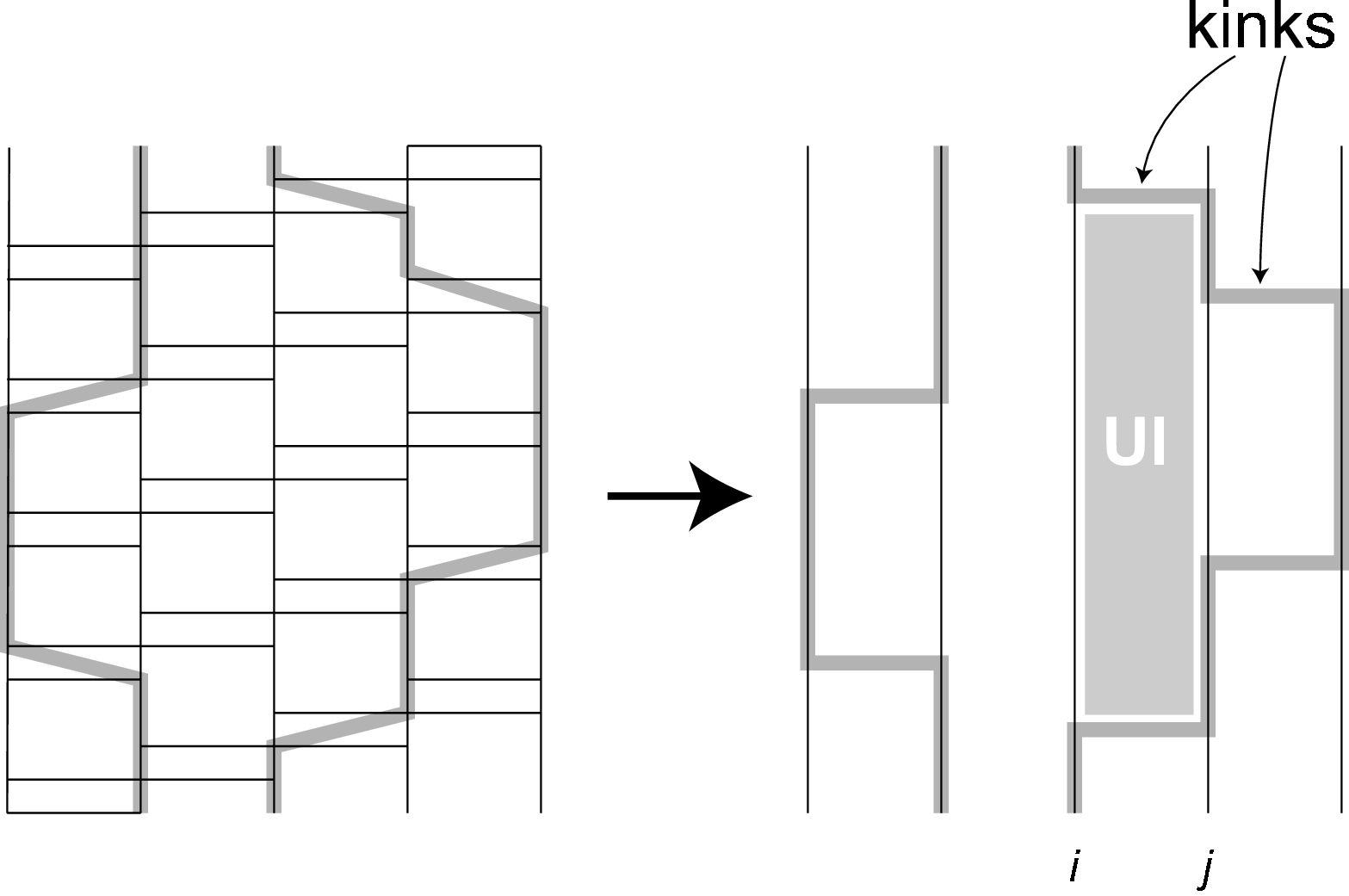}{0.4}{
The correspondence between the world-line configuration 
in the discrete time and that in the continuous time.
When, a loop algorithm is used,
a real-time ``animation'' on the computer screen would
look the same for both the representations as long as the
imaginary time step $\Delta\tau$ in the discrete representation 
is small enough.
A uniform interval (UI) is indicated as a lightly shaded region.
}

Another advantage of the continuous-imaginary-time algorithms
is that we do not have to deal with the fine structure of the `space-time'.
For example, the time ordering of the plaquettes with different
$b$ in each layer (such as the one shown in \Fig{DiscreteTimeWorldLine})
can be arbitrary, because
in the continuous-time limit, individual plaquettes do not appear
and therefore the order of them does not matter at all.

Since we consider the $\Delta\tau\to 0$ limit in what follows,
the height of a plaquette is zero, i.e., a plaquette 
in the discrete time corresponds to a horizontal line.
We call the horizontal line (plaquette) 
on which a non-trivial graph is placed a {\it vertex}.
The four corners of a plaquette are called {\it legs}.
(See \Fig{LatticeAndObjects} for the names of
various objects.)

It may be helpful to summarize here the procedure of one Monte
Carlo step with the continuous-imaginary-time loop algorithm.
Starting from an arbitrary pair of $S$ and $G$ that match each other,
first we remove all the vertices (i.e., graph elements)
at which there is no kink.
Next, for each pair of the nearest neighbor sites $(ij)$, 
we decompose the interval $(0,\beta)$ into uniform intervals (UI). 
(Here, a UI for a pair of sites $(i,j)$, is an imaginary-time interval 
delimited by two kinks that involves one or both of $i$ and $j$.
(See \Fig{ContinuousTime}))
For each UI, which we denote as $I$, and for each kind of graph elements,
which we denoted as $g$,
we generate an integer $n$ with the Poisson distribution of mean $Ia(g)$,
and place $n$ graph elements of the type $g$ uniform-randomly in $I$.
When this is done for all types of graphs, all the uniform intervals, 
and all the nearest neighbor pairs, 
we identify loops, or clusters.
Finally we flip each loop (cluster) with probability $1/2$.

\subsection{Large Spins}
\ssclabel{LargeSpins}
The generalization of the loop algorithm to larger 
(i.e., higher) spins can be done by replacing each spin operator 
by the sum of $2s$ $s=1/2$ spins\cite{KawashimaG1994}.
That is, we replace the spin operators in \Eq{XYZHamiltonian} as 
\begin{equation}
  S^{\alpha}_i \Rightarrow 
  \tilde S^{\alpha}_i \equiv \sum_{\mu=1}^{2s} \sigma^{\alpha}_{i,\mu}
  \qquad (\alpha = x,y,z), \eqlabel{SpinDecomposition}
\end{equation}
where each spin operator $\vect{\sigma}_i$ carries $s=1/2$.
Accordingly, a basis vector is specified by eigenvalues of 
the $2sN$ operators $\{ \sigma^z_{i,\mu} \}$ ($i=1,2,\cdots,N$
and $\mu=1,2,\cdots,2s$).
In what follows, we identify the label $\psi_b(k)$
with a set of $2sN$ variables $\{ \sigma_{i,\mu} \}$,
where $\sigma_{i,\mu} = \pm\frac12$ denotes 
an eigenvalue of $\sigma^z_{i,\mu}$.
The new Hilbert space spanned by these vectors
has the dimension $2^{2sN}$,
somewhat larger than the original one which 
is spanned by only $(2s+1)^N$ basis vectors.
Therefore, we have to eliminate many states in order to obtain the correct
partition function of the original model.
This can be achieved by introducing the projection operator $\hat P$
\cite{HaradaTK1998, TodoK2001}, i.e.,
\begin{equation}
  Z 
= \Tr{ e^{-\beta\Ham\left(\left\{\vect{S}_i\right\}\right)} }
= \Tr{ \hat P e^{-\beta\Ham\left(\left\{\tilde{\vect{S}}_i\right\}\right)} }.
\end{equation}
The projection operator $\hat P$ eliminates all the states
that do not have corresponding states in the original problem,
such as the singlet states in the $s=1$ problem.

When the original spins are split into $s=1/2$ spins,
the pair Hamiltonian $\Ham_{ij}$ can be written as
$$
  \Ham_{ij} = \sum_{\mu=1}^{2s}\sum_{\nu=1}^{2s} \Ham_{i\mu,j\nu},
$$
where $\Ham_{i\mu,j\nu}$ is the pair Hamiltonian that can be obtained
by replacing $\vect{S}_i$ and $\vect{S}_j$ by 
$\vect{\sigma}_{i\mu}$ and $\vect{\sigma}_{j\nu}$, respectively,
in the definition \Eq{XYZHamiltonian}.
The pair Hamiltonian $\Ham_{i\mu,j\nu}$ is nothing but the 
pair Hamiltonian of the $s=1/2$ model discussed above.
It is thus obvious that we can apply the general prescription 
described in \mssc{PathIntegral}, \mssc{LoopUpdate} and
\mssc{ContinuousImaginaryTimeLimit}
simply by re-interpreting $b$ as $((i,j,\mu,\nu),g)$
instead of $((i,j),g)$.
As a result we have 
$2s$ vertical lines for each site $i$ as illustrated
in \Fig{LoopAlgorithmS1}.
The graph-assignment procedure must be repeated 
$(2s)^2$ times corresponding to $(2s)^2$ pairs of the indices
$\mu$ and $\nu$.
The procedure is otherwise identical to the one for the $s=1/2$ model.
The types of the graphs and the graph-assignment density are
exactly the same as the corresponding $s=1/2$ model.

We can handle the projection operator $\hat P$
through a graphical decomposition as we do for the Hamiltonian.
It should be noted that the operator $\hat P$ projects the extended
Hilbert space onto the sub-space that is isomorphic to
the original Hilbert space.
This sub-space consists of the simultaneous eigenstates
of the Cashmir operators $(\vect{S}_i)^2$.
The states must be symmetric in the $\mu$ space for each site.
In other words, the state is invariant under any permutation
of the split spins 
$(\sigma_{i,1}, \sigma_{i,2}, \cdots, \sigma_{i,2s})$ for each $i$.
Therefore, the projection operator is
a product of local projection operators, and each local
projection operator can be expressed as the sum of permutation operators;
$$
  \hat P = \prod_i \hat P_i, \qquad
  \hat P_i = \frac{1}{(2s)!} \sum_{\pi} \hat \Delta_{\pi}.
$$
Here, the summation is taken over the set of permutations among
the split spin indices $\mu = 1,2,\cdots,2s$,
and $\hat \Delta_{\pi}$ is an operator that generates the permutation $\pi$.
Specifically,
$$
  \left\langle \tilde\psi' \right| \hat\Delta_{\pi} \left| \tilde\psi \right\rangle
  \equiv \prod_{\mu =1}^{2s} \delta_{\sigma'_{i,\pi(\mu)},\sigma_{i,\mu}},
$$
where $\tilde\psi = (\sigma_{i,1},\sigma_{i,2},\cdots,\sigma_{i,2s})$.
This operator corresponds to a graph element that connects
a point on the vertical line specified by $(i,\mu)$ to
a point on the other line specified by $(i,\pi(\mu))$.
This correspondence is similar to that of the operator 
$\hat\Delta_{ij}(g)$ and the graph element $g$ as we see above.
Therefore, in order to take the projection operator into account,
we have only to include the following step in the updating procedure.
That is, after assigning graph elements to the vertices, we
assign a special graph element to the end points of the
$2s$ vertical lines for each site $i$.
Each graph element represents a particular permutation of
$2s$ spins and connects the end points of the $2s$ vertical lines
at $\tau = \beta$ to those at $\tau = 0$.
The graph element is chosen with equal probability from the ones
that are compatible to the current spin configuration
(at $\tau = \beta$ and $\tau = 0$).

Among a number of applications of the split-spin algorithm described in
this section, we briefly mention the calculation done by Todo and Kato,
\cite{TodoK2001} since it is illustrative of the high efficiency of the
algorithm. 
They computed the energy gap $\Delta$ between the ground state and
the first excited state, and the correlation length $\xi$
of the antiferromagnetic Heisenberg model in one dimension at $T=0$ for
$s=1,2$ and $3$. This system is known to exhibit the Haldane gap 
and is disordered even at zero temperature.
The correlation length for $s=1$ is about 6. 
Therefore, one can use the exact diagonalization for obtaining 
a rather accurate estimates of various quantities in this case.
However, since the inverse gap and the correlation length diverge
exponentially as the spin length increases,
it is increasingly difficult to obtain accurate estimates for larger spins.
They obtained the following estimates:
\begin{equation}
  \begin{array}{lll}
    \xi = 6.0153(3), & \Delta = 0.41048(6) & (s=1), \\
    \xi = 49.49(1),  & \Delta = 0.08917(4) & (s=2), \\
    \xi = 637(1),    & \Delta = 0.01002(3) & (s=3).
  \end{array}
\end{equation}
It is obvious from this result that we cannot compute
these quantities with the exact diagonalization method
for $s=2$ and $3$.
To our knowledge, these numbers are very difficult to compute
by any other methods than the ones described in this article.
The estimates for $s=2$, for example, are the best estimates known so far.
For the estimates for $s=3$, we are not aware of any other methods
that can compute them.

\deffig{LoopAlgorithmS1}{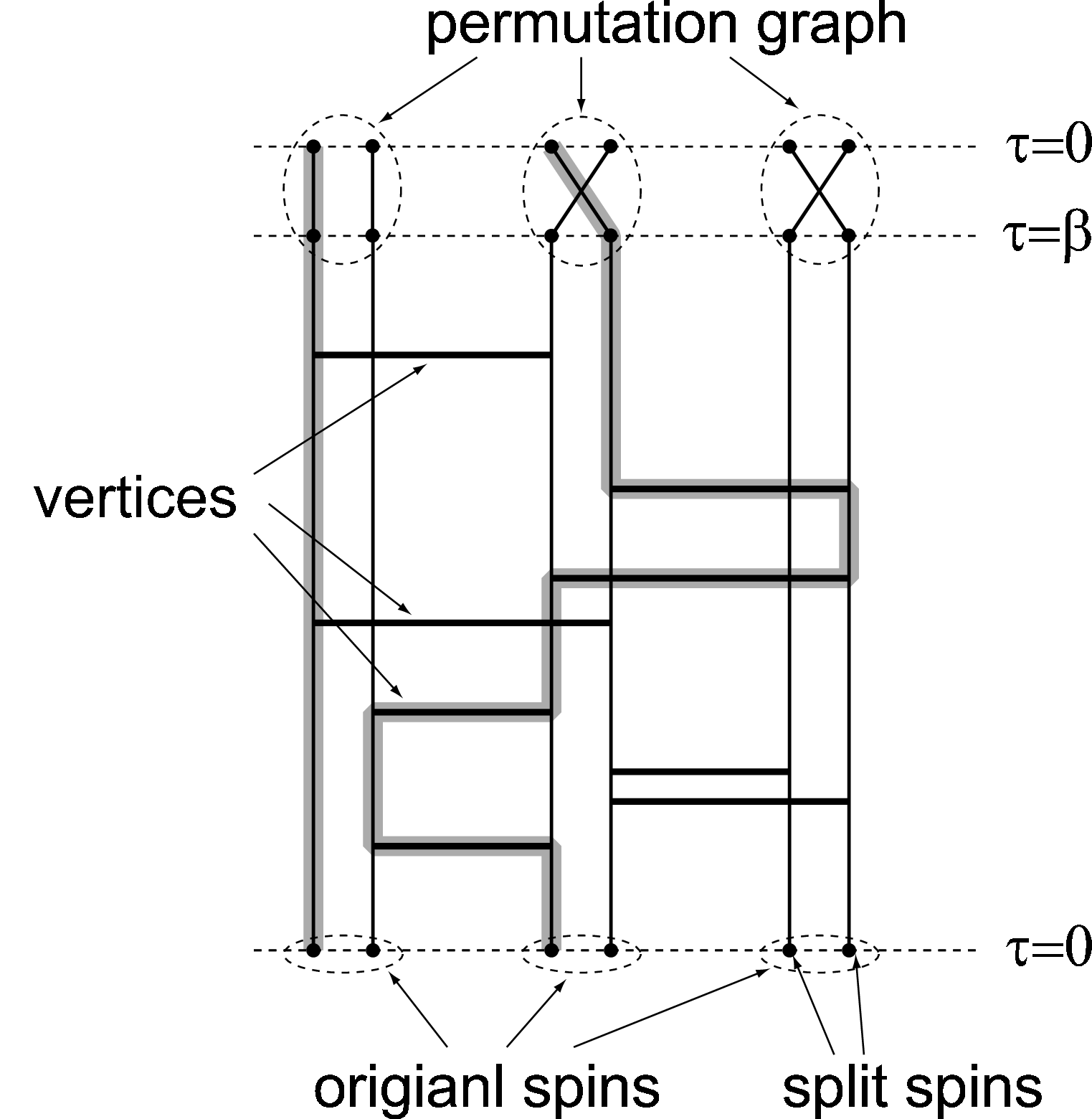}{0.4}{ 
  The split-spin representation of a world-line configuration
  for the $s=1$ quantum spin system.
}

\subsection{Loop Algorithms with Non-Binary Loops}
\ssclabel{NonBinaryLoops}
In some applications, it is advantageous to use non-binary loop variables.
For example, let us consider the bilinear-biquadratic interaction model
with $s=1$\cite{BIQ},
\begin{equation}
  \Ham = J \sum_{(ij)} (
  (\cos\theta) \vect{S}_i\cdot\vect{S}_j
  + (\sin\theta) (\vect{S}_i\cdot\vect{S}_j)^2
  ).
  \eqlabel{BilinearBiquadraticModel}
\end{equation}
Simulation of this model can be done with the split-spin method
described in \mssc{LargeSpins} with or without the coarse-graining
in \mssc{CoarseGrainedAlgorithm} and the
details can be found in \msec{Recipe}.
(For an application, see Harada and Kawashima\bigcite{HaradaK2002}.)
In what follows, however, we consider an alternative algorithm which
is particularly useful in dealing with special cases with higher symmetry.

The model \Eq{BilinearBiquadraticModel} obviously has the SU(2) symmetry.
At $\theta = \pm\pi/2$ and $\theta = \pm\pi/4$, however,
it possesses a higher symmetry than is obvious from the definition.
Here we consider the case $\theta = -\pi/2$ for which
\begin{equation}
  \Ham = - J \sum_{(ij)} ((\vect{S}_i\cdot\vect{S}_j)^2 - 1),
  \eqlabel{SU3}
\end{equation}
where the constant $-1$ is added for convenience.
The Hamiltonian \Eq{SU3}, as well as the Hamiltonian at other
special values of $\theta$, has the SU(3) symmetry.
Using the ordinary $S^z$ representation basis 
$|\sigma_i\rangle$ $(\sigma_i =-1,0,+1)$
$$
  S^z_i | \sigma_i \rangle = \sigma_i | \sigma_i \rangle,
$$
the pair Hamiltonian can be re-written as
\begin{equation}
  - \Ham_{ij} = J \times \hat\Sigma \times \hat\Delta(g_{\rm H}).
  \eqlabel{GraphDecompositionSU3}
\end{equation}
Here, $\hat\Sigma$ is the operator that carries the sign,
whose matrix element is
$+1$ or $-1$, and is $-1$ if and only if one of the initial state 
$(\sigma_i,\sigma_j)$ and the final state $(\sigma_i',\sigma_j')$ is 
$(0,0)$ and the other is $(1,-1)$ or $(-1,1)$.
It is easy to see that this sign is irrelevant since
the negative signs always occur in pairs leaving the sign 
of the whole system positive.
Therefore, we can simply neglect the operator $\hat\Sigma$,
as we do in what follows.
The operator $\hat\Delta(g_{\rm H})$ is defined by its
matrix elements, which we denote by $\Delta(S_u,g_{\rm H})$
and are defined as
\begin{eqnarray*}
  \Delta(S_u, g_{\rm H}) 
  & \equiv &
  \langle \sigma_i', \sigma_j' | \hat\Delta(g_{\rm H}) 
  | \sigma_i, \sigma_j \rangle  \\
  & = &
  \left\{ \begin{array}{ll}
  1 & \mbox{(if $\sigma_i+\sigma_j = \sigma_i'+\sigma_j' = 0$)} \\
  0 & \mbox{(otherwise)}
  \end{array}\right..
\end{eqnarray*}
This is almost identical to \Eq{MatrixElementOfHorizontalGraph}.
The only difference is that the present operator is defined on
a larger Hilbert space 
($\sigma_i=-1,0,1$) than the previous one 
($\sigma_i=-1/2,1/2$).
It is therefore obvious that the present problem is a
generalization of the ordinary SU(2) antiferromagnetic 
Heisenberg model.
The constraint imposed by $\hat\Delta(g_{\rm H})$ upon the
world-line configuration can be expressed 
by the same graph element as the one in the SU(2) case, i.e.,
the horizontal graph in \Tab{GraphElements}.
The local spins bound by the graph must take the values
complementary to each other (such as $+1$ and $-1$, or $0$ and $0$).
Therefore, once a loop has been formed, the local spin value must be
$0$ everywhere along the loop or it must alternate between $+1$ and $-1$.
As in the SU(2) case, choosing a local spin state at one point of
the loop determines the state of the whole loop.
The difference is simply that 
every loop can take three possible states rather than two.
The loop flipping process must be altered
accordingly when we consider the loop algorithm 
for the present model;
we must choose one state among three possible ones
with equal probability for each loop.
All the rest of the procedure remains the same.
For example, the graph assignment is done in the same way
as the SU(2) case; the horizontal graph elements are assigned 
with the density $J$ between two nearest neighbor sites
if the local spin values at the two sites are complementary 
to each other.

A similar algorithm can be constructed in other cases with
lower symmetry, i.e., the SU(2) symmetry, for the parameter
region $-3\pi/4 \le \theta \le -\pi/2$.
We start from the expression\cite{HaradaK2001}
\begin{equation}
  - \Ham_{ij} = 
  J \left(
    ( -\sin\theta+\cos\theta ) \hat\Delta(g_{\rm H})
    - ( \cos\theta ) \hat\Delta(g_{\rm C}) 
  \right),
  \eqlabel{GraphDecompositionBiq}
\end{equation}
where an irrelevant sign and an additive constant have been omitted.
The symbol $\hat\Delta(g_{\rm C})$ corresponds to
the cross graph in \Tab{GraphElements}.
To be more specific, its matrix elements are
$$
  \Delta(S_u, g_{\rm C}) \equiv 
  \left\{ \begin{array}{ll}
  1 & \mbox{(if $\sigma_i = \sigma_j'$ and $\sigma_j = \sigma_i'$)} \\
  0 & \mbox{(otherwise)}
  \end{array}\right..
$$
The loop construction and the loop flipping 
can be done in much the same way as described above.

These algorithms can be easily generalized to the case where each
local spin variables takes $N$ possible values, i.e., 
$\sigma_i = (-N+1)/2, (-N+3)/2, \cdots, (N-1)/2$.  
Of particular interest is the Hamiltonian that consists of 
$\hat\Delta(g_{\rm H})$ only, which possesses the SU($N$) symmetry.  
See the reference\bigcite{HaradaKT2003} for results of a numerical simulation.
It should be also pointed out that
the algorithm presented here is similar to 
the Swendsen-Wang algorithm\cite{SwendsenW1987} 
for the classical antiferromagnetic $N$-state Potts model, 
in which a cluster is constructed in much the same
way as the SW algorithm for the Ising model,
and each cluster can take $N$ different states.  

\subsection{Magnetic Field}
For a number of models, 
the loop algorithm described above is 
the most efficient algorithm among the ones described in the present article.
For instance, the easy-axis $XXZ$ model with general spin size $s$ can be 
best handled with the loop algorithm.
The easy-plane $XXZ$ models can also be simulated
most efficiently with the loop algorithm
if there is no external magnetic field parallel to the 
diagonalization axis, namely, the $z$ axis in the present case.
However, if we have such an external magnetic field and 
it is competing with the spin-spin couplings, 
the loop algorithm does not work.\cite{KashurnikovPST1999}

To see this, we first describe a simple loop algorithm for a case
with magnetic field in the $z$-direction, 
and see what makes it inefficient.
In the simple algorithm, we deal with the magnetic field separately;
we simply neglect the external field while assigning graph elements.
Then, in flipping clusters or loops, we take it into account.
This can be formally justified as follows.
First decompose the Hamiltonian into the field-free part $\Ham^{(0)}$ and
the field part $\Ham^{(1)}$.
$$
  \Ham = \sum_b \Ham^{(0)}_b + \sum_i \Ham^{(1)}_i.
$$
Then, we have 
\begin{eqnarray*}
  W_L(S) 
  & = &
  \prod_{b,k}
  \left\langle \psi_{b+1}(k) \left| 
  e^{ -\Delta\tau\, \Ham^{(0)}_b }\,
  \right| \psi_b(k) \right\rangle \\
  &  &
  \times
  \prod_{i,k}
  \left\langle \psi_1(k) \left| 
  e^{ -\Delta\tau\, \Ham^{(1)}_i }
  \right| \psi_1(k) \right\rangle.
  \\
  & = &
  \sum_G W^{(0)}_L(S,G) V(S),
  \\
\end{eqnarray*}
where we have assumed that the field part is diagonal.
The factor $V(S)$ is the contribution from the magnetic field term
defined as
$$
  V(S) \equiv
  e^{ - \sum_{k,i} \Delta\tau\,
  \left\langle \psi_1(k) \left| \Ham^{(1)}_i \right| \psi_1(k) \right\rangle
  }
  =
  e^{ \int_0^{\beta} d\tau\, \sum_i F_i(\tau) },
$$
where
$$
  F_i(k\,\Delta\tau)
  \equiv
  \left\langle \psi_1(k) \left| \left(-\Ham^{(1)}_i\right) 
  \right| \psi_1(k) \right\rangle.
$$
The factor $V(S)$ can be rewritten in terms of clusters as
$$
  V(S) = \prod_c V_c(S),
$$
where $V_c(S)$ is defined for each cluster $c$ in $G$ as
$$
  V_c(S) \equiv e^{\int_c dX F(X)}.
$$
Here $\int_c dX$ is the $(d+1)$-dimensional integral in the space-time
over the cluster $c$.

It is obvious from this form that the magnetic-field term
does not affect the graph-assignment probability \Eq{TransitionProbabilities}
whereas it modifies the cluster-flipping probability.
Substituting $W(S,G) = W^{(0)}_L(S,G) V(S)$ for
\Eq{TransitionProbabilities}, we obtain
$$
  T(S|G) = \Delta(S,G)
  \prod_c \frac{V(S_c)}{\sum_{S'_c} V(S'_c)},
$$
where $S_c$ is the specifier of the state of the cluster $c$.
(For the $s=1/2$ spin models $S_c$ is a binary variable.)
This indicates that we have to choose the cluster state
$S_c$ with the probability $V(S_c)/\sum_{S'_c} V(S'_c)$ for each $c$.
When the external field is zero, 
this reduces the random unbiased choice between two possible 
cluster states as already explained above.

This procedure works well when the magnetic field is cooperative with
the spin-spin couplings
as is the case with the easy-axis ferromagnetic $XXZ$ model with a uniform
magnetic field parallel to the axis.
However, if the field is competitive against the spin-spin couplings,
as is the case with the antiferromagnet with a uniform field,
the procedure becomes increasingly inefficient as the temperature is lowered.
To see this, we consider a small system that consists of only two spins
coupled with each other by an antiferromagnetic interaction.
Let us first suppose that spins are totally aligned in the direction of the
magnetic field.
Because of the graph-assignment role presented above,
the density with which we assign the non-trivial graph elements is zero.
The resulting graph is therefore a trivial one that consists of
two loops, i.e.,
two vertical lines going from the bottom of the system to the top.
In order to visit a different spin configuration, 
we have to flip at least one of these two loops.
However, the flipping must be done against the magnetic field,
and the flipping probability is roughly proportional to $e^{-\beta H}$
according to the simple procedure discussed above.
Here $H$ is the magnetic-field strength.
Therefore, flipping seldom takes place at a low temperature
regardless of the magnitude of $J$ relative to $H$.
However, we need to visit other states frequently,
particularly when $J$ is much larger than $H$.
When $H$ is much larger than $J$, on the other hand,
we need to visit the completely aligned state frequently.
However, it hardly happens if we start from the 
anti-parallel state in which one of the two spins is up and the other
is down because the transition probability to the completely
aligned state is exponentially small at a low temperature.
This is because we cannot change the total magnetization 
unless we flip a loop whose temporal winding number is not zero;
but such a loop can be formed only when no
non-trivial graph elements are assigned to the system.
Such an event takes place with an exponentially small probability 
proportional to $e^{-\beta J/2}$, regardless of $H$.
In short, when the magnetic field competes with the other couplings,
the transition probability from one value of magnetization
to another becomes very small at low temperatures,
making the simulation extremely slow.

\subsection{Worm Algorithm}
\ssclabel{WormAlgorithm}
There are cases where one can avoid the freezing problem 
due to the magnetic field
by using the worm algorithm\cite{ProkofevST1998,KashurnikovPST1999}.
Updates of the world-line configuration in the worm algorithm
is done through stochastic movements of two discontinuity points
at which the conservation rule is violated.
In the case of particle-hole problems or $s=1/2$ quantum spin problems,
a world-line terminates at these points.
Only one of the two points moves around in our implementation, and 
we call the mobile one the {\it head} of the worm and 
the other stationary one the {\it tail}.
A {\it worm} is the pair of these two points.
(A worm in the present paper does not have a `body', 
in contrast to real ones.\cite{ladybug})
The spin configuration is modified as the head moves around.
There can be several types of heads depending on the change in the local
state caused by them.
In many applications, however, we only consider two types;
the one for which the local state above the head is 
one higher than that below (positive head),
and the one for which the opposite is true (negative head).
The types of the tail are defined likewise.

\deffig{WormMovements}{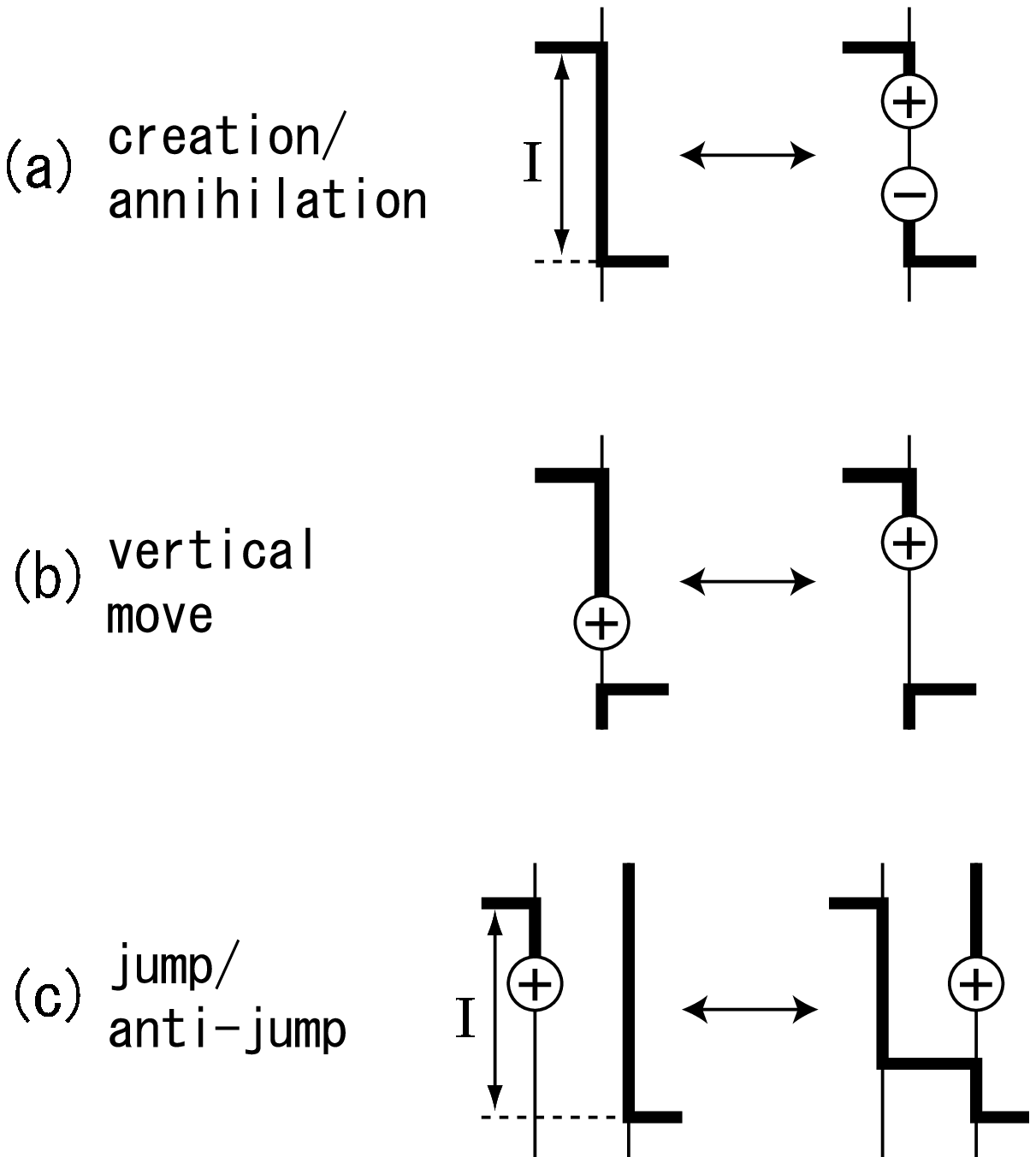}{0.4}{ 
Three elementary movements and their anti-movements of the head.
At the positive head labeled `+', the local spin variable
increases by one whereas at the negative one labeled `$-$',
it decreases by one.
The thick line, therefore, is the part where the spin value
is greater relative to the thin part.
}

One step in the worm algorithm consists of 
three elementary movements and their anti-movements:
the creation/annihilation of a worm,
the vertical movement, and the jump/anti-jump,
as illustrated in \Fig{WormMovements}.
Each movement is a stochastic transition that satisfies the
detailed balance condition with respect to the weight in
\Eq{BoltzmannWeightQ} with an additional contribution from the source term.
That is, we consider the Hamiltonian $\Ham - \eta\,Q$,
where the operator $Q$ represents the source and is the sum of
local operators, $Q = \sum_i Q_i.$
The partition function is expressed in the discrete imaginary time as
\begin{eqnarray}
  Z & = & \lim_{L\to\infty} \sum_{S} \tilde W_L(S),
  \eqlabel{PartitionFunctionWithSourceTerm}\\
  \tilde W_L(S) & \equiv & 
  \prod_u \left\langle \psi'_u \left| \left(
    1-(\Delta\tau) \Ham_u
  \right) \right| \psi_u \right\rangle 
  \nonumber \\
  & & \times
  \prod_v \left\langle \psi'_v \left| \left(
    1+(\Delta\tau) \eta Q_v
  \right) \right| \psi_v \right\rangle,
  \eqlabel{WeightWithSourceTerm}
\end{eqnarray}
where $v$ specifies a local unit defined on a vertical line $i$
so that every layer contains exactly one unit for each vertical lines.
The symbol $Q_v$ stands for $Q_i$ as $\Ham_u$ does for $\Ham_b$
in \Eq{BoltzmannWeightQ}.
The right-hand side of 
\Eq{WeightWithSourceTerm} consists of three parts;
the contribution from the diagonal matrix
elements of the Hamiltonian, the contribution from the
off-diagonal matrix elements, and the contribution from the
source term.
Specifically, denoting the number of kinks as $N_{\rm kink}$
and the number of discontinuity points as $N_{\rm dc}$, we can rewrite 
the weight as
\begin{eqnarray}
  \tilde W_L(S) & = & W_D \times (\Delta\tau)^{N_{\rm kink}} W_K 
                      \times (\Delta\tau)^{N_{\rm dc}} W_W,
                      \nonumber
    \\
  W_D & \equiv & 
    \prod_{u:{\rm non-kink}} \left\langle \psi'_u \left| \left(
      1-(\Delta\tau) \Ham_u 
    \right) \right| \psi_u \right\rangle 
    \nonumber
  \\
  & = &
    \exp\left(-\int_0^{\beta} d\tau\,
      \sum_b \langle \psi(\tau) | \Ham_b | \psi(\tau) \rangle 
    \right),
    \nonumber
  \\
  W_K & \equiv & 
    \prod_{u:{\rm kink}} 
    \langle \psi'_u | \left(-\Ham_u\right) | \psi_u \rangle,
    \nonumber
  \\
  W_W & \equiv & 
    {\prod_{v:{\rm dc}}}' 
    \langle \psi'_v | \,\eta Q_v\, | \psi_v \rangle.
    \nonumber
\end{eqnarray}
Since we only need up to the second order in $\eta$,
we truncate the last product at the second order,
which is indicated by the prime in $\prod'_{v:{\rm dc}}$.
In what follows, we consider only configurations
with no worm or those with exactly one worm.

The detailed balance condition
\begin{equation}
  T(S'|S) \tilde W_L(S) = T(S|S') \tilde W_L(S')
  \eqlabel{DBWorm}
\end{equation}
has to be satisfied by the transition matrices expressing 
three elementary movements of the head.

In the creation process (\Fig{WormMovements}(a)),
we first choose a site $i$, a uniform interval of it, 
say $I$, and two temporal positions in $I$, $\tau$ and $\tau'$, 
for placing the worm.
Then we decide if the proposed placement is accepted.
In this process,
$W_W$ and $W_D$ are altered while $W_K$ remains the same.
Let the probability of choosing $I$ be $P(I)$,
the probability of choosing $x$ and $y$ in the intervals
$dx$ and $dy$ respectively, $P(x,y)\,dx\,dy$,
and the probability of acceptance, $P_{\rm create}$.
For the inverse process, namely, the annihilation, we do not have
to choose $I$, $\tau$ or $\tau'$.
We simply decide whether we erase the worm or not 
with probability $P_{\rm annihilate}$.
Thus the detailed balance \Eq{DBWorm} in this case
can be rewritten as
\begin{eqnarray}
  & & \frac12 P_{\rm create}\times P(I)\,\left(d\tau\,d\tau'\,P(\tau,\tau')\right) W_D(S) 
  \nonumber \\
  & & = P_{\rm annihilate}\times W_D(S')\,\left(d\tau\,d\tau' W_W(S')\right),
  \eqlabel{DBCCreateAnnihilate}
\end{eqnarray}
where $S$ is the state with no worm and $S'$ is the state
with a worm whose head is at $x$ and the tail $y$.
The factor $1/2$ on the left-hand side is due to the two
possibilities concerning the initial type of the worm.
Here we obviously have many degrees of freedom.
One of many possible choices, though it may not be the optimal, 
is given by setting $P(I) = |I|/(N\beta)$,
which corresponds to choosing the interval $I$ by ``throwing a dart''.
Then, we obtain
$$
  \frac{P_{\rm create}}{P_{\rm annihilate}}
  = R \equiv
  \frac{2N\beta}{|I|}
  \frac{\int_I dx\,dy\, W_D(S') W_W(S')}{W_D(S)}
$$
and 
\begin{equation}
  dx\,dy\,P(x,y) = 
  \frac{dx\,dy\, W_D(S') W_W(S')}{\int_I dx\,dy\, W_D(S') W_W(S')}.
  \eqlabel{CreationAnnihilation}
\end{equation}
To be more specific, the acceptance probabilities can be chosen as
\begin{equation}
  P_{\rm create} = \min(1,R),\quad P_{\rm annihilate} = \min(1,R^{-1}),
  \eqlabel{PCreate}
\end{equation}
and the free parameter $\eta$ is adjusted so that neither
of $P_{\rm create}$ nor $P_{\rm annihilate}$ is too small.

In practice, it is often too cumbersome to compute
\Eq{CreationAnnihilation} every time a creation or an annihilation
is proposed.
Therefore, an alternative may be used. That is,
\begin{equation}
  dx\,dy\,P(x,y) = 
  \frac{dx\,dy\,\exp\left(-|x-y|\times \overline{\Delta V}\right)}
  {\int_I dx\,dy\,\exp\left(-|x-y|\times \overline{\Delta V}\right)},
  \eqlabel{CreationAnnihilationAlternative}
\end{equation}
where $\overline{\Delta V}$ is the average excess action (per unit time)
caused by the creation of the worm,
$$
  \overline{\Delta V} \equiv
  - \frac1{|I|} \ln\left( \frac{W_D(S(I))}{W_D(S)} \right),
$$
where $S(I)$ is the world-line configuration that results from
creating the worm with the tail at the bottom and the head 
at the top of the interval $I$.
When this alternative is used, $R$ in \Eq{PCreate} must be
modified accordingly, so that the detailed balance condition
\Eq{DBCCreateAnnihilate} is satisfied.
(As a result, the new $R$ depends on the times $\tau$ and $\tau'$.)

The vertical movement (\Fig{WormMovements} (b)) is much simpler.
The head moves to another point of the
vertical line on which it is currently located.
The new position is chosen from the interval $I$
that contains no kink in it and is delimited by two kinks.
The choice is made with an appropriate density so that the detailed
balance is satisfied.
Since the kink contribution $W_K$ and the worm contribution $W_W$ 
are the same for the initial and the final state of the move,
we have only to consider the diagonal part $W_D$ for the detailed
balance.
Namely, the detailed balance \Eq{DBWorm} is satisfied if the probability 
$d\tau P_{\rm vertical}(\tau)$
of choosing the new position of the head in the interval $d\tau$ is
$d\tau P_{\rm vertical}(\tau) \propto W_D(S')$
where $S'$ is the state after the head position is moved to $\tau$.
Therefore, $P_{\rm vertical}(\tau)$ should be
$$
  d\tau P_{\rm vertical}(\tau) =
  \frac{d\tau\, W_D(S')}{\int_I d\tau\, W_D(S')}.
$$

Finally we consider the jump and the anti-jump (\Fig{WormMovements}(c)).
A jump is a movement in which the head changes its
spatial position while the temporal position is kept.
At the same time, a kink is created in a jump process between the 
two vertical lines.
There are two kinds of jumps according to the temporal location of the kink
to be created; whether it is above the head or below.
In the original article\cite{KashurnikovPST1999}, 
one of the two is called a reconnection.
We do not distinguish the two, since both the movement
can be done in exactly the same way.
The anti-jump, too, has two kinds according to the position of the kink
relative to the head.
The detailed balance in the jump process can be worked out 
in a fashion similar to the two cases discusses above.
This time, all three of $W_D$, $W_K$ and $W_W$ change.
The detailed balance condition is
\begin{eqnarray}
  & & P_{\rm jump}\times (P(x)\,dx) W_W(S) W_D(S) W_K(S) \\
  & & = 
      P_{\rm anti-jump}\times W_W(S') W_D(S') (W_K(S')\,dx),
  \eqlabel{JumpAndAntiJump}
\end{eqnarray}
where $S'$ is the state after the jump.
$P_{\rm jump}$ and $P_{\rm anti-jump}$ are the probabilities
of accepting a proposed jump and anti-jump, respectively,
and $P(x) dx$ is the probability of choosing the position
of the new kink in the infinitesimal interval $dx$ around $x$.
We choose
$$
  dx\,P(x) = 
  \frac{dx\, W_D(S') W_K(S')}{\int_I dx\, W_D(S') W_K(S')}.
$$
Then, the acceptance probabilities must be chosen so that
$$
  \frac{P_{\rm jump}}{P_{\rm anti-jump}}
  = R \equiv
  \frac{W_W(S') \int_I dx\, W_D(S') W_K(S')}{W_W(S) W_D(S) W_K(S)}
$$
is satisfied.
Then, one possible choice for the acceptance probability is
$$
  P_{\rm jump} = \min(1,R),\quad P_{\rm anti-jump} = \min(1,R^{-1}).
$$

A few comments on the worm weight may be appropriate here.
In general, we can assign a non-trivial weights to the head and the tail.
A frequent choice is
\begin{equation}
  \langle \psi'_v | Q_v | \psi_v \rangle,
  \eqlabel{TheWormWeight}
\end{equation}
where $\psi_v$ and $\psi'_v$ are the local spin states 
just below the head (or the tail) and the above, respectively,
and $Q_v$ is an operator that represents
the order parameter relevant for the model.
For example, for the $XY$ model $Q_v=S_i^x$ is used.
In the $s=1/2$ case, in particular, the weight is a constant.
The reason for the choice \Eq{TheWormWeight}
is obvious, considering the relationship
between the head's trajectory and Green's function
$\Gamma_Q(X'-X) \equiv \langle Q(X') Q(X) \rangle$,
with $X$ and $X'$ specifying space-time points.
(See \mssc{EstimatorsAndEfficiency} for estimators of various quantities.)
When the worm is assigned the above-mentioned weight,
it can be shown that $\Gamma_Q(X)$ is proportional to the 
frequency with which the head visits a location specified by
$X$ relative to the head's original location.
Therefore, if the range where $\Gamma_Q(X)$ of $O(1)$
is determined by the system's correlation length,
the head's trajectory extends a region
whose linear size is roughly equal to the correlation length.

This is desirable since this guarantees that the scale of the 
update coincides with the correlation length.
This is also the case with the loop algorithm
with no external field.
However, the worm algorithm works better than the loop algorithm
when a competing external field exists.
This is because the effect of the field is reflected 
in choosing each local movement of the head.
Therefore, a typical trajectory of the head strongly depends on
the strength of the field.
In the loop algorithm, on the other hand, the loop construction 
is done with no reference to the external field,
making the typical loop, 
which corresponds to the trajectory of the head,
depends on the external field only indirectly.
As a result, the acceptance ratio of the loop flipping can be
extremely small in the loop algorithm whereas the 
acceptance is always unity in the worm algorithm 
(the local spin state along the trajectory is already changed
when the head finishes its journey).

\subsection{Directed-Loop Algorithm}
\ssclabel{DirectedLoopAlgorithm}
The directed-loop algorithm\cite{Sandvik1999, SyljuasenS2002} can be thought 
of as a hybrid of the loop algorithm and the worm algorithm.
While it has an advantage of the worm algorithm, we do not need 
to do integrations for obtaining the transition probabilities.
In addition, although the directed-loop algorithm becomes identical
to the loop algorithm when the external magnetic field is zero,
it does not have the freezing problem even when the field is turned on.

The directed-loop algorithm can be formulated in much the same way as
the formulation of the loop algorithm.
Therefore, we start with \Eq{Quantum}
(or \Eq{BoltzmannWeightQ}).
In the loop algorithm, we have decomposed the local Hamiltonian
into several terms, each corresponding to a particular graph element.
In addition, we split each original spin into $2s$ Pauli spins
in the case of $s>1/2$.
Therefore, $b$ in \Eq{HamiltonianDecomposition} is equivalent to
$((ij),(\mu\nu),g)$.
In the directed-loop algorithm,
we do not decompose the local Hamiltonian at all.
Accordingly, $b$ in \Eq{HamiltonianDecomposition} must be 
regarded simply as $(ij)$.
Then, the procedure of updating $G$ follows from the general prescription
in \mssc{PathIntegral}.
For example, we set $G_u = 1$ for a given $u$ with probability
$(\Delta\tau) w(S_u) \equiv$
$(\Delta\tau)\langle \psi'_u | (-\Ham_{ij}) | \psi_u \rangle$
when $\psi'_u = \psi_u$.
This means, in the continuous-time formulation, that
vertices (which correspond to $-\Ham_{ij}$ here,
rather than $\hat\Delta(g)$ in the loop algorithm) 
are placed with the density
$\langle \psi_u | (-\Ham_{ij}) | \psi_u \rangle$ 
in uniform intervals.
In addition, a vertex is placed on every kink.

The updating procedure for $S$, on the other hand, is quite different from
that in the multi-cluster variant of the loop algorithm
discussed in the previous subsections.
There, clusters are formed naturally as a result
of the graph assignment because the Hamiltonian has been decomposed 
into graph elements.
Since we do not have graph elements in the directed-loop algorithm,
loop (cluster) must be formed in the $S$-updating process rather than
in the $G$-updating process.

While the $S$-updating is done with a worm in the directed-loop algorithm
similarly to the worm algorithm,
the head of the worm in the directed-loop algorithm cannot choose the positions
at which it creates kinks unlike the worm algorithm.
This is because $G$ has been fixed (i.e., all the vertices are fixed)
before the worm is created, and we cannot have a kink 
at a plaquette on which there is no vertex, i.e., $G_u = 0$.
Therefore, new kinks can be made only at the vertices
which are fixed during the worm's life-time.
However, this is not an essential difference because
one can easily generalize\cite{SyljuasenS2002} the algorithm so that
the vertices are generated dynamically during the head's motion.
Another (probably more important) difference between the 
directed-loop algorithm and the worm algorithm arises from 
the direction of the head's motion.
In the worm algorithm, the direction of the head's motion is biased only by
the weight $W(S,G)$ and there is no algorithmically preferred direction.
In the directed-loop algorithm, on the other hand,
the head has a ``moment of inertia'' and can go only in the
direction that is the same as in the previous step.
The head can change its direction of motion only when it is 
scattered by a scatterer, i.e., a vertex.
Therefore, $G_u=1$ can be interpreted as having a scattering object at $u$.
This is a clear advantage of this method compared to the worm 
algorithm, because in the worm algorithm, a head in general goes back and
forth along a vertical line, sometimes unnecessarily.
When applied to the $s=1/2$ antiferromagnetic Heisenberg model, 
for example, the trajectory of the head is roughly the same
as the loop in the loop algorithm when the field is absent.
Therefore, the head's motion in the worm algorithm
is a random walk along a loop.
While it takes a time proportional to the squared length of the loop
for a head to finish its travel in the worm algorithm,
it takes only a time proportional to the length in the directed-loop algorithm.

\deffig{Scattering}{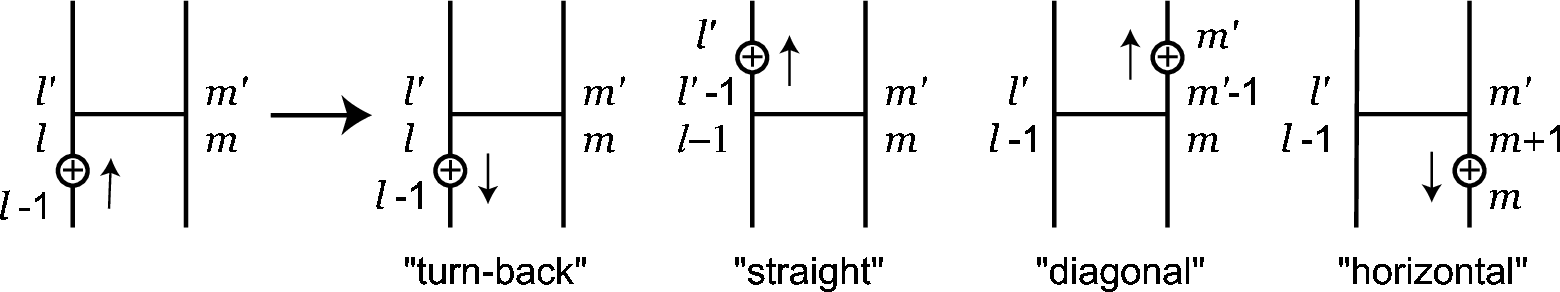}{0.45}{
The four events that can happen when a head hits a vertex.
Arrows indicate the directions of the heads' motion.
The numbers indicate the local spin states.
Namely, $l=0,1,\cdots,2s$ correspond to
the local spin states $S^z_i = -s, -s+1, \cdots, s$, respectively.
}

When the head arrives at a vertex,
it may or may not change its location as well as the direction of motion.
It has four possibilities as to the location after the scattering,
namely, the four legs of the vertex (\Fig{Scattering}).
The choice among the four is made probabilistically.
However, unlike all the cases discussed above, we cannot
use the detailed balance condition for determining the probability
due to the direction of the head's motion.
It is obvious that the probability of having the left-most state
in \Fig{Scattering} as the final state is zero 
when the initial state is one of the four states on the right,
because of the direction of motion. 
(The head is moving away from the vertex, not coming in.)
Instead of the detailed balance, we use the time-reversal symmetry
condition as we discuss below.

\deffig{DirectedLoopProcess}{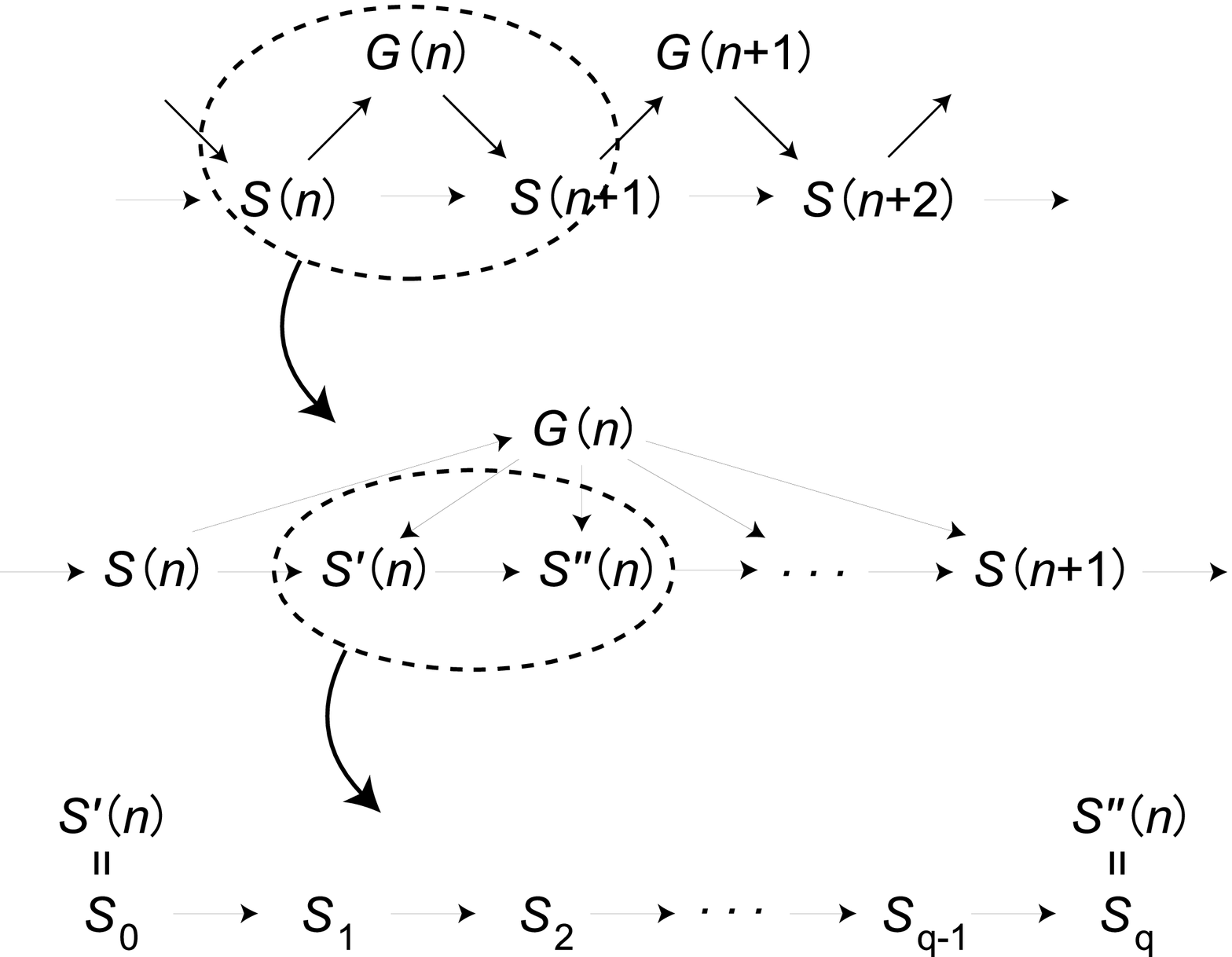}{0.45}{
A schematic illustration of the directed-loop Monte Carlo.
The top, middle, and bottom parts show an overview,
one step, and one cycle, respectively
}

The stochastic process of the directed-loop algorithm
can be formally viewed as the stochastic process in the extended state space.
The extension of the state is done in two ways.
As mentioned above,
the first extension is due to the introduction of the auxiliary variable $G$,
and the other is due to the introduction of a worm.
Since the directed-loop algorithm is a kind of single-cluster
algorithm similar to the Wolff algorithm,
the whole stochastic process is not a simple alternating Markov
chain as in the loop algorithm (\Fig{DualMonteCarlo}).
As illustrated in the top part of \Fig{DirectedLoopProcess},
the probability of generating a new state $S(n+1)$ depends not only
on the current graph $G(n)$, but also on the current state $S(n)$.
This is in contrast to the multiple-cluster variant of
the loop algorithm that corresponds to \Fig{DualMonteCarlo}.
This updating process of the spin configuration is achieved
by a number of worm creation/annihilation cycles.
Each cycle starts with a state that contains no worm and ends
with another worm-free state.
Let us denote the initial state $S_0$ and the final state $S_q$
where $q$ stands for the number of elementary motions of the head
during the life-time.
Each state between the two, $S_1, S_2, \cdots,$ or $S_{q-1}$,
contains a worm.

Let us denote the transition probability that governs the
elementary head motion as $T_{\rm w}(S'|S)$.
(Here we have dropped the dependence on $G$ of the transition probability
because it is fixed throughout the cycle.)
Instead of the detailed balance condition, this transition probability
is chosen so that it satisfies the time-reversal symmetry condition
\begin{equation}
  T_{\rm w}(S_{k+1}|S_k) W_L(S_k,G) 
  = T_{\rm w}(\bar S_k|\bar S_{k+1}) W_L(S_{k+1},G),
  \eqlabel{TimeInversionSymmetry}
\end{equation}
where $\bar S$ is the state identical to $S$ except the direction of the
head's motion.
In other words, $\bar S$ is the time-inversion of $S$.
Note that the weight of a state does not depend on the direction of a head.
Once \Eq{TimeInversionSymmetry} is satisfied, 
the ordinary detailed balance condition
is recovered in the process from $S_0$ to $S_q$, i.e.,
$$
  T(S_q|S_0) W_L(S_0,G) =   T(S_0|S_q) W_L(S_q,G).
$$
This can be seen easily as follows. 
First we note that
\begin{eqnarray*}
  & & T(S_q|S_0)W(S_0,G) \\
  & & \qquad = \sum_q \Big( \sum_{S_1}  \sum_{S_2}  \cdots \sum_{S_{q-1}} \\
  & & \qquad \quad T_{\rm w}(S_q|S_{q-1})  T_{\rm w}(S_{q-1}|S_{q-2}) \cdots \\
  & & \qquad \quad T_{\rm w}(S_1|S_0) W_L(S_0,G) \Big).
\end{eqnarray*}
But because of the direction independence of the weight,
by using the time-reversal invariance of the transition matrix
\Eq{TimeInversionSymmetry} repeatedly,
we obtain
\begin{eqnarray*}
  & & 
    T_{\rm w}(S_q|S_{q-1})  T_{\rm w}(S_{q-1}|S_{q-2}) \cdots  T_{\rm w}(S_1|S_0) W_L(S_0,G) 
    \nonumber 
  \\
  & & \quad = 
    T_{\rm w}(\bar S_0|\bar S_1) T_{\rm w}(\bar S_1|\bar S_2) 
    \cdots  T_{\rm w}(\bar S_{q-1}|\bar S_q) W_L(S_q,G).
    \nonumber
\end{eqnarray*}
Thus, the detailed balance is recovered for every individual path
that leads from $S_0$ to $S_q$.

Here we consider the weight of the states with worms.
Since the worm is an artifact for the algorithm,
in principle we can assign any weight to the 
states with a worm.\cite{AletWT2003}
The most natural definition, however, is 
to use the same expression as \Eq{BoltzmannWeightQ} with 
an additional factor for the worm,\cite{AletWT2003}
\begin{equation}
  w_{\rm w}(S_{\rm x})
  = \eta\, \langle \psi'_{\rm x} | S^x_i | \psi_{\rm x} \rangle,
  \eqlabel{WormWeight}
\end{equation}
where x is h or t corresponding to the head or the tail, respectively.
The local state $S_{\rm x}$ is defined as $(\psi'_{\rm x}, \psi_{\rm x})$, 
where $\psi_{\rm x}$ and $\psi'_{\rm x}$ are the local spin states 
just below and above the discontinuity point, respectively.
The constant $\eta$ is included for adjusting the worm creation
and annihilation probabilities.
A similar factor for the tail is also included.
The weight of a state with a worm altogether becomes
\begin{eqnarray*}
  \tilde W_L(S) 
  & \equiv & 
    w_{\rm w}(S_{\rm h}) w_{\rm w}(S_{\rm t}) \nonumber \\
  & & 
    \times \prod_u
    \left\langle \psi'_u \left|\left(
      1-(\Delta\tau) \Ham_u 
    \right)\right| \psi_u \right\rangle,
  \nonumber
\end{eqnarray*}
where $S_{\rm h}$ and $S_{\rm t}$ are the local states around
the head and the tail, respectively.
The product is taken over all the vertices (plaquettes).
Note that the weight does not depend on the direction of the head's
motion.

Next, we consider how to define $T_{\rm w}(S'|S)$ so that it satisfies
eq.\ \Eq{TimeInversionSymmetry}.
Three cases must be considered;
(i) the scattering of the head at the vertex,
(ii) the pair creation, and 
(iii) the pair annihilation.
We first look into the case (i).
For the scattering process, \Eq{TimeInversionSymmetry} can be written as
\begin{equation}
  T_{\rm w}(S'|S) \tilde w(S_{u+w})
  =
  T_{\rm w}(\bar S|\bar S') \tilde w(S'_{u+w}),
  \eqlabel{DirectedLoopEquation}
\end{equation}
where 
$$
  \tilde w(S_{u+w}) \equiv w(S_u) w_{\rm w}(S_{\rm w}).
$$
Here, $S_w$ is the local state around the head and
$S_{u+w}$ stands for $(S_u, S_w)$.
Remember that there are only four possible final states for each initial state.
Suppose that $S_u^{(1)}$ is the initial local state of the vertex.
The state $\bar S_u^{(1)}$ is obviously one of the four possible final states 
because if the head turns back at the vertex, the state
$\bar S_u^{(1)}$ is the final state.
Let us denote the inverse of the other three possible final states as
$S_u^{(2)}$, $S_u^{(3)}$, and $S_u^{(4)}$.
Then, the four states $S_u^{(k)}$ $(k=1,2,3,4)$ form a closed set, i.e.,
if the initial state is among the four
the final state is always the inverse of one of the four.
Therefore, eq.\ \Eq{DirectedLoopEquation} can be generally decomposed into
several closed sets of four equations.

In order to find a solution to one of these quartets, let us suppose
that a matrix $w_{ij}$ exists and satisfies the properties
\begin{equation}
  w_i (\equiv \tilde w(S_u^{(i)})) = \sum_{j=1}^4 w_{ji}
  \eqlabel{DirectedLoopWeightEquation}
\end{equation}
and
\begin{equation}
  w_{ij} = w_{ji}.
  \eqlabel{SymmetryCondition}
\end{equation}
Then, it is easy to verify that
$$
  t_{ij} (\equiv T_{\rm w}(S_u^{(i)}|S_u^{(j)})) \equiv \frac{w_{ij}}{w_j}
$$
satisfies the property \Eq{DirectedLoopEquation}.
Therefore, the problem of solving \Eq{DirectedLoopEquation} has been
reduced to finding a symmetric matrix that satisfies
\Eq{DirectedLoopWeightEquation} with given $w_{i}$.

The following solution is always available for any model:
\begin{equation}
  w_{ij} \equiv \frac{w_iw_j}{\sum_{k} w_k}.
  \eqlabel{heat-bath}
\end{equation}
The final state is chosen simply proportional to the weight of the final 
state if we use this solution;
hence the name ``heat-bath'' type solution.
However, it has been known that this solution yields an inefficient 
algorithm in many cases.

In \Eq{DirectedLoopWeightEquation},
we have ten free parameters and only four equations.
However, the bounce-free condition $w_{ii} = 0$ is often imposed
for obtaining better efficiency.
In the case of the quantum $s=1/2$ $XXZ$ model, in particular,
the solution becomes unique with this additional constraints.
Still, we have six free parameters left.
While little is known about the general principle for obtaining solutions 
that lead to efficient algorithms,
good solutions are known for many important cases.
In the next section, we show such a solution for the $XXZ$ quantum
spin model with an arbitrary $s$.

As for the pair creation/annihilation process, we have to consider
the detailed balance between a state with a worm and a state without.
Specifically, the relation
$$
  T(S'|S) 
  = 
  T(S|S') w_{\rm w}(S'_{\rm h}) w_{\rm w}(S'_{\rm t})
$$
must hold for the transition probability $T$ where 
$S'$ and $S$ are the states with and without a worm,
respectively.
Note that $S$ and $S'$ are identical except that $S'$ contains a worm.
The symbols $S'_{\rm h}$ represents the local state
around the head, just before the collision of the head and the tail,
whereas $S'_{\rm t}$ is the state around the tail.
It should be noted here that the creation of the worm
consists of two steps; the selection of the position of the
creation and the rejection/acceptance of the proposed creation.
In the discrete-time representation there are $NL$ positions
at which we can place a worm.
Therefore, denoting the acceptance probability for the 
creation by $P_{\rm create}$, we can write $T(S'|S)$ as 
$$
  T(S'|S) = \frac{1}{NL} P_{\rm create}(S_v),
$$
where the $S_v$ is the local state around the proposed point of
creation before the creation.
On the other hand, there is no position selection
in the annihilation process. Therefore,
$$
  T(S|S') = P_{\rm annihilate}(S_v).
$$
(Note $S'_v = S_v$.)
The detailed balance condition becomes
$$
  \frac{1}{NL} P_{\rm create}(S_v)
  = 
  P_{\rm annihilate}(S_v)
  w_{\rm w}(S'_{\rm h}) w_{\rm w}(S'_{\rm t}).
$$
This yields the choice of the acceptance probabilities
$$
  P_{\rm create} = \min(1,R),\quad P_{\rm annihilate} = \min(1,R^{-1})
$$
with
$$
  R \equiv NL w_{\rm w}(S'_{\rm h}) w_{\rm w}(S'_{\rm t}).
$$
In particular, when the worm weight is the matrix element of $S^x_i$,
we obtain
$$
  R = NL \eta^2
  \langle \psi'_{\rm h} | S_i^x | \psi_{\rm h} \rangle
  \langle \psi'_{\rm t} | S_i^x | \psi_{\rm t} \rangle.
$$
As we did in \mssc{WormAlgorithm},
we can use $\eta$ for adjusting the transition probabilities.
In general, we should choose $\eta$ so that
none of the transition probabilities is too small.
If the worm weight does not depend on the local state, 
as is the case with the $s=1/2$ and $s=1$ spin systems,
we can choose the free parameter $\eta$ so that $R = 1$,
which is obviously the optimal choice.
In general, however, no such choice exists and 
the creation probability and/or the annihilation probability 
is smaller than 1 at least in some cases.
In \mssc{CoarseGrainedAlgorithm}, we present an
example of the choice for the $XXZ$ model.

\deffig{worm-possibility}{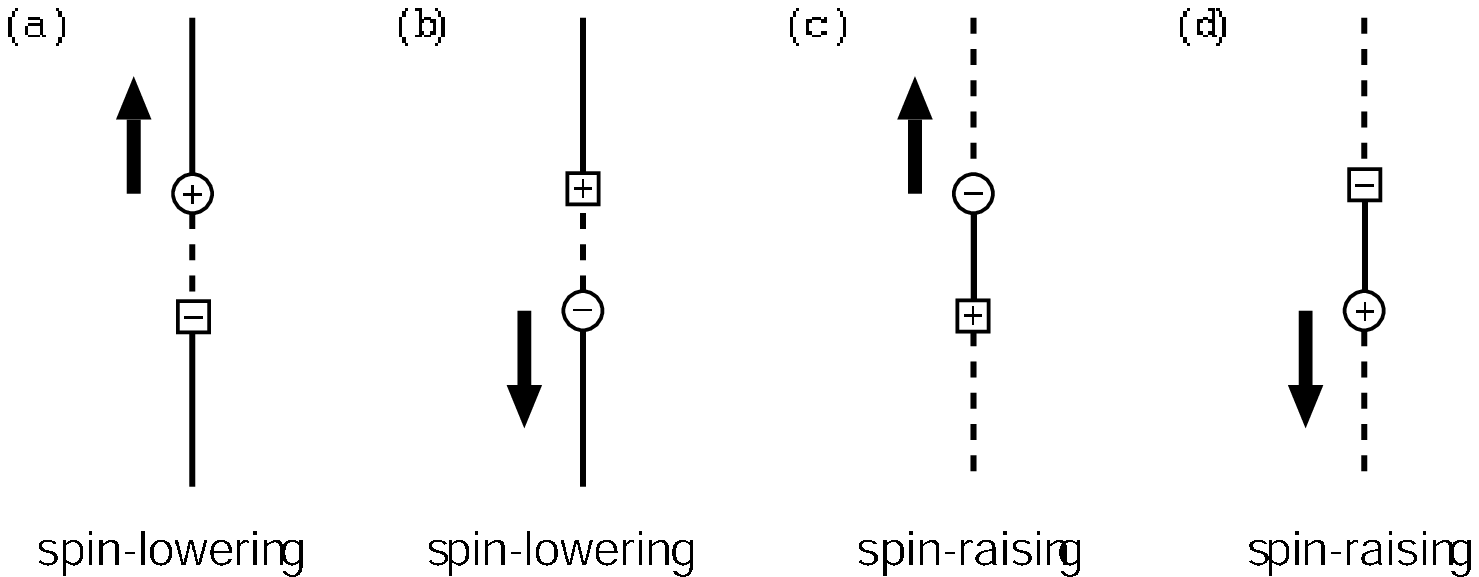}{0.5}{
  The four initial states of the worm. 
  Circles denote the heads and squares the tails.
  A spin state on a dashed line is lower 
  than that on a solid one. 
  In the case $s=1/2$, in particular, 
  a solid line and a dashed line correspond
  up and down spins, respectively. 
  An arrow indicates the direction of the head's initial motion.
}

It may be useful to consider here the case of the $s=1/2$
$XXZ$ spin model to make the description concrete.
In this case, the pair creation/annihilation is simple as discussed above.
The pair creation is always accepted at any proposed 
position and the pair annihilation takes place whenever the head meets the tail.
When the worm is created at a point where the local spin
is up, the upper discontinuity point is positive where 
the lower one is negative (see \Fig{worm-possibility}).
For a point with a down spin, the types of the created worm
should be the opposite.
The vertex density, as stated at the beginning of the
present subsection, is the negated diagonal matrix element of the 
pair Hamiltonian.
For example, it is $J/2$ for the antiferromagnetic Heisenberg model.
The probabilities that governs the scattering of the head at vertices
can be derived from solving the quartets of the equations
discussed above. The result depends on the anisotropy.
The solution is presented in \msec{Recipe} for various cases.
The resulting algorithm is rather simple for the antiferromagnetic 
Heisenberg model;
whenever the head hits a vertex we let it make the horizontal scattering.

\deffig{DirectedLoopUpdate}{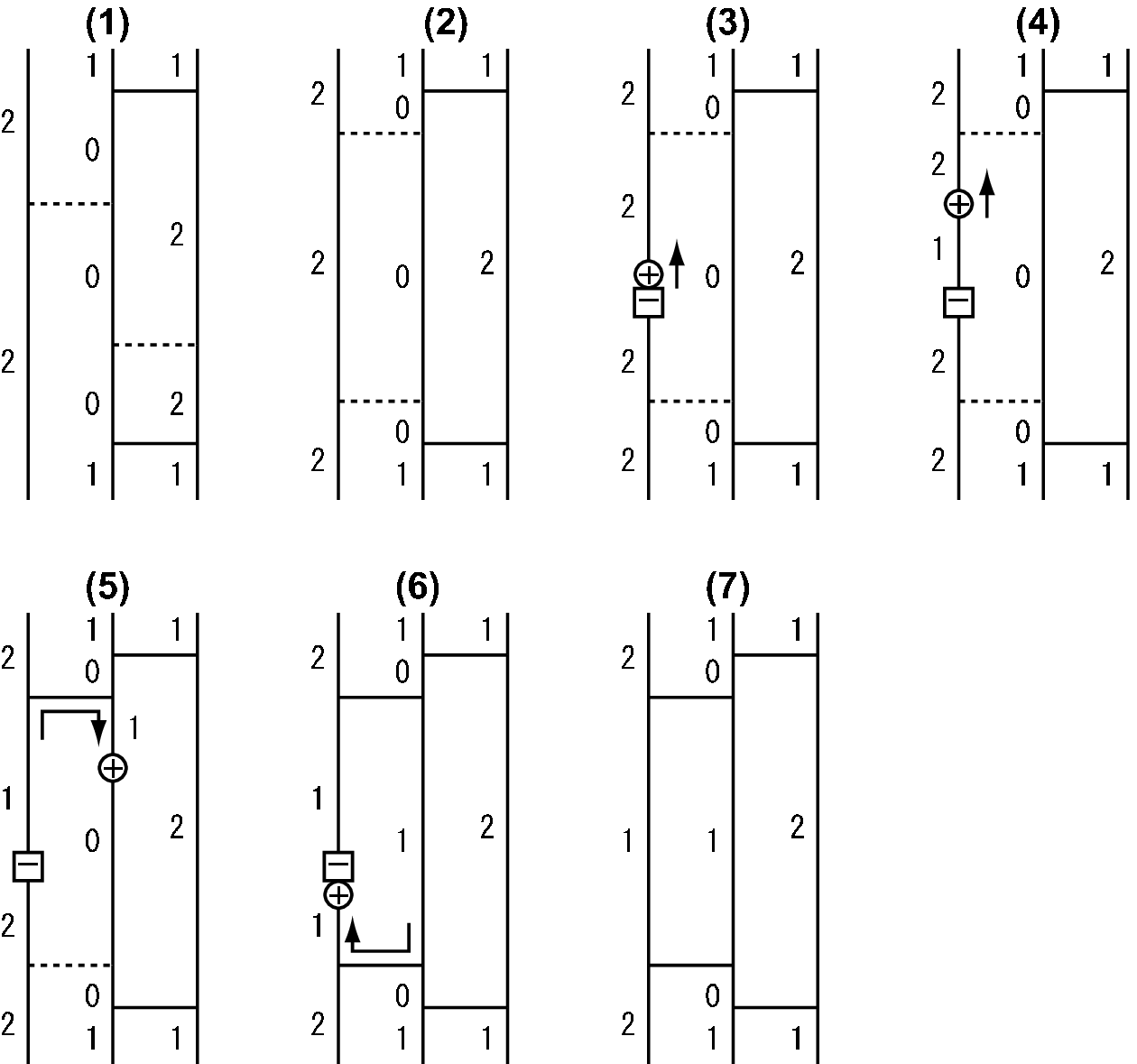}{0.4}{ 
  Assignment of vertices and a cycle of the worm update.
  Kinks are indicated by solid horizontal lines whereas
  other vertices are indicated by dotted lines.
  First, all the existing non-kink vertices are removed 
  while new vertices are assigned (1 to 2).
  Then, a worm is created (3).
  The head starts moving and it changes the local spin state (4).
  Every time the head hits a vertex, one of four
  possible events in \Fig{Scattering} is chosen stochastically.
  In this figure, both the scatterings happen to be the {\it horizontal} ones
  (5 and 6).
  When the head comes back to the original position,
  it annihilates with some probability
  leaving a worm-less configuration (7).
  The cycle (such as the one from (3) through (7)) is repeated 
  a number of times
  before the vertices are updated.
}

In the original paper by Sylju\r{a}sen and Sandvik\cite{SyljuasenS2002},
we can find a good example that shows the utility of the directed-loop
algorithm.
In \Fig{SS} (b), the integrated auto-correlation time defined for the
magnetization is shown (the middle panel) as a function of 
the magnetic field.
The magnetization itself (a) and the average loop size 
(c) are also shown in the same figure.
As has been discussed above, the presence of the magnetic field
competing against the exchange couplings makes the configuration
freeze in simulations with the loop algorithm.
As a result, it is impossible to observe a magnetization curve
such as \Fig{SS} (a).
By using the directed-loop algorithm, one can obtain 
the curve within a reasonable amount of computational time.
However, the difficulty has not been completely removed
as can be seen in \Fig{SS} (b).
The figure shows that the auto-correlation time diverges between two
successive plateaus in the magnetization curve.
So far, a solution to this problem is not known.

\deffig{SS}{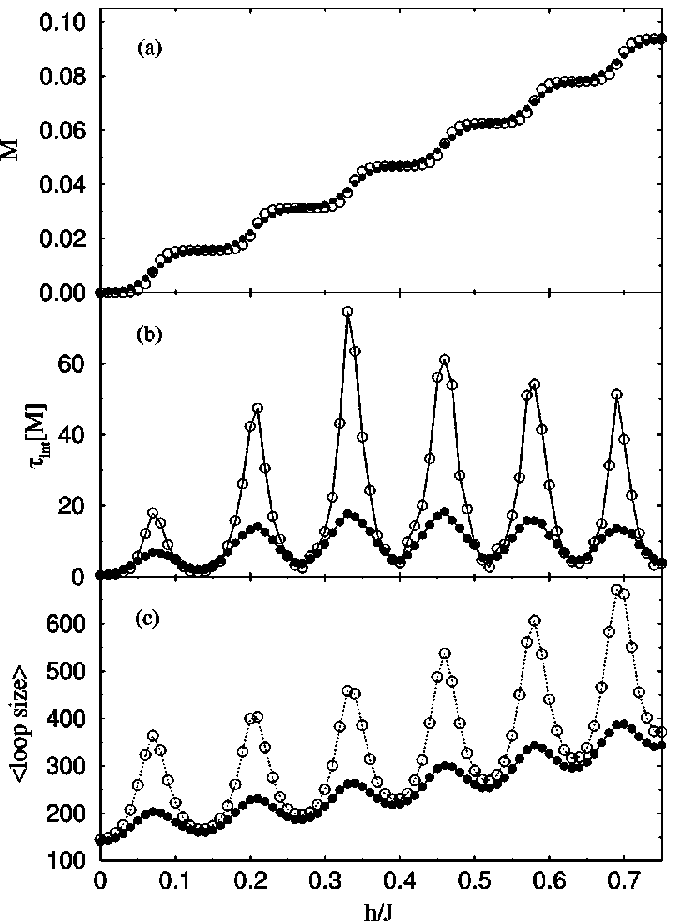}{0.4}{
The magnetization (a), the integrated auto-correlation time (b),
and the average length of the loop (i.e., the number of 
the visited vertices) (c), plotted against the magnetic field $h/J$
for the $s=1/2$ antiferromagnetic Heisenberg model in one dimension.
The solid and open circles are for $\beta=64$ and $\beta=128$,
respectively. The linear dimension of the system is $64$ lattice spacings.
(Adopted from Sylju\r{a}sen and Sandvik\cite{SyljuasenS2002}.)
}

\subsection{Coarse-Grained Algorithm}
\ssclabel{CoarseGrainedAlgorithm}
In general, the solution of the time-reversal-symmetry 
equation \Eq{TimeInversionSymmetry} is not unique.
In addition, the choice of the worm weight is arbitrary.
However, the efficiency of the resulting algorithm largely depends on 
these choices.
While one can obtain the solution by solving 
the equation \Eq{TimeInversionSymmetry} numerically in general,
there is no automatic way to choose a good one.
It is up to the practitioner's physical insight, experience,
and, to a certain extent, luck to find a solution and worm weights
that lead to an efficient algorithm.
Therefore, it is worthwhile to present some efficient solutions for
models of particular importance.
We here consider the $XXZ$ model with general $s$.
For this model, a set of simple formulas for such solutions 
are known\cite{HaradaK2002}. 
It includes the single-cluster variant of the loop algorithm
for the $s=1/2$ case.
Therefore, the algorithm can be viewed as a natural generalization of the
loop algorithm
to the cases with larger spins and with a uniform magnetic field.
While the solution was found in a way quite different from solving the
time-reversal symmetry condition \Eq{TimeInversionSymmetry},
we can show that the resulting solution satisfies
\Eq{TimeInversionSymmetry}.\cite{AletWT2003}
Below, we briefly describe the procedure for obtaining the solution.
The explicit formulas for the head-scattering probability and the
vertex density of the $XXZ$ model are presented in \Sec{Recipe}.

The idea is based on the split-spin representation.
As discussed in \mssc{LargeSpins},
it is in general possible to reformulate the model with $s>1/2$ 
in terms of the $2s$ Pauli spins: 
$\vect{S}_i \Rightarrow \sum_{\mu} \vect{\sigma}_{i,\mu}$.
We would obtain the algorithm in which a head moves around
in the space-time that consists of $2s$ vertical lines for each site.
What, then, would happen if we look at the real-time animation of the 
simulation on a low-resolution monitor?
The $2s$ lines are blurred and they appear to be a single thick line.
In the blurred image on the monitor, 
we cannot tell on which one of $2s$ lines, namely $\mu$, the head is. 
The only that we can tell is on which site, $i$, and at what time, $\tau$, it is.
Similarly, we cannot tell on which one of $2s$ lines a 
particular vertex is footed
while we can tell the site and the time.
Suppose also that the single line in the blurred image look brighter
when we have more up-spins in the $2s$ lines in the original image.
Then, there are $2s+1$ levels of brightness distinguishable in the
blurred image.
As the head moves, it changes the brightness level of the line by one.

It was pointed out\cite{HaradaK2002}
that such a blurred animation can be generated
with a set of transition matrix defined directly in terms of the brightness, 
without constructing the original sharp image.
We should note that we only need the blurred animation for our original purpose
to compute various physical quantities.
In short, the split-spin representation is not necessary for
describing the algorithm or writing computer codes while it is
useful in deriving them, as we see below.

To see how the head-scattering probability can be derived,
let us consider the general $s=1$ antiferromagnetic Heisenberg model
for example.
Suppose that the head has just hit a vertex
that is in the state $S_u$
(the first diagram in \Fig{CoarseGrain}).
The probability of obtaining the last diagram as the final
state of the scattering can be given as
\begin{equation}
  T(S'_u|S_u) = \sum_{\Sigma_u} \sum_{\Sigma'_u}
  T(S'_u|\Sigma'_u) T(\Sigma'_u|\Sigma_u) T(\Sigma_u|S_u).
  \eqlabel{CoarseGraining}
\end{equation}
The symbol $T(\Sigma_u|S_u)$ is the probability
that the original (sharp) image of the blurred image $S_u$ is $\Sigma_u$.
It is proportional to the weight of the original image, i.e.,
$$
  T(\Sigma_u|S_u) = 
  \frac{w(\Sigma_u,1)\Delta(\Sigma_u,S_u)}
  {\sum_{\Sigma_u}w(\Sigma_u,1)\Delta(\Sigma_u,S_u)},
$$
where $\Delta(\Sigma_u,S_u)=1$ if and only if $S_u$ is the blurred image of $\Sigma_u$.
The weight $w(\Sigma_u,1)$ is the one in the split-spin representation,
$$
  w(\Sigma_u,1) = \langle \sigma'_{i\mu}\sigma'_{j\nu} | 
  \left( -\Ham_{i\mu,j\nu} \right)
  | \sigma_{i\mu} \sigma_{j\nu} \rangle.
$$
The second factor $T(\Sigma'_u|\Sigma_u)$ in \Eq{CoarseGraining}
is the scattering probability in the split-spin algorithm,
i.e., the scattering probability in the case of $s=1/2$.
The third factor $T(S'_u|\Sigma'_u)$ only represents the compatibility
of the final state $S'_u$ with its original image $\Sigma'_u$, i.e.,
$$
  T(S'_u|\Sigma'_u) = \Delta(\Sigma'_u,S'_u).
$$

It should be noted here that we do not explicitly introduce the worm weight.
In fact, it was pointed out\cite{AletWT2003} that the present
algorithm agrees with the directed-loop algorithm
that follows from a special solution to \Eq{DirectedLoopWeightEquation}
with the choice of the worm weight:
$w(S_{\rm x}) \propto \langle \psi'_{\rm x} | S^x_i | \psi_{\rm x} \rangle$ 
$({\rm x}= {\rm h} , {\rm t})$.

The worm creation/annihilation probabilities can also be obtained
from the blurring (or coarse-graining).
In what follows, we express the local spin state by an integer
$l=0,1,2,\cdots,2s$, which corresponds to the 
$2s+1$ eigenvalues of $S^z_i$, $-s, -s+1, -s+2, \cdots, +s$,
respectively.
In the split-spin representation, we choose a point 
uniform-randomly from the space-time, and if the local spin state
at the chosen point is up,
we place a positive discontinuity point above the negative one.
We do the opposite if the local spin state is down.
When coarse-grained, this yields the following;
when the local spin state at the chosen point is $l$,
the probability of creating a positive discontinuity point
above the negative one is $l/2s$.
For the worm annihilation,
if a positive discontinuity point
is above a negative one before the ``rendezvous''
and the spin state between the two is $l$,
the probability that the two are on the same line
in the split-spin representation is $(2s-l)^{-1}$.
Therefore, the annihilation takes place with the probability
$(2s-l)^{-1}$ in the coarse-grained algorithm.
If the relative location of the head and the tail is the opposite,
the probability is $l^{-1}$.

Finally, the vertex assignment density can be derived as follows.
Let us consider an interval in which a local spin state is $l$ on one of
the two sites and $m$ on the other.
In the original (split-spin) image, 
we assign vertices with the density
$\langle \sigma'_{i\mu},\sigma'_{j\nu} | (- \Ham_{i\mu,j\nu}) | 
\sigma_{i\mu},\sigma_{j\nu} \rangle$ 
between the two vertical lines
specified by $(i\mu)$ and $(j\nu)$.
Therefore, in the blurred image, we assign vertices with the density
\begin{eqnarray*}
  \rho & = &
  \sum_{\mu,\nu} \langle \sigma_{i\mu},\sigma_{j\nu} | 
  \left( - \Ham_{i\mu,j\nu} \right)
  | \sigma_{i\mu},\sigma_{j\nu} \rangle 
  \nonumber \\
  & = &
  lm \rho_{++} + l\bar m \rho_{+-} + \bar l m \rho_{-+} + \bar l \bar m \rho_{--},
\end{eqnarray*}
where $\rho_{\pm\pm}$ is the vertex density
for the $s=1/2$ model with the local 
spin state $(\pm\frac12,\pm\frac12)$.

\deffig{CoarseGrain}{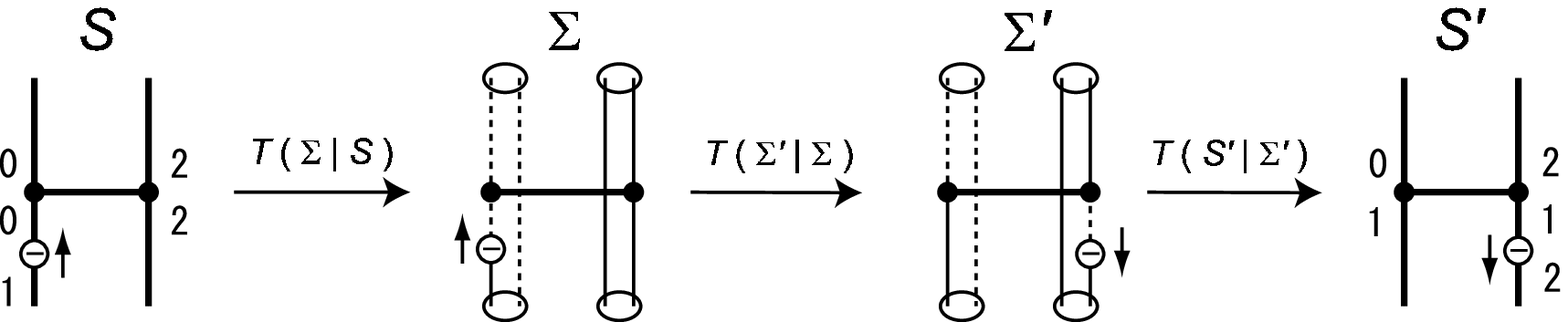}{0.45}{
The derivation of the scattering probability of the head
in the ``blurred'' algorithm.
$S_u$ and $S'_u$ are the initial and the final states in the blurred image
whereas $\Sigma_u$ and $\Sigma'_u$ are the corresponding states in the
original sharp image.
(The suffix $u$ is dropped in the figure for clarity.)
The numbers represent the ``brightness'' of the line.
}

Below we see an example that shows the efficiency of the coarse-grained
algorithm. Although the algorithm can be applied to an arbitrary $s$,
we only show the case for the $s=1$ antiferromagnetic Heisenberg chain 
in a uniform magnetic field.
This model has the freezing problem when simulated with the loop algorithm,
and it was one of the primary motivations for developing the coarse-grained
algorithm.
In \Fig{xxz-error}, we can see the performance of the coarse-grained
algorithm.
For comparison, we exploited the degrees of freedom in the
time-reversal symmetry equation and obtained many solutions.
Algorithms 1--4 in \Fig{xxz-error} are the ones chosen 
(in an ad-hoc manner) from them. 
(See the paper\cite{HaradaK2002} for how these were chosen.)
Plotted in \Fig{xxz-error} is $\Delta (M_{\pi}^2)N_v^{1/2}/L$,
where $\Delta (M_{\pi}^2)$ is the estimated statistical error  of the
squared staggered magnetization and $N_v$ is the average number of 
the vertices visited by the head during its lifetime.  
Here (only in this paragraph and in \Fig{xxz-error}) $L$ is the system size,
not the Trotter number or the order of the expansion.
Since the scattering process is the most time-consuming
part of the code,  the total CPU time is roughly proportional to the
total number of scattering events of heads, including the ``straight''
scatterings.  Therefore, the CPU time is proportional to $N_v$.  This
is why the statistical error should be multiplied by $N_v^{1/2}$ in
order to make the comparison fair.
In \Fig{xxz-error}, we can clearly see that the coarse-grained algorithm
performs as well as the best algorithm among the the other four (i.e.,
Algorithm 1).  Obviously, there is no exponential slowing-down in the
coarse-grained algorithm and Algorithm 1, as was the case 
with Sylju\r{a}sen and Sandvik's solution for $s=1/2$.

\deffig{xxz-error}{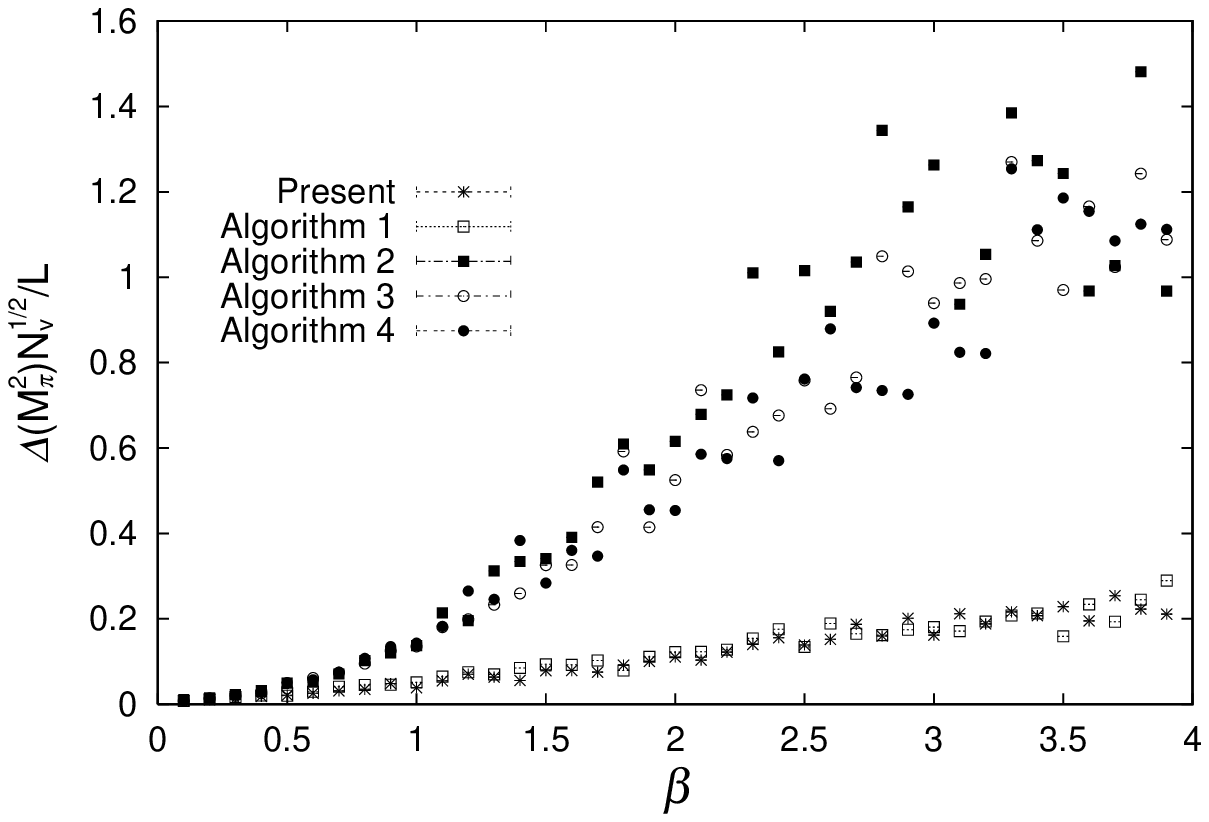}{0.5}{
The statistical error in the estimate of the squared staggered 
magnetization multiplied by the square root of the average number of 
scattering events during the lifetime of a worm.
The ``present'' algorithm is the coarse-grained algorithm 
described in the text.
The system is the $s=1$ antiferromagnetic
Heisenberg chain of the length of $64$ lattice spacings 
in a uniform magnetic field $H=0.1$.
Each point is a result of 50 sets of simulations, where each set 
consists of 20000 creations and annihilations of worms.
(Adopted from Harada and Kawashima\cite{HaradaK2002}.)
}

\subsection{Algorithms for Bosons}
In this section, we present an algorithm for simulating bosonic systems.
The algorithm\cite{SmakovHK2003} is based on mapping of bosonic models
to spin models and the coarse-graining discussed in \mssc{CoarseGrainedAlgorithm}.
The result is similar to the worm algorithm, as we see below.
While the ordinary directed-loop algorithm can also be used for the boson models
directly, a problem arises from the fact that the boson occupation number is
unbounded in general.
An artificial bound must be set to make the resulting
solution to the detailed balance equation \Eq{DirectedLoopWeightEquation}
meaningful.
The limitation is, however, undesirable
since the range of values that the occupation number 
takes on in the equilibrium is not known {\it a priori}.
While this is not a serious problem in a uniform model
where a typical value as well as a typical fluctuation in the
occupation number is known,
it can be serious in some cases, such as the soft-core boson 
model with random chemical potential;
the typical occupation number may largely vary from site to site
in the inhomogeneous potential.
The algorithm presented below is free from this problem.

In order to explain the idea, we consider a simple model of
non-interacting soft-core bosons on a $d$-dimensional hyper-cubic
lattice.
The Hamiltonian is
\begin{equation}
\eqlabel{ham_boson}
H=-\frac{t}{2}\sum_{(ij)}(b_i^{\dagger}b_j+b_j^{\dagger}b_i)-
\mu\sum_i b_i^{\dagger}b_i,
\end{equation}
where $t$ is the (positive) hopping amplitude, $\mu$ is the chemical
potential, and $b_i^{\dagger}$ and $b_i$ are the boson creation and
annihilation operators, respectively.
In addition, the chemical potential must satisfy $\mu \le -dt$.
In order to map the boson model to the spin model, we use
the Holstein-Primakoff transformation\cite{HolsteinP1940},
\begin{eqnarray*}
  S_i^+ & = & b_i^{\dagger}(2s - b_i^{\dagger} b_i)^{1/2}, \\
  S_i^- & = & (2s - b_i^{\dagger}b_i)^{1/2}b_i, \\
  S_i^z & = & b_i^\dagger b_i - S, 
\end{eqnarray*}
where $S_i^+, S_i^-$ and $S^{z}_i$ are spin operators on the $i$th site.  
With this transformation, the model of \Eq{ham_boson} is
approximately transformed to an XY spin model,
\begin{equation}
H=-\frac{t}{4S}\sum_{(ij)}\left(
S_i^{+}S_j^{-}+S_j^{+}S_i^{-}\right)-
\mu\sum_iS_i^z.
\end{equation}
In the limit of infinite $s$, this mapping is exact.
Therefore, if the infinite $s$ limit of the coarse-grained algorithm
of the spin system exists, it serves as an exact algorithm for
the boson system.
In the following, therefore, we consider the infinite $s$ limit of
the coarse-grained algorithm discussed in \mssc{CoarseGrainedAlgorithm}.

We first consider the beginning and the ending of a cycle;
the creation and the annihilation of a worm.
In the coarse-grained algorithm,
a spin-lowering worm
(i.e., the positive head (tail) above the negative tail (head)) 
is created with the probability $l/2s$.
Here, $l=0,1,2,\cdots,2s$ specifies the local spin state,
which corresponds to the number of bosons in the bosonic algorithm
presented below.
Since the number of particles is finite, 
the probability is zero in the limit of infinite $s$. 
Therefore, we always start a cycle with a spin-raising or
boson-creating worm (i.e., the negative head (tail) above the
positive tail (head)).
On the other hand, when the head meets the tail, 
it may annihilate with its partner or simply pass it. 
The probability of the annihilation depends on the type of the head.
If the head is of such a type that 
its passage increases the occupation number by one,
i.e., if it is positive and moving downward or
if it is negative and moving upward,
then the annihilation probability is $1/(2s-l)$
where $l$ is the local spin state between the the head and the tail
just before they come to the same location.
The probability is zero in the infinite $s$ limit.
Therefore, the annihilation takes place only for
a head whose passage decreases the occupation number by one,
and it happens with the probability $1/l$.

Next we consider the vertex assignment and the scatterings of 
the head at the vertices.
Since the density of vertices are proportional to $s$, at first
glance, assigning the vertices in the coarse-grained algorithm is
impossible in the limit of infinite $s$.
However, the head goes straight through most of the vertices.
The probability that the head changes the direction of motion 
at a vertex is proportional to $1/s$.
Therefore, the density of real scattering events, 
which is the product of the density of vertices and 
the scattering probability at a vertex, remains finite. 
With this density of the scattering event,
the imaginary time at which the next scattering happens 
can be generated by a Poisson process,
similar to what we do in taking the continuous-imaginary-time limit
in the loop algorithm (see \mssc{ContinuousImaginaryTimeLimit}).
In this way, we can make the head move and scatter
with a finite number of procedures.
(See \mssc{WormAlgorithmRecipe} for the details of the procedure.)

In \Fig{rho_s}, the result of the numerical simulation using
the present algorithm is shown together with the 
exact result.
Plotted is the superfluid density $\rho_s$ at $t \beta = 2$
as a function of the average occupation number $n$.
The transition point is around $n_{\rm c} \sim 0.6$.
With the present method, there is no major difficulty in
performing simulations near the critical point
as can be seen in \Fig{rho_s}.
Although not shown in the figure, we also tried a direct
application of the directed-loop algorithm with the
heat-bath-type solution discussed in \mssc{DirectedLoopAlgorithm}
to the present problem.
For the reason mentioned at the beginning of the present
section, we have to set the upper bound for the
occupation number.
When we set it to be 20, which is close to the minimal to 
perform a non-biased simulation, we observed that
the simulation becomes very slow at $n\sim n_c$.
It was practically impossible to do a simulation 
when the occupation number exceeds the critical value.

\deffig{rho_s}{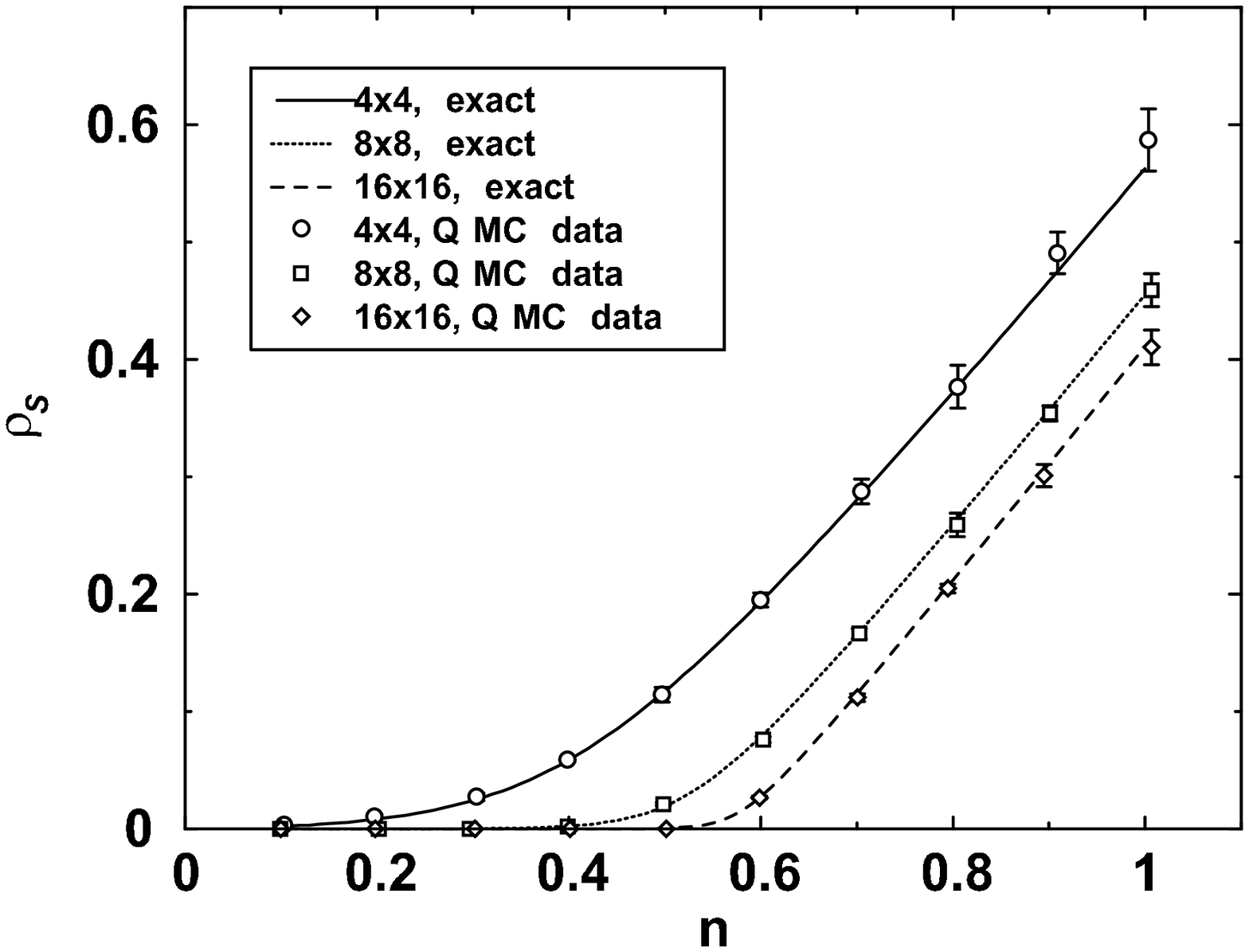}{0.4}{
  The superfluid density plotted against
  the average occupation number for the three-dimensional free 
  lattice-boson system. The lines are the exact analytical values.
  (Adopted from \v{S}makov et al\cite{SmakovHK2003}.)
}

\subsection{Negative-Sign Problem and Meron Algorithm}
The negative-sign problem is unarguably the worst obstacle
in numerical simulations of quantum models.
It is originated in the negative matrix elements of the Hamiltonian.
When some of the off-diagonal matrix elements of the local Hamiltonian
$\Ham_u$ are negative, in general the weight \Eq{BoltzmannWeightQ} 
can be negative for some of the states $S$.
In such a case, we perform a Monte Carlo simulation of which the
target distribution is $|W(S)|$, not $W(S)$.
Then, we estimate an arbitrary quantity $Q$ using the identity
\begin{eqnarray}
  \langle Q \rangle_{\rm thermal} 
  & = &
  \frac{\sum_S |W(S)| \sgn{S} Q(S)}{\sum_S |W(S)| \sgn{S}} 
  \nonumber \\
  & = &
  \frac{\sum_S |W(S)| \sgn{S} Q(S) / \sum_S |W(S)|}{\sum_S |W(S)| \sgn{S} / \sum_S |W(S)|}
  \nonumber \\
  & = &
  \frac{\langle \sgn{S} Q(S) \rangle_{\rm MC}}{\langle \sgn{S} \rangle_{\rm MC}},
  \eqlabel{NegativeSign}
\end{eqnarray}
where $\sgn{S} = \pm 1$ is an abbreviation of $\sgn{W(S)}$
and $\langle \cdots \rangle_{\rm MC}$ is
the Monte Carlo average with the weight $|W(S)|$.

The negative contribution to the partition function, $Z_-$, 
cancels out with a part of the positive contribution, $Z_+$, 
and the total $Z = Z_+ - Z_-$ must be always positive.
In fact, in many cases, the negative contribution 
{\it almost completely} cancels out the positive one.
This can be seen\cite{HatanoS1992}
by considering a fictitious Hamiltonian $\Ham'$,
whose matrix elements are the absolute value of the corresponding
matrix elements of the original Hamiltonian, $\Ham$;
$$
  \langle \psi' | \Ham' | \psi \rangle \equiv
  \left\{\begin{array}{ll}
  \langle \psi' | \Ham | \psi \rangle & (\psi' = \psi) \\
  \left| \langle \psi' | \Ham | \psi \rangle \right| & (\psi' \ne \psi)
  \end{array}\right..
$$
The difference between the free energy per site 
$\Delta f \equiv (F' - F)/N$ is of $O(N^0)$,
when the difference between the two Hamiltonians is extensive.
It follows that
$$
  \frac{Z_+ - Z_-}{Z_+ + Z_-} = \frac{Z}{Z'}
  = e^{-\beta N \Delta f},
$$
where $Z'$ is the partition function of the fictitious system.
It is clear from this expression that the positive and the negative
part cancel out almost completely at a low temperature and/or for a large size,
and it becomes practically impossible to estimate
the denominator (as well as the numerator) in \Eq{NegativeSign}
because of the statistical error.

In some special cases, the loop algorithm is useful for 
solving the negative-sign problem in fermionic systems.
Considering that there are not many cases where the negative-sign
problem is overcome with or without the loop algorithm,
it may be worthwhile to mention here the meron algorithm
\cite{ChandrasekharanW1999,ChandrasekharanCOW2002}
by which we can overcome the negative-sign problem in some cases.
Let us consider the simplest non-trivial model for the fermion,
\begin{eqnarray*}
  \Ham & = & 
  - t \sum_{(ij)} ( c_i^{\dagger} c_j + c_j^{\dagger} c_i ) \\
  & &       + U \sum_{(ij)} ( n_i - 1/2 ) ( n_j - 1/2 ),
\end{eqnarray*}
where $t > 0$ and $ U \ge 2t $. 
The spin degrees of freedom are absent.
We first consider the case $ U = 2t $ and then describe how to
generalize the algorithm to the case $ U > 2t $.

Apart from the fermion sign due to the exchange of particles,
when $U=2t$, this problem is identical to the
$s=1/2$ antiferromagnetic Heisenberg model
(eq.\ \Eq{XYZHamiltonian} with $J_x = J_y = J_z$).
Therefore, the graph decomposition of the Boltzmann weight can be
done only with the horizontal graph $g_{\rm H}$ (See \Tab{GraphElements}).
Using the world-line representation discussed above,
we can express the partition function as
$$
  Z = \sum_{S} \sgn{S} W_L(S).
$$
(Here, the absolute value of the weight in \Eq{BoltzmannWeightQ}
is simply written as $W_L(S)$.)
The sign is positive if and only if the world-lines of the fermions 
corresponds to an even permutation of particles.
The weight $W_L(S)$ is the same as the one for the $s=1/2$
antiferromagnetic Heisenberg model.
Applying the graphical decomposition of the weight\Eq{GraphDecomposition},
we can rewrite it as
$$
  Z = \sum_{S,G} \sgn{S} W_L(S,G).
$$
Let us choose an arbitrary $G$ and fix it, then
consider a loop variable $\sigma_l = 0,1$ assigned to every loop $l$ in $G$.
(We can specify a state $S$ compatible to $G$ by specifying
the values of these variables.)
The following properties can be shown for 
the sign of the compatible states to $G$:
(i) The sign $\sgn{S}$ can be expressed as a product of factors each of which is
    a function of only one of the loop variables, and
(ii) if the sign $\sgn{S}$ does not depend on the choice of the
   loop variables, $\sgn{S} = 1$.

The example of the factorization (the property (i)) is shown in \Fig{Meron}.
There are two loops in the middle diagram, c and d, whose flipping changes
the sign of the whole configuration.
Note that these loops always cause a change in the sign 
no matter what the initial state may be.
On the other hand, all the other loops do not cause any sign change.
Generally, whether a flip of a loop causes a sign change or not depends
only on the geometric feature of the loop, not on the initial spin
configuration.
Therefore, one can determine if a given loop may cause
the sign-change or not without knowing the state $S$.
The loops that cause sign change are called {\it merons}.
In \Fig{Meron}, there are two merons, c and d.

It is easy to show the positivity of the state with no merons
(the property (ii)).
It suffices to show that there is at least one positive configuration
among the ones compatible to $G$.
But the sign of the Ne\'el state is positive and 
the Ne\'el state is compatible with any $G$.
To see the latter, let us notice that
the constraint imposed by any graph is simply that
when we trace a loop the local spin state must 
alternate everytime we make the horizontal (spatial) move.
The Ne\'el state obviously satisfies this condition
and is compatible to any $G$.

Because of the two properties, the sign can be expressed as
$$
  \sgn{S} = \prod_{l} (\epsilon_l)^{\sigma_l},
$$
where $\epsilon_l = -1$ for merons whereas $\epsilon_l = 1$ for the other loops.
The variables denoted as $\sigma_l$ are the loop variables;
each of them takes $0$ or $1$.
They are defined so that $\sigma_l = 0$ for all the loops $l$ if 
the whole system is in one of the two Ne\'el states.
(The choice of the Ne\'el state does not matter because
there are always an even number of merons.)
The whole partition function now becomes
$$
  Z = \sum_{G} \sum_{\{\sigma_l\}} W(S(\{\sigma_l\}),G) 
  \prod_{l} \epsilon_l^{\sigma_l},
$$
where $S(\{\sigma_l\})$ is the state specified by the loop variable $\sigma_l$.
But the summation over $\{\sigma_l\}$ is zero if there is a meron in $G$.
Therefore, the summation over all $G$ can be replaced by the summation over
all meron-free $G$;
$$
  Z = \sum_{G \atop \mbox{{\scriptsize (no meron)}}} V(G) 2^{N_{\rm C}(G)}
$$
where
$$
  V(G) \equiv \sum_{\{\sigma_l\}} W(S(\{\sigma_l\}),G).
$$
Thus, it becomes apparent that we can avoid the negative sign problem 
if it is possible to perform a simulation in the restricted phase space
of the meron-free graphs.

This can be achieved in the following way.
Every time insertion or removal of a graph element is attempted,
we check whether the attempt would create merons.
If it would, such an attempt is rejected.
The actual insertion or removal is executed only when it does not 
create merons.
This checking procedure requires a relatively large amount of
computational time and raises the complexity of the algorithm.
This additional complication, however, is of the algebraic type
at most and pays off considering the exponential complexity of the 
simulation with negative signs.
In an actual simulation, because of the necessity of computing the 
susceptibility, we soften the condition of no merons.
Namely, we need to sample the graphs with two merons in order to estimate
the susceptibility.
Therefore, sampling of graphs is usually done for two-meron graphs as
well as meron-free graphs.
In this case, the insertion and the removal are rejected only 
when an attempt is made to create the third meron.
For the same purpose, instead of placing a strict upper limit in the
meron number, we can also use an extended-ensemble method 
with respect to the meron number, in which we consider a fictitious
weight $W(n_{\rm meron})$ for controlling the meron number.
(For the extended-ensemble method, see \mssc{ExtendedEnsembleMethods}.)

The algorithm can be easily generalized to the case where 
the easy-axis anisotropy exists, i.e., the case where $U>2t$ ($J_z > J_x$).
In the easy-axis case, we have to introduce the binding graph $g_{\rm HB}$
that bind two loops into one cluster. (See \Tab{GraphElements}.)
For instance, if two merons are bound they form a non-meron.
Insertion or removal of these binding graphs is performed
in the same way as above; first count the number of new merons
that would be created or annihilated by the attempt and then
accept or reject the attempt according to the restriction or the
fictitious weight mentioned above.
\deffig{Meron}{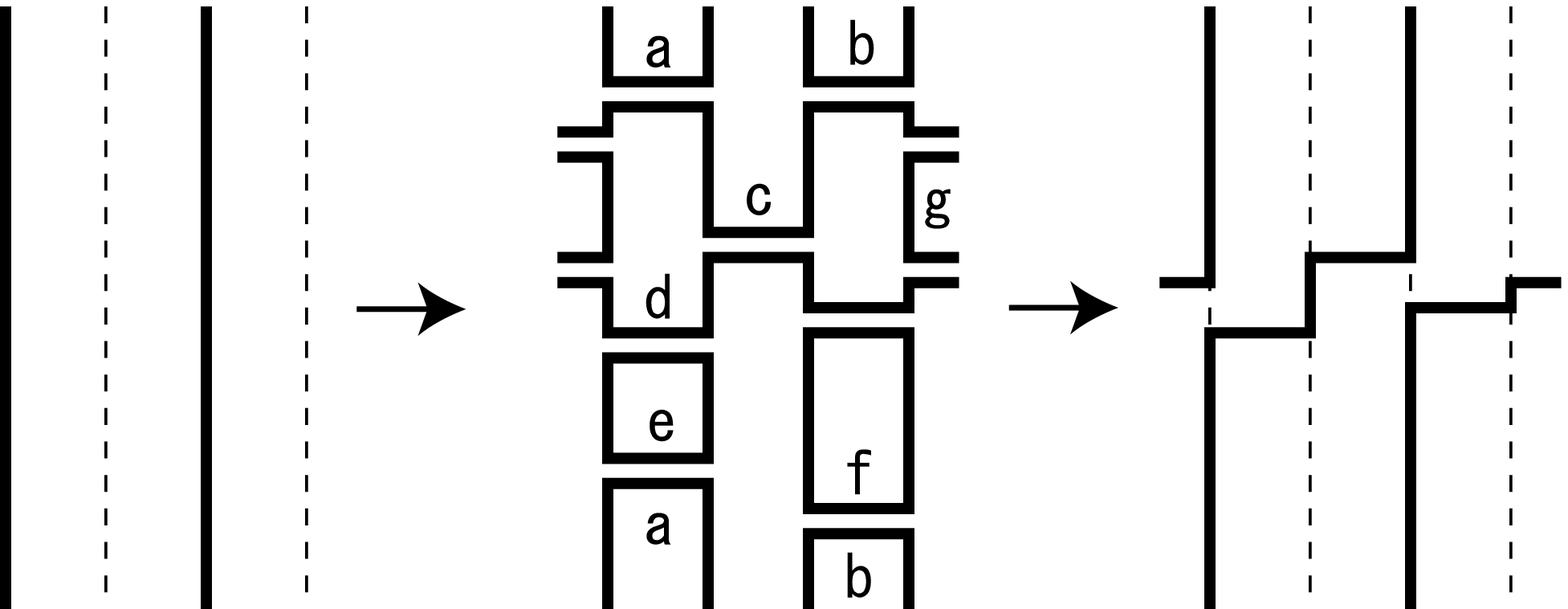}{0.38}{
A loop decomposition (from the left to the middle)
and a loop flip (from the middle to the right) 
in a four-spin chain of the $s=1/2$ antiferromagnetic 
Heisenberg model with the periodic boundary condition.
The initial spin configuration is the perfect Ne\'el state.
The only non-trivial graph elements are the horizontal graphs
in \Tab{GraphElements}.
(All the graphs are drawn with solid lines for clarity
whereas dashed lines are used in \Tab{GraphElements} 
for emphasizing the relative spin orientations)
The loops, $c$ and $d$, in the middle diagram are merons.
From the middle to the right, the loop $d$ is flipped.
As a result, the two particles are exchanged in the left diagram,
which makes the sign of the whole diagram negative.
}

\subsection{$T=0$ Simulation}
The ground state properties are of particular interest in low-dimensional models.
While the ground state properties can be deduced, in principle,
by extrapolating the finite temperature results to zero temperature,
it is prone to the systematic error due to the extrapolation 
especially when we do not have a solid knowledge about how low
the temperature must be to obtain a reasonable extrapolation.
Evertz and von der Linden\cite{EvertzL2001} proposed a method 
for directly obtaining the zero-temperature expectation values 
for a given model without extrapolation.
The method is a special application of a more general method
(proposed also by them)
for obtaining results for an infinite lattice.
(Note that a system at zero temperature is regarded as
a system with an infinite length in the imaginary time direction.)
Since this idea can be most clearly described using an example of the
ferromagnetic Ising model, let us consider this case first.

The method is closely related to Wolff's single-cluster algorithm\cite{Wolff1989}.
In Wolff's single-cluster algorithm, we start the construction of 
a cluster from a null cluster that contains no spin.
Then, we choose a spin at random from the whole system and 
include it in the cluster.
Next, we assign bonds between the spins sorrounding the cluster 
and the ones already in the cluster in the same way as
the SW algorithm.
If a new spin is bound to a spin that is already in the cluster,
the new spin is also included in the cluster.
This procedure is continued until there is no surrounding spins
to be checked.
When the cluster construction is finished, 
the spins in the cluster are flipped.
The Wolff's procedure is statistically the same as
constructing clusters by the SW algorithm for the whole system,
choosing a point at random, and then flipping the cluster containing
the chosen point.
However, the Wolff's procedure is obviously better in efficiency
for constructing a single cluster because bond assignments 
to the spins outside of the formed cluster are a waste of time.

Roughly speaking the method we discuss here is the Wolff's
procedure applied to an infinite system.
The difference arises from the fact that choosing a point
at random from an infinite system does not make much practical sense.
Therefore, we simply stick to the same point, which we call the origin,
throughout the simulation in the new method.
In addition, we have to make sure that the cluster does not percolate,
since if it does we cannot finish the cluster construction within a
finite computational time.
To this end, we starts from the Ne\'el state
as the initial spin configuration.
Due to this choice of the initial spin configuration,
the first round of the cluster construction is destined to be very short
since we can assign no bond that connects the origin to its nearest neighbor.
After flipping this first cluster, i.e., the one that consists of the
origin only, we start the second round.
We can go a little further this time.
Since the spin at the origin is now aligned with its neighbors,
we assign bonds between them with a non-zero probability.
As a result, the cluster that we obtain from the second round
tends to be larger than the first one.
In this way, we can go on.
It should be noted that we only have to store the spin configuration
of the part that is modified in the simulation at least once
and do not have to keep the spin configuration beyond the frontier.
It must be also noted that the frequency of visiting a site
whose distance from the origin is $R$ is proportional to
the correlation function $\Gamma(R) \equiv \langle S(R) S(0) \rangle$.
Therefore, the `already visited' region does not grow too fast
after its size reaches the correlation length.
This means that the whole process is manageable as long as the 
system is in the disordered state.
Since the distribution of the spin configuration within the region 
approaches the equilibrium one, one can in principle compute any
correlation function that can be defined within this region.

The procedure can be easily generalized for studying the 
disordered state of the quantum system.
For example, in the case of the $s=1$ antiferromagnetic 
Heisenberg model, which is disordered even at $T=0$ due to
the Haldane gap, let us consider the split-spin loop algorithm.
The initial spin configuration is the complete Ne\'el state
where the up-spin straight world-lines and the down-spin ones
alternate.
Initially, only the local spin configuration at the origin is 
stored in the computer memory.
The head is placed on the origin and the direction of the
initial motion is set to be upward or downward (with probability 1/2).
We then generate the scattering object in this direction
at some stochastically chosen distance from the origin.
The distance is generated following the Poissonian process
with the density $J/2$ for each nearest neighbor site, 
as we have seen in \mssc{DirectedLoopAlgorithm}.
At the collision, the head changes its direction of motion
as well as the location, as it does in the directed-loop algorithm.
The same procedure is repeated until the head comes back to the origin.
The spin configuration is stored only for the part 
that has been visited by the head.

When the system has zero energy gap, a naive application of this
method would fail,
since the `already visited' region would glow endlessly.
This can be avoided, however, by using a system whose 
spatial size is finite while its temporal length is 
infinite ($\beta = \infty$).
This is because there is usually a finite gap associated 
with the system's spatial dimension.
Therefore, we can at least perform a zero-temperature simulation
for finite systems.
This is still somewhat advantageous compared 
to ordinary simulations for which one needs two kinds of extrapolations,
the one with respect to the size and the other with respect to the temperature.

\subsection{Extended-Ensemble Methods}
\ssclabel{ExtendedEnsembleMethods}
While extending the state space by introducing the auxiliary variables $G$
is a powerful way of speeding up the simulation as we have seen,
there is another general strategy for efficient simulations.
That is, we can overcome a slow relaxation by using a
fictitious target weight, $\tilde W(S)$ instead of the given weight $W(S)$.
Simulation methods based on this idea are called extended-ensemble methods.
In most of the extended-ensemble methods, 
one does not fix the target weight $\tilde W(S)$ 
but use it as an adjustable function.
One of the successful examples is the multi-canonical Monte Carlo (MCMC) method.
\cite{BergN1991,BergN1992,Lee1993}
In the MCMC method, the weight depends on the state $S$ only through
the energy $E(S)$. Therefore, we have
$$
  \tilde W(S) = \tilde w(E(S)),
$$
and the function $\tilde w(E)$ is adjusted adaptively to make the frequency of having 
the event $E=E(S)$ independent of $E$.
To this end, several sets of simulation may be performed.
For the first set, the initial guess of the appropriate weight
that is often used is $\tilde w(E) = {\rm const}$.
In every set, the histogram $h(E)$, i.e.,
the number of times $E(S)$ takes the value $E$, is recorded.
In other words, every time the configuration is changed we do
\begin{equation}
  h(E(S)) := h(E(S)) + 1. \eqlabel{HistogramUpdate}
\end{equation}
At the end of each set, 
the fictitious weight $\tilde w(E)$ is updated as
$$
  \tilde w(E) := \tilde w(E) / h(E),
$$
and the new weight is used for the next set.
In each set, the Metropolis single-spin-flip Monte Carlo method is used with
the target weight $\tilde W(S) = \tilde w(E(S))$ 
for updating the spin configuration.

At the end of the last set, one can obtain the density of states $g(E)$, as
$$
  g(E) \equiv \sum_S \delta_{E,E(S)} \approx {\rm const} \times (\tilde w(E))^{-1}.
$$
With this $g(E)$, we can compute the canonical average 
of an arbitrary quantity $Q$ as 
$$
  Q(T) \equiv \sum_E g(E) w(E) Q(E) \Big/ \sum_E g(E) w(E),
$$
where $Q(E)$ is the micro-canonical average of $Q$ at the energy $E$,
which we can compute in the last set of simulation.

While this procedure turned out to be quite useful in studying various
systems with slow dynamics, one drawback was noticed.
That is, the visited range of $E$ do not widen much in each set of simulation,
partially due to the poor initial guess of the weight $\tilde w(E)$ and 
also to the slow diffusive nature of the random walk in the energy space.
This drawback was removed in Wang and Landau's variant of the
MCMC method\cite{WangL2001}.
In their method, $\tilde w$ is updated every time an attempt is made at 
changing the spin configuration.
The update is done as
\begin{equation}
  \tilde w(E) := r \times \tilde w(E)
  \eqlabel{WLmethod}
\end{equation}
where $0 < r < 1$ is the reduction factor fixed throughout each set of 
simulation.
At the beginning of each set, the histogram is reset, i.e., $h(E) := 0$
for all $E$, and the reduction factor is updated 
as $r := r^\alpha$ where $0< \alpha <1$,
while the weight $\tilde w(E)$ is kept unchanged.
Each set does not have a prefixed duration, but is terminated when
the histogram becomes approximately flat.
Since the dynamic update \Eq{WLmethod} strongly penalizes 
the random walker's staying at the same value of the energy,
the already visited region widens much faster than the ordinary MCMC.

Since these extended-ensemble methods are complementary to the 
algorithms described in previous subsections, 
it is natural to consider the combination of the two.
However, since the weight $\tilde W(S)$ is not the function of the
energy observable in a typical quantum Monte Carlo method,
there is no good reason to assign a special role to the energy observable
in the quantum case.
In addition, the value of the energy observable is not a multiple of
a single constant, which is inconvenient for the present purpose.
In the loop algorithm (and in the directed-loop algorithm),
therefore, the Wang-Landau method was used
\cite{YamaguchiK2002,TroyerWA2003,YamaguchiKO2002}
with the histogram of the number of vertices (or graph elements),
$n(G) = \sum_u G_u$, rather than the energy.
(The idea was originally proposed by Janke and Kappler
\cite{JankeK} and applied to the Ising model.)

Two types of the implementation are possible;
in one, the pair Hamiltonian is decomposed graphically,
\cite{YamaguchiK2002} as in the loop algorithm, 
and in the other, it is used undivided
\bigcite{TroyerWA2003} as in the directed-loop algorithm.
We present below a continuous-time variant of 
an algorithm closer to the latter.

\deffig{WL}{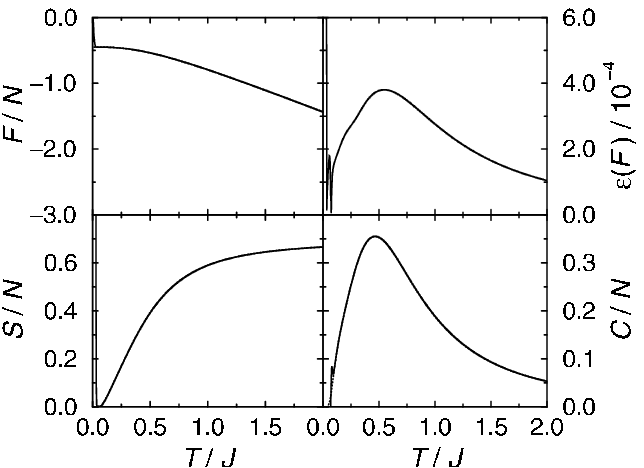}{0.4}{
The free energy $F$, the entropy $S$, and the specific heat $C$ of
the antiferromagnetic Heisenberg model in one dimension.
The system's linear dimension is $10$ lattice spacings.
The relative error in the free energy $\epsilon(F) \equiv$
$|(F-F_{\rm exact})/F_{\rm exact}|$ is also shown.
(Adopted from Troyer, Wessel and Alet\cite{TroyerWA2003}.)
}

We start from \Eq{TheWeight} with $u$ being $(i,j)$.
We control the histogram by introducing an adjustable weight
$f(n)$ and replacing $(\Delta\tau)^n \equiv \beta^n / L^n$ 
by $f(n) / L^n$ in \Eq{TheWeight}, which yields
\begin{eqnarray}
  Z & = & \lim_{L\to \infty} \sum_S W_L(S), \nonumber \\
  W_L(S) & = & \sum_G W_L(S,G), \nonumber \\
  W_L(S,G) & = & \frac{f(n(G))}{L^n} \prod_u 
  \langle \psi'_u | (-\Ham_u)^{G_u} | \psi_u \rangle. \nonumber
\end{eqnarray}
We then perform a Monte Carlo simulation in the $(S,G)$ space
so that the detailed balance condition is satisfied with 
respect to the weight $W_L(S,G)$.

One sweep of the simulation consists of two steps,
as in the ordinary directed-loop algorithm.
In the first step we update $G$.
To do this, we decompose the whole system into 
uniform intervals (UIs) with no kink.
For each possible local state, $S_u$, 
we consider the class of the UIs on which the local state is $S_u$.
Let the sum of the lengths of all the UIs in this class be $I(S_u)$
or simply $I$.
We have $IL$ local units in this class.
(Here we use the convention in which the total length
of the lattice along the imaginary time direction is 1.)
Our task, then, is to set variables $G_u$ for 
all of these local units.
To do this, we first remove all the existing vertices from the UIs.
There are ${}_{IL}C_m$ ways of assigning $m$ vertices to $IL$ units.
Therefore, the probability of placing $m$ vertices becomes
$$
  P(m) \propto 
  \left(\begin{array}{c} IL \\ m \end{array}\right)
  \frac{f(n_0 + m)}{L^m} w^m,
$$
where $w$ is the local weight 
$w \equiv \langle \psi_u | (-\Ham_u) | \psi_u \rangle$
for the currently chosen type of UIs,
and $n_0$ is the total number of vertices before placing $m$ vertices.
In the limit $L\to\infty$ this leads to
$$
  P(m) = \frac1{A} \frac{(Iw)^m f(n_0+m)}{m!},
$$
where $A$ is the normalization constant.
The procedure of assigning graphs to the UIs follows 
directly from this expression.
That is, we first generate a uniform random number $r$ ($0<r<1$)
and find the integer $m$ that satisfies
$$
  X(m-1) < r < X(m),
$$
where $X(m) \equiv \sum_{m'=0}^m P(m')$.
We then choose $m$ points uniform-randomly from the intervals 
belonging to the current class and place non-trivial graph elements there.
We repeat this procedure for all the classes.

In the next step we update the spin configuration.
However, this step can be done in exactly the same way as
described in \mssc{DirectedLoopAlgorithm} using worms.

It should be noted here that in the procedure given above,
the histogram updating \Eq{HistogramUpdate} (with $E(S)$ replaced
by $n(G)$) can be done only once in a sweep 
because there is no intermediate state in the procedure. 
On the other hand, in other methods such as the 
original algorithm proposed by Janke and Kappler\cite{JankeK},
the histogram updating can be done after every local assignment 
of a vertex.
This difference would make the statistical error greater than
it should be for the present method.
This apparent disadvantage can be easily avoided by adding
$P(n)$ to the histogram, rather than 0 or 1.
Specifically, we should replace \Eq{HistogramUpdate} by
\cite{YamaguchiK2002}
$$
  h(n) := h(n) + P(n).
$$

The utility of the extended-ensemble methods can be demonstrated 
best in the computation of the quantities directly related to the
density of states, such as free energy, entropy and specific heat. 
These quantities can be computed easily with the aid of the extended-ensemble methods.
In \Fig{WL}, some results\cite{TroyerWA2003} of the computation 
with the extended-ensemble method is shown.
The method used to obtained the results
is based on the discrete-time formulation and
the details are different from the one discussed here.
However, the basic idea is the same 
and the efficiency is believed to be similar.
Because of the cut-off (the maximum order of the expansion), 
the results deviate from the exact results at low temperatures
($T/J<0.1$) whereas they are indistinguishable from the exact results
at high temperatures ($T/J>0.1$) in the scale shown in \Fig{WL}.
(Note, however, that the accessible temperature range can be easily widen
by modifying the cut-off.)

It was shown recently\cite{YamaguchiKO2002} that
a further improvement can be made by employing the broad-histogram method
\cite{Oliveira,WangTS}.
Compared to the other extended-ensemble methods,
this method is unique in that it is based on the exact relation between
the expectation value of the transition matrix and the density of states.
While one estimates the density of states directly from the histogram itself
in the ordinary extended-ensemble methods,
the micro-canonical average of the transition matrix
is used in the broad-histogram method.
This estimator generally gives a better result than the histogram itself
because of its macroscopic nature.
An improvement of more than one order of magnitude in the relative
error was reported\cite{YamaguchiKO2003} in the case shown in \Fig{WL}.

\subsection{Estimators and Efficiency of Algorithms}
\ssclabel{EstimatorsAndEfficiency}
As discussed in \mssc{PathIntegral}, the local updating algorithm
is slow to approach the equilibrium because the size of the
region modified at a time is fixed and can be much smaller than the
correlation length.
On the other hand, the size of the modified region is 
roughly equal to the correlation length (both in space and in time)
in the three algorithms discussed in this article;
the loop, the worm and the directed-loop algorithm.
This can be seen by considering the estimators of Green's functions.
In this section, we therefore discuss estimators of various quantities 
including Green's functions.

Let us consider a quantity $\hat Q$ that can be decomposed in the same way
as the Hamiltonian is in \Eq{HamiltonianDecomposition};
$$
  \hat Q = \sum_b \hat Q_b.
$$
Then, its canonical average can be expressed with the source term as
\begin{equation}
  \langle \hat Q \rangle \equiv 
  \frac1{\beta} \left[ \left. \frac{d}{dH} Z(H) \right/ Z(H) \right]_{H\to0}
  \eqlabel{measurement}
\end{equation}
where
\begin{equation}
  Z(H)  \equiv \Tr{ e^{-\beta (\Ham - H \hat Q) } }.
  \eqlabel{extended-Z}
\end{equation}
Accordingly, the weight \Eq{BoltzmannWeightQ} is modified as
\begin{equation}
  W'_L(S) \equiv \prod_u
  \left\langle \psi'_u \left|
      \left[ 1 -\Delta\tau \left(\Ham_u - H Q_u\right) \right]
    \right| \psi_u \right\rangle.
\end{equation}
Substituting \Eq{extended-Z} for \Eq{measurement}, 
we can express the thermal average of $\hat Q$ as the
Monte Carlo average of the estimator $Q(S)$ as
\begin{equation}
  \langle \hat Q \rangle_{\rm thermal} = 
  \langle Q(S) \rangle_{\rm MC}
  \eqlabel{rep-of-Q}
\end{equation}
with
\begin{eqnarray}
  Q(S) & \equiv & \frac1{\beta} \sum_u q(S_u), \nonumber \\
  q(S_u) & \equiv &
  \frac{\left\langle \psi'_{u} \left| \Delta\tau Q_u \right| \psi_{u} \right\rangle}
{\left\langle \psi'_{u} \left| 1 -\Delta \tau \Ham_u \right| \psi_{u} \right\rangle}.
  \eqlabel{rep-of-Q-1}
\end{eqnarray}

In the limit of $L\to\infty$, eq.\ \Eq{rep-of-Q-1} becomes
\begin{equation}
  q(S_u)
  = \left\{\begin{array}{cc}
      \Delta\tau \left\langle \psi_{u} \left| Q_u \right| \psi_{u} \right\rangle
    & (\psi'_{u} = \psi_{u})\\
      \frac{\left\langle \psi'_{u} \left| Q_u
      \right| \psi_{u} \right\rangle}
      {\left\langle \psi'_{u} \left| (-\Ham_u) \right| \psi_{u} \right\rangle}
      & (\psi'_{u} \ne \psi_{u})
    \end{array}
  \right..
\end{equation}
Therefore, if the $Q$ is expressed as a diagonal operator, 
the above equation is reduced to
\begin{equation}
 Q(S) 
  = \frac1{\beta} \int_0^\beta d\tau  Q(\tau) \quad (L\to\infty),
\end{equation}
where $Q(\tau) \equiv \langle \psi(\tau)| Q | \psi(\tau) \rangle$.
For example, the magnetization $Q \equiv \sum_i S^z_i$
can be easily evaluated with this formula.
In general, however,
we cannot ignore the contribution of kinks ($\psi'_{u} \ne
\psi_{u}$) in $Q(S)$. 
For example, the contribution of a kink
to the total energy $\langle \Ham \rangle$ is $O(-1/\beta)$. 

Next, suppose that the operator $Q_u$ has a non-zero 
off-diagonal matrix element for some local state whereas
the corresponding matrix element of the Hamiltonian is zero.
In such a case, the estimator $q(S_u)$ diverges,
and we cannot define $Q(S)$ to start with.
An example is the measurement of the squared magnetization 
in the transverse direction, $Q \equiv (\sum_i S^x_i)^2$,
with the Hamiltonian being that of the $XXZ$ Hamiltonian.

By using the loop algorithm, 
such non-diagonal quantities may be measured\cite{BrowerCW1998}.
The idea is based on the improved estimator.
An improved estimator is an estimator defined in terms of the
graph degrees of freedom rather than the original spin 
(or occupation number) degrees of freedom.
A classical example is Wolff's estimator
\cite{Wolff1989} of the susceptibility
of the ferromagnetic Ising model.
An improved estimator can be generally derived as follows.
Consider first the case where an ordinary estimator $Q(S)$ 
can be defined for a quantity $\hat Q$.
The thermal average can be expressed as
\begin{eqnarray*}
  & & 
    \langle Q \rangle_{\rm thermal} =  \frac{\sum_S W(S) Q(S)}{\sum_S W(S)} \\
  & & \quad = 
    \frac{\sum_{S,G} W(S,G) Q(S)}{\sum_{S,G} W(S,G)} 
  =  \frac{\sum_{G} W(G) Q(G)}{\sum_{G} W(G)},
\end{eqnarray*}
where $Q(G)$ is the fixed-graph average of $Q(S)$:
\begin{equation}
  Q(G) \equiv \left. \sum_S W(S,G) Q(S) \right/ W(G).
  \eqlabel{ClassicalImprovedEstimator}
\end{equation}

In the case of the magnetic susceptibility of the Ising model,
we take $Q(S) \equiv N^{-1}(\sum_i S_i)^2$.
Then, eq.\ \Eq{ClassicalImprovedEstimator}, with $W(S,G)$ 
defined by \Eq{ExtendedWeight} and \Eq{LocalExtendedWeight},
is reduced to
$$
  Q(G) = 2^{-N_{\rm C}(G)} \sum_S \Delta(S,G) Q(S),
$$
where $\Delta(S,G)$ is defined in \Eq{ClusterFlippingProbability}.
Let us define cluster variables $\sigma_c$ so that 
$S_i = \sigma_{c(i)}$ with $c(i)$ being the specifier of the
cluster to which $i$ belongs.
Then, $Q(S) = N^{-1} \sum_{i,j} S_i S_j$ can be rewritten as
$Q(S) = N^{-1} \sum_{c,c'} V_c V_{c'} \sigma_c \sigma_{c'}$
where $V_c$ is the total number of sites in the cluster $c$.
Therefore, we obtain
\begin{equation}
  Q(G) = N^{-1} \sum_c (V_c)^2
  \eqlabel{IEClassicalSusceptibility}
\end{equation}
as the graphical estimator of the squared magnetization.

A similar argument can be used for deriving an improved
estimator of the uniform magnetic susceptibility of a quantum spin models,
$$
  Q \equiv \chi_{zz}
  = {N}^{-1}\int_0^{\beta} d\tau 
  \langle M_z(\tau) M_z(0) \rangle.
$$
The ordinary estimator of this quantity is
$$
  Q(S) = (N\beta)^{-1} \left(\int dX\, S^z(X)\right)^2,
$$
where $X \equiv (i,\tau)$ and the integral $\int dX$
stands for $\sum_i \int d\tau$.
Then, similar to the case of the Ising model,
the improved estimator can be expressed
by \Eq{IEClassicalSusceptibility} with $V_c$ replaced by the
cluster magnetization $M_c$,
$$
  M_c \equiv \int_c dX\, S^z(X).
$$

Now, let us consider the case where $Q(S)$ cannot be defined,
as is the case with off-diagonal quantities such as
the transverse susceptibility
$$
  Q = (N\beta)^{-1} {\cal T} \int dX dY S^x(X) S^x(Y),
$$
where 
$X = (i,\tau)$ and $Y = (j,\tau')$ specify two points
in the space-time.
The symbol ${\cal T}$ indicates the time-ordered product.
We can express its thermal average as
\begin{eqnarray*}
  \chi_{xx}
  & \equiv & 
  \langle Q \rangle_{\rm thermal} \\
  & = &
  (N\beta)^{-1} \int dX dY Z^{-1} {\sum_S}' W'(S) \\
  & &
  \times\langle \psi'(X) | S^x_i | \psi(X) \rangle
  \langle \psi'(Y) | S^x_j | \psi(Y) \rangle,
\end{eqnarray*}
where $\psi'(X)$ and $\psi(X)$ are the states just above and below
the point $X$, respectively.
The prime in $\sum'_S$ indicates that the
summation is taken over all the states that have 
discontinuities at $X$ and $Y$.
Such a state is shown in \Fig{OffDiagonalGreensFunction}.
The prime in $W'(S)$ indicates that it allows the discontinuities
at the two points.
(Note $W(S) = 0$ when the state $S$ has any discontinuities.)

\deffig{OffDiagonalGreensFunction}{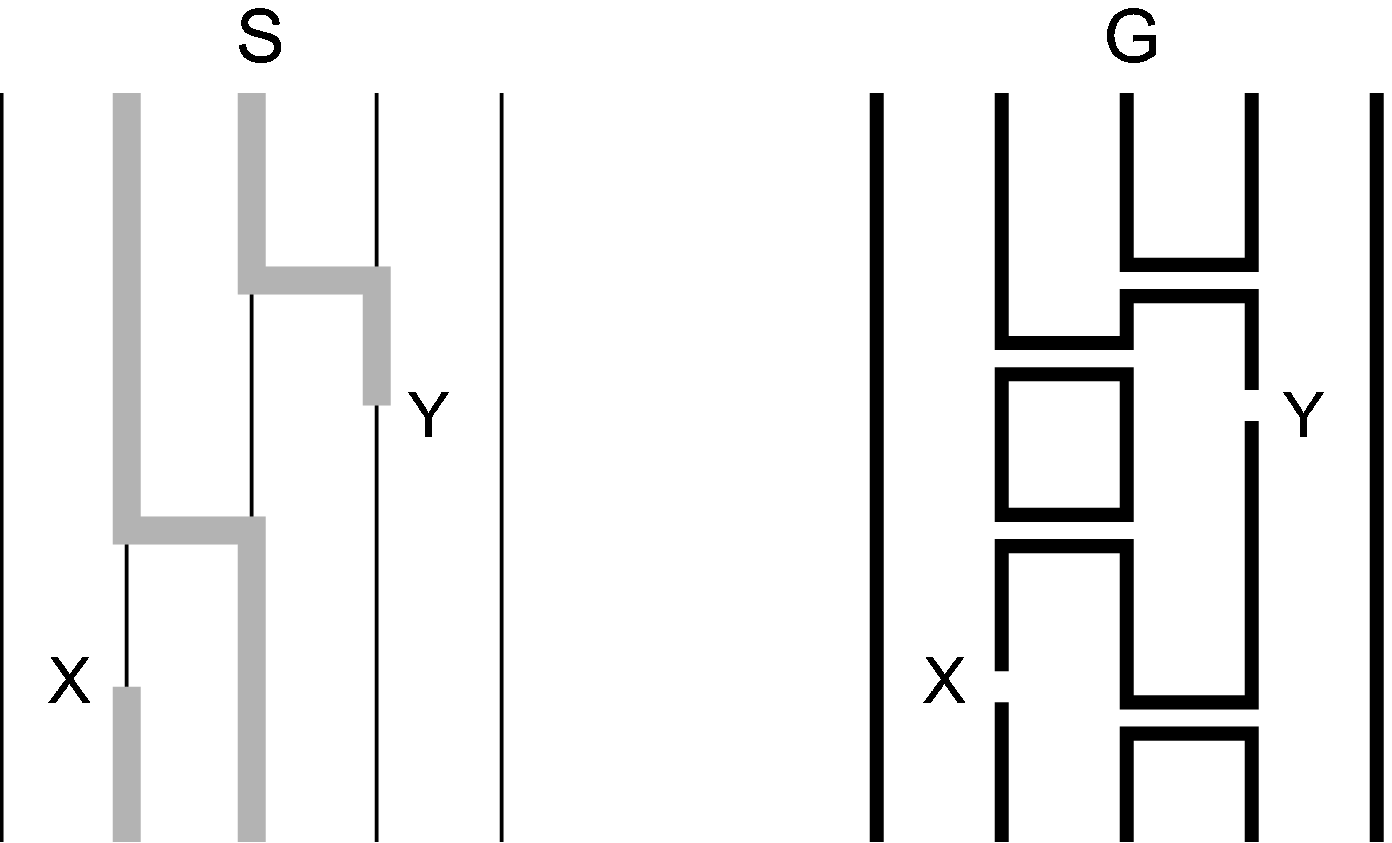}{0.4}{
The spin configuration $S$ and the graph $G$ that appear in the
computation of an off-diagonal Green's function.
Discontinuities in the world-lines and in the graph 
at two points $X$ and $Y$ are tolerated.
}

Then, we introduce graphs in the above expression to yield
\begin{eqnarray}
  & & \chi_{xx}
  =
  (N\beta)^{-1} \int dX dY Z^{-1}
  {\sum_{S,G}}' V(G) \Delta(S,G) \frac14 \nonumber \\
  & & \ 
  (1- \delta_{\psi(i,\tau +0),\psi(i,\tau-0)})
  (1- \delta_{\psi(j,\tau'+0),\psi(j,\tau-0)}).
  \eqlabel{ChiX}
\end{eqnarray}
We can take the summation over $S$.
Note here that there is no way to assign local spin values
along a loop so that the value is discontinuous at one and
only one point in the loop.
It means that the summation is zero unless the two points
$X$ and $Y$ are connected by the graph $G$.
Therefore, the result of the summation becomes
$$
  \chi_{xx}
  =
  (N\beta)^{-1} \int dX dY\, 
  \frac{ \sum_{G'} 2^{n_{\rm C}(G')} V(G') \frac14 \delta_{l(X),l(Y)} }
       { \sum_{G } 2^{n_{\rm C}(G )} V(G ) },
$$
where $l(X)$ and $l(Y)$ are the specifiers of the loops
to which the points $X$ and $Y$ belong respectively.
The summation in the numerator is over the set of graphs 
that yields a non-zero term in the summation \Eq{ChiX},
i.e., graphs that have at least one matching spin configuration
with two discontinuity points.
For most quantities and models, this coincides with 
the set of graphs over which the summation in the denominator
is taken, i.e., the graphs that have at least one matching 
spin configuration with no discontinuity points.
(In some pathological cases, however, this is not the case, and
the present estimator does not work in such cases.)
If such is the case, the above expression can be simply rewritten as
\begin{equation}
  \chi =
  \frac14 (N\beta)^{-1} \left\langle \sum_l V_l^2 \right\rangle_{\rm MC}.
  \eqlabel{ImprovedEstimatorQuantumSusceptibility}
\end{equation}

Thus we have obtained estimators of the $z$-component
susceptibility \Eq{IEClassicalSusceptibility} 
for the Ising model and the $x$-component susceptibility
\Eq{ImprovedEstimatorQuantumSusceptibility} for the quantum model.
In both cases, the estimator is expressed as the average
cluster (or loop) size, $\bar{V_c} \equiv \sum_c (V_c)^2 / \sum_c V_c$.
This means that the typical size of clusters correctly reflects the
correlation length in the loop algorithm.
This is the reason why the loop algorithm works very efficiently
in reducing the critical slowing-down.

The worm algorithm and the directed-loop algorithm
are also efficient near the critical point for a similar reason.
In these cases, the estimator of Green's functions is
the frequency of the head's visiting a certain location.
For example, in order to compute the correlation function
$\Gamma(X,Y) \equiv \langle S^x(X) S^x(Y) \rangle$, we have only to count 
the number of times the head passes the position $X-Y$ 
(relative to the original point).
This is quite natural, considering that
the trajectory of the head in the directed-loop algorithm 
is statistically identical to
a loop in the loop algorithm in the cases where the two
algorithms coincide.
In what follows, we see that the estimator is valid in general
even when the directed-loop algorithm does not coincide with the
loop algorithm.

We start with eq.\ \Eq{WeightWithSourceTerm}.
When the worm weight is chosen as \Eq{WormWeight}, 
the correlation function can be expressed as
$$
  (\Delta\tau\eta)^2 \Gamma(X,Y) = 
  \frac{ {\displaystyle \sum_{S' : X,Y           } \tilde{W}_L(S')} }
       { {\displaystyle \sum_{S' : \mbox{\scriptsize no worm}} \tilde{W}_L(S')} },
$$
where the summation in the numerator is over the states with
the head at $X$ and the tail at $Y$ or the other way around.
The number of times that we encounter a state $S$ in the 
Monte Carlo simulation is proportional to $\tilde{W}_L(S)$.
Therefore, the above expression can be rewritten as
$$
  (\eta)^2 \Gamma(X,Y)\, dXdY= 
  \frac{\langle \Delta_{X,Y}(S)\, dXdY \rangle_{\rm MC}}
  {\langle \Delta_{\emptyset}(S) \rangle_{\rm MC}},
$$
where $\Delta_{X,Y}(S)\,dXdY = 1$ if and only if one discontinuity point
is in the interval $dX$ centered at $X$ and 
the other in the interval $dY$ centered at $Y$.
The other function $\Delta_{\emptyset}(S)$ is $1$ if there is no worm in $S$.
Now, we obtain
\begin{eqnarray*}
  \Gamma(R) 
  & \equiv & 
    \frac1{N\beta} \int dX dY \Gamma(X,Y) \delta(R-(X-Y)) \\
  & = &
    \frac1{N\beta\eta^2} 
    \frac{ \int dX \langle \Delta_{X,X+R}(S) \rangle_{\rm MC} }
         { \langle \Delta_{\emptyset}(S) \rangle_{\rm MC} } \\
  & = &
    \frac1{N\beta\eta^2} 
    \langle n(R) \rangle_{\rm MC},
\end{eqnarray*}
where $n(R)$ is the average number of times the head passes,
during a cycle, the point whose distance from the origin is $R$.


\section{Numerical Recipe}
\seclabel{Recipe}
%
In \msec{Theory}, the mathematical framework of several algorithms have 
been described.
In the present section, we present detailed descriptions of
the procedures for realizing the algorithms.
Since the efficiencies of the algorithms have been compared 
only for a very limited number of models,
there is not much hope of presenting here a flawless recommendation 
as to which algorithm should be used for a given instance.
However, there are some properties that we know already.
For example, the loop algorithm is the best of all the
algorithms, or at least not much worse than any other algorithm, 
in the cases where the relationship (such as 
\Eq{ImprovedEstimatorQuantumSusceptibility}) 
between the relevant susceptibility and the cluster size holds.
The $XXZ$ model of an easy-axis anisotropy with no magnetic field,
and the bilinear-biquadratic Heisenberg model are among these cases.
When a finite magnetic field is present, however,
the cluster size in the loop algorithm does not in general reflect
the correlation length correctly, and it may perform poorer
even than the local updating algorithm.
The directed-loop algorithm and the worm algorithm
work much better for such cases.
It should be also pointed out that applications of these two 
algorithms are usually more straight-forward than that of the 
loop algorithm; for the loop algorithm 
we first need to find a good graphical decomposition of the
Hamiltonian, which largely depends on the Hamiltonian under consideration.
However, the two algorithms become essentially equivalent to 
the local updating algorithm
for the models with the Ising-like anisotropy,
and hence are not very efficient for reducing the critical
slowing-down in the Ising-like models.

The description of the procedures in the following subsections
is given mostly in graphical terms,
such as {\it segments} and {\it vertices} defined in the first subsection.
Therefore, we have to translate it into one of the computer languages.
While the actual coding with a specific computer
language is out of the scope of the present article, 
a remark on the data structure may be helpful here.
The coding can be done, in principle, with any commonly-used computer language.
However, the graphical objects that we have to deal with 
are created and annihilated during the simulation and their number varies.
In addition, when we work with the infinite $L$ limit,
there is no discrete lattice that usually provides us with
the index system of arrays. 
Therefore, we feel that some sort of a linked-list data structure is necessary,
and that this data structure naturally fit in the object-oriented programming.
While the object-oriented programming is possible with any language,
working with object-oriented languages such as the C$++$ and the Java
might make the programming easier.
For example, the above-mentioned graphical entities may be most 
conveniently defined as objects, such as a ``class'' in the C$++$ language.
A segment may be defined as an object that has
some (or all) of the following member variables (i.e., properties):
the local spin state (see \mssc{GraphicalTerms} below),
the spatial location, the beginning time, the ending time,
(the pointers to) the vertices that delimit the segment,
and a variable used for cluster identification (see Appendix B).
In a sample program which may be found in our web site\cite{WebSite},
the graphical objects are handled through a container 
that has a built-in linked-list structure and is provided
as a part of a standard library of the language.

\subsection{Graphical Terms for World-Line Monte Carlo}
\ssclabel{GraphicalTerms}

Using the path-integral representation, our Monte Carlo simulation
can be formulated as a Markov process in the space of
graphical objects called {\it world-lines}.
A world-line is a curve on a $(d+1)$-dimensional space-time lattice, 
where $d$ denotes the real-space dimension. 
The additional dimension, depicted as the vertical dimension
throughout the present article, is called the imaginary-time dimension. 
The periodic boundary condition is imposed in this direction,
while an arbitrary boundary condition may be used for the other
(spatial) dimension.
The height of the system, i.e., the system size in the 
imaginary-time direction is the inverse temperature 
$\beta = 1/k_{\rm B}T$.
Therefore, each site $i$ in the real space is represented by a
vertical line of the length $\beta$.
Along each vertical line, an integral-valued function $n(i,\tau)$ is defined,
which is constant (as a function of $\tau$) almost everywhere and 
is discontinuous only at kinks.
The value of the function is called the {\it local state} of the 
space-time point $(i,\tau)$.
A point at which the function is discontinuous is called a {\it kink}.
A spin {\it configuration} (or simply a configuration) is
the set of the functions along all the vertical lines.
In other words, a configuration is equivalent to an assignment
of integers to all the space-time points.

In particular, when the local state is binary, say 0 or 1,
and the sum of the variables over the real space is conserved in 
the imaginary-time direction, which is the case with the $s=1/2$ $XXZ$ model, 
a configuration can be represented by a set of lines that connect
the space-time points at which the local state is 1.
These are the world-lines.
In \Fig{LatticeAndObjects}, such a world-line configuration of a
one-dimensional quantum spin model is shown on a $(1+1)$-dimensional
space-time lattice.

For a quantum spin model of the spin size $s$, 
the integral variable $n(i,\tau)$ takes $2s+1$ 
values from $0$ to $2s$.
The eigenstate of the operator $S_i^z+s$ is customarily
used as the local spin variable, where $S_i^z$ is the
$z$ component of the spin at the site $i$.
We may also call it the {\it number of particles}, 
even when we are considering a spin model.
Although configurations in the case $s>1/2$ cannot be
represented by the world-lines, 
we still refer to the configuration as  a ``world-line configuration''.

In addition to the local spin variables,
we introduce auxiliary variables that can be represented 
conveniently by a graph that consists of
vertical lines called {\it segments} and
objects called {\it vertices} that delimit the segments.
A vertex is represented by a graph element in the loop algorithm,
whereas it is represented simply by a single horizontal line
in the directed-loop algorithm.
An example in the case of the loop algorithm
is shown in \Fig{LatticeAndObjects},
where a vertex is represented by a pair of two horizontal lines.

\deffig{LatticeAndObjects}{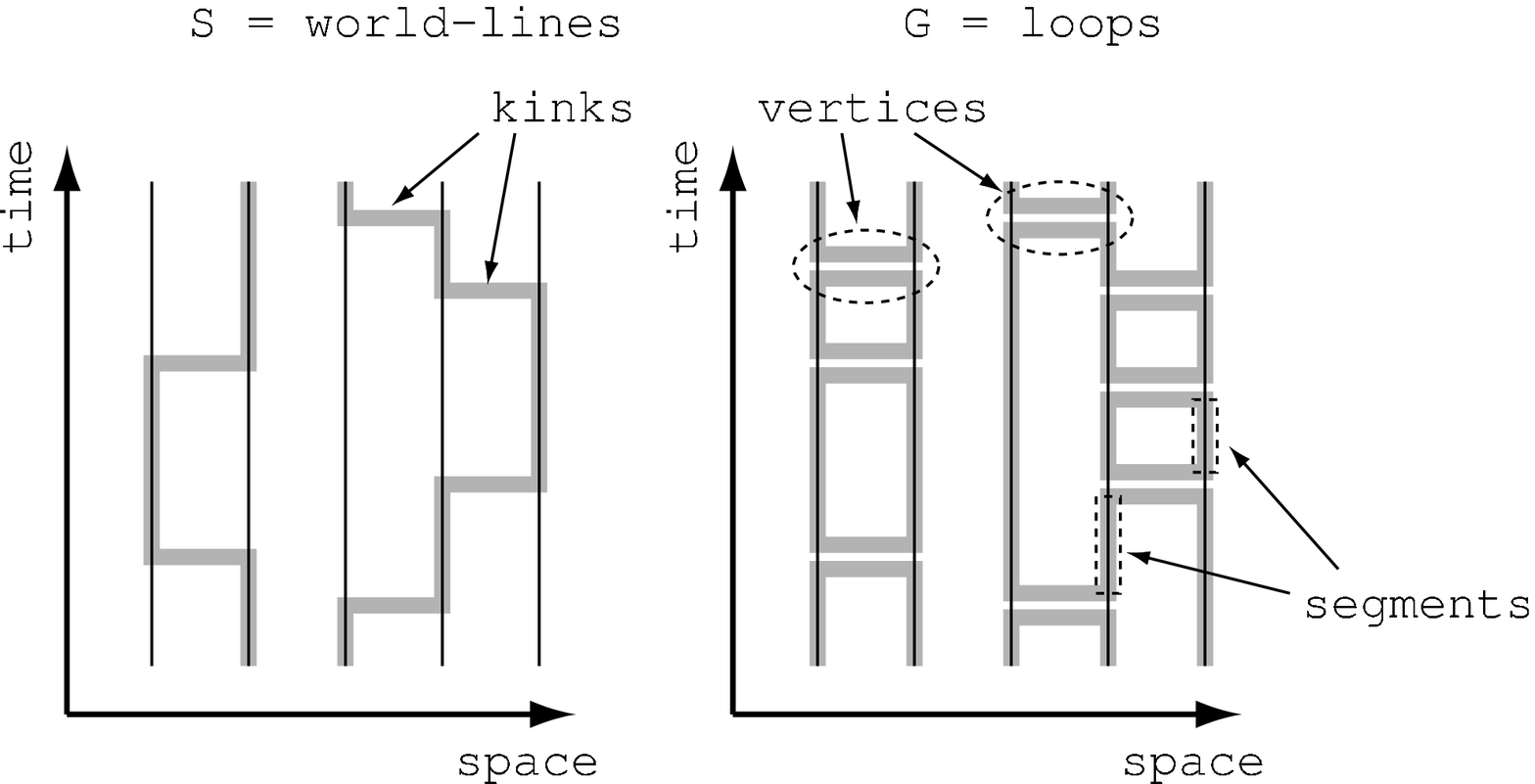}{0.4}
{
Various graphical objects associated with 
a world-line configuration $S$ and a graph $G$ that match each other.
Shown is the case of a one-dimensional system.
A {\it vertex}, in this case, is the graph element that consists 
of a pair of two horizontal lines.
A {\it kink} is located on the vertex at which the local state changes.
A {\it segment} is a part of a vertical line that is delimited by
two vertices.
}

%
%
%
\subsection{Loop/cluster Algorithm}
The loop algorithm can be applied to various quantum spin models and gives
much better performance than traditional quantum Monte Carlo
methods\cite{Evertz2003}.
The models that can be handled efficiently with this algorithm include
the $XXZ$ model with no magnetic field, the bilinear-biquadratic model,
and the SU($N$) symmetric model as discussed below.
Note, however, that it may not be efficient in the cases
where terms in the Hamiltonian conflict with each other,
such as the case of the anti-ferromagnetic $XXZ$ model
in an uniform magnetic field. 
In such cases, one should consider using the worm or the directed-loop 
algorithm discussed in the following subsections.

One cycle in a loop algorithm generally consists of the following operations: 
(i) assigning graphs to a given world-line configuration probabilistically;
(ii) decomposing the world-line configuration into loops or clusters defined 
by graphs; 
(iii) assigning new values to integral variables on each loop or cluster 
probabilistically.
The types of the graphs to be assigned and the probability of
the assignment depend on the details of the model.

In this subsection, we describe the loop algorithm in detail
for some characteristic models.
Discussed in the following are 
(i) the quantum $s=1/2$ $XYZ$ spin model\cite{EvertzLM1993,Kawashima1996}, 
(ii) the quantum $s \ge 1$ $XYZ$ spin model, 
(iii) the transverse field Ising model\cite{RiegerK1999}, 
(iv) the quantum $s=1$ bilinear-biquadratic model\cite{HaradaK2001}, and 
(v) the quantum SU($N$) model\cite{HaradaKT2003}.

\subsubsection{Quantum $s=1/2$ $XYZ$ spin model\cite{EvertzLM1993,Kawashima1996}}
\ssslabel{LoopAlgorithmXYZ}

First we consider the loop algorithm for the $s=1/2$ $XYZ$ model
\begin{equation}
  \eqlabel{hamiltonian-XYZ-spin-one-half}
  {\cal H} =  -\sum_{(ij)}
  \big(J_x\sigma_i^x \sigma_j^x  +J_y\sigma_i^y \sigma_j^y
  +J_z \sigma_i^z \sigma_j^z\big) - B\sum_i \sigma_i^z,
\end{equation}
where $\sigma_i^\alpha$ is an $s=1/2$ spin operator in the $\alpha$
direction at the $i$th site, whose eigenvalues are $\pm 1/2$, 
and the summation is over all the pairs of interacting spins. 
In what follows, we consider the case, $J_x \ge J_y \ge 0$.
(Other cases can be transformed to this case if the lattice is
a bipartite lattice.)
\Figure{LatticeAndObjects} shows a world-line configuration of this model. 
On the world-line (the thick gray lines on the left panel of 
\Fig{LatticeAndObjects}), 
the integral variable $n(i,\tau)=\sigma_i^z+1/2$ takes 1,
whereas it takes 0 elsewhere.

One sweep of the simulation with the loop algorithm consists
of two phases: the graph assignment and the cluster flip.
The graph assignment is done by 
(i) deleting the existing graph,
(ii) assigning the graph elements to the uniform intervals (UI)
with a certain density, and (iii) connecting the graph elements by
vertical lines to form loops/clusters.
Here, a UI for a pair of sites $(ij)$
is an interval in which no change takes place in the local spin
state variable $n(i,\tau)$ or in $n(j,\tau)$.
The cluster flip is done by (i) identifying
the clusters, and (ii) flipping the clusters.
The cluster identification is done as the assignment of
a number to every segment so that the number uniquely specifies
the cluster to which the segment belongs.
The flipping of a cluster is simply the simultaneous
inversion of the local spin states (from 0 to 1 or vice versa)
for all the segments in the cluster.
As shown in \Tab{XYZ-graph-1}, 
we need three types of graph elements
for the loop algorithm for the $XYZ$ model.
The graph elements are called {\it cross}, {\it horizontal} and {\it binding}
from the top to the bottom.
Each graph element represents
a certain constraint on the integral variables at four 
space-time points, namely, {\it legs}. 
The integral variables on the points connected by a line
belong to the same loop (or cluster) 
and must be changed simultaneously when
a loop is flipped.  
Although the graph elements in \Tab{XYZ-graph-1} are drawn 
as if they had a finite height for clarity of the illustration,
they have in fact no temporal extension.
The types of the graph elements as well as
the density and the probability of the graph assignment 
depend on the anisotropy.
There are four cases to be considered separately: 
(i) $J_z \ge J_x \ge 0$;
(ii) $J_x \ge J_z \ge 0$;
(iii) $J_z \le 0, |J_z| \ge J_x$; 
(iv) $J_z \le 0, J_x \ge |J_z| $.  

To be specific, the updating procedure 
of the integral variables $n(i,\tau)$ is as follows
(see also \Fig{steps-XYZ}):
\begin{liststep}
\item 
  Delete the whole graph.
\item 
  For each pair of interacting sites $(ij)$, do the following.
  Decompose the interval $(0,\beta)$ into UIs.
  Then for each UI, place graph elements randomly 
  with the density given in
  Table \ref{tab:XYZ-graph-1}-\ref{tab:XYZ-graph-4}.
  (See Appendix A for the procedure of
  generating the temporal positions for the placement
  of the graph elements.)
  \label{st:XYZ-1}
\item 
  Place a graph element on every kink with the probability given in
  Table \ref{tab:XYZ-graph-1}-\ref{tab:XYZ-graph-4}.
\item 
  Draw vertical lines between two graph elements and
  connect two legs (one from each graph element).
  We refer to the resulting graph as $G$.
\item 
  Identify clusters.
  (For the identification algorithm, see Appendix B.)
\item 
  For each cluster,
  flip the values of the variables
  $n(i,\tau)$ (i.e., $n(i,\tau) := 1-n(i,\tau)$)
  for all the segments on it,
  with the probability $p(c)=\exp(-Bm_c)/(\exp(Bm_c)+\exp(-Bm_c))$.
  Here $m_c$ is the cluster magnetization of the cluster $c$ defined as
  \begin{equation}
    \eqlabel{mag-cluster}
    m_c \equiv \int_{c} dX \left( n(X) - \frac12 \right),
  \end{equation}
  where $\int_c dX$ stands for the sum-integral on the
  cluster (the summation with respect to the site index and 
  the integration with respect to the time).
\item 
  Do measurements.
  (For the estimators, see \msss{Estimators}.)
\end{liststep}

\begin{figure}[htbp]
  \centering
  \includegraphics[width=3cm]{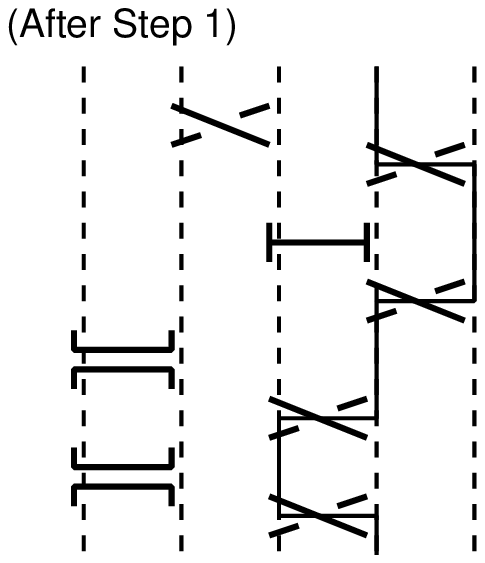}\hspace{0.4cm}
  \includegraphics[width=3cm]{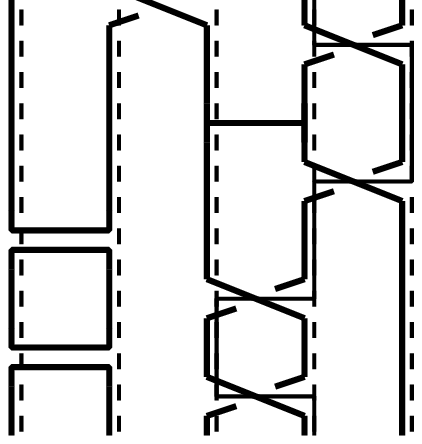}\vspace{0.2cm}\\
  \includegraphics[width=3cm]{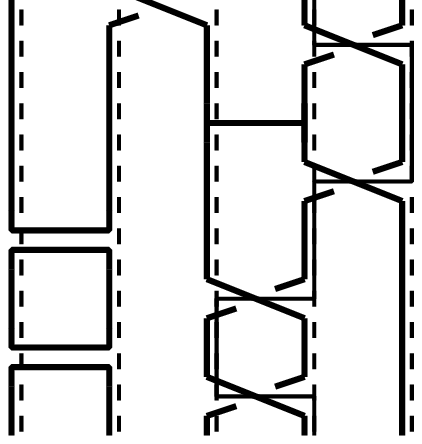}\hspace{0.4cm}
  \includegraphics[width=3cm]{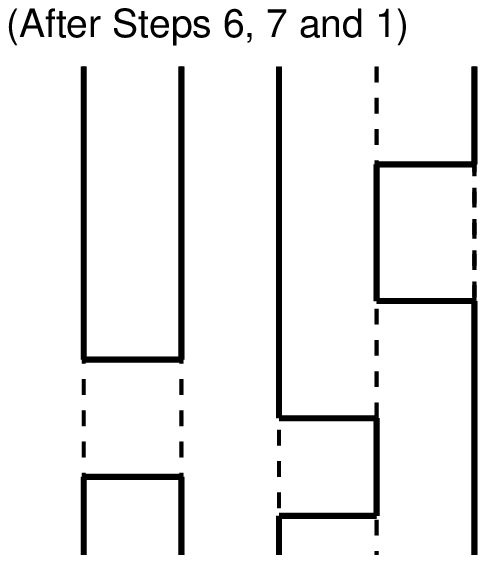}
  \caption{The procedures of one step in the loop algorithm for 
    the $s=1/2$ $XYZ$ model ($0 < J_x < J_z$).}
\figlabel{steps-XYZ}
\end{figure}

\begin{table}[htbp]
  \centering
  \caption{
    The density of graph elements for uniform intervals (the second
    and the third columns),
    and the probability of choosing graph elements for kinks
    (the forth and the last),
    in the case where $0 \le J_x \le J_z$.
    The top row shows the local spin states, in which 
    a solid (dashed) line denote an up (down) spin.
    The densities and the probabilities for all the other local states 
    can be derived easily using the symmetries with respect to 
    the time inversion, the exchange of the two sites, and the spin inversion,
    The first column shows graphs.
    The spins connected by a solid (dashed) line
    must be parallel (anti-parallel).
    The second and the third columns show the density
    of graph elements.
    The forth and the fifth column shows
    the probability of choosing the graph element on a kink.
  }
  \tablabel{XYZ-graph-1}

\ \\
  \begin{tabular}{|c|c|c||c|c|}
    \hline
    \rule{0cm}{1.2cm}
    \includegraphics[width=1.0cm,height=1.0cm]{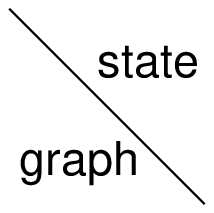}
    & \includegraphics[width=1.0cm]{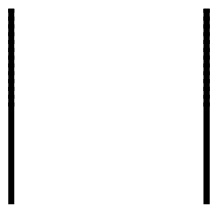}
    & \includegraphics[width=1.0cm]{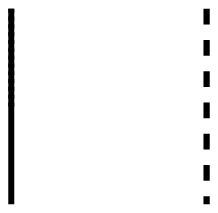}
    & \includegraphics[width=1.0cm]{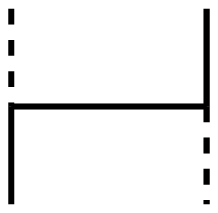}
    & \includegraphics[width=1.0cm]{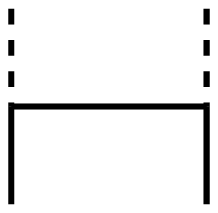}\\
    \hline
    \parbox{1.0cm}{\rule{0cm}{1.2cm}\includegraphics[width=1.0cm,height=1.0cm]{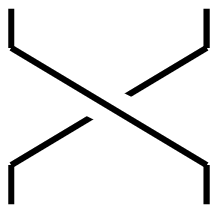}}
    &  $\frac14(J_x+J_y)$ &  0 &  1 &  0\\
    \hline
    \parbox{1.0cm}{\rule{0cm}{1.2cm}\includegraphics[width=1.0cm,height=1.0cm]{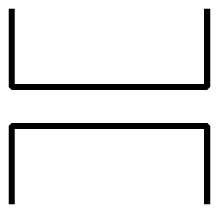}}
    &  $\frac14(J_x-J_y)$ &  0 &  0 &  1\\
    \hline
    \parbox{1.0cm}{\rule{0cm}{1.2cm}\includegraphics[width=1.0cm,height=1.0cm]{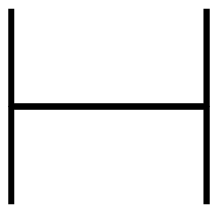}}
    &  $\frac12(J_z-J_x)$ &  0 &  0 &  0\\
    \hline
  \end{tabular}
\end{table}

\begin{table}[htbp]
  \centering
  \caption{
    The same as Table \ref{tab:XYZ-graph-1} for 
    the case $0 \le J_z \le J_x$.
  }
  \tablabel{XYZ-graph-2}

\ \\
  \begin{tabular}{|c|c|c||c|c|}
    \hline
    \rule{0cm}{1.2cm}
    \includegraphics[width=1.0cm,height=1.0cm]{FIG/table-head.eps}
    & \includegraphics[width=1.0cm]{FIG/state-XYZ-1111.eps}
    & \includegraphics[width=1.0cm]{FIG/state-XYZ-1010.eps}
    & \includegraphics[width=1.0cm]{FIG/state-XYZ-0110.eps}
    & \includegraphics[width=1.0cm]{FIG/state-XYZ-0011.eps}\\
    \hline
    \parbox{1.0cm}{\rule{0cm}{1.2cm}\includegraphics[width=1.0cm,height=1.0cm]{FIG/cross-graph.eps}}
    &  $\frac14(J_z+J_y)$ &  0 &  $\frac{J_z+J_y}{J_x+J_y}$ &  0\\
    \hline
    \parbox{1.0cm}{\rule{0cm}{1.2cm}\includegraphics[width=1.0cm,height=1.0cm]{FIG/horizontal-graph.eps}}
    &  $\frac14(J_x-J_y)$ &  0  & 0  &  1\\
    \hline
    \parbox{1.0cm}{\rule{0cm}{1.2cm}\includegraphics[width=1.0cm,height=1.0cm]{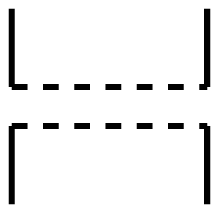}}
    &  0 &  $\frac14(J_x-J_z)$ & $\frac{J_x-J_z}{J_x+J_y}$  &  0\\
    \hline
  \end{tabular}
\end{table}

\begin{table}[htbp]
  \centering
  \caption{
    The same as Table \ref{tab:XYZ-graph-1} for 
    the case $0 \ge J_z, \quad  J_x \le |J_z|$.
  }
  \tablabel{XYZ-graph-3}

  \begin{tabular}{|c|c|c||c|c|}
    \hline
    \rule{0cm}{1.2cm}
    \includegraphics[width=1.0cm,height=1.0cm]{FIG/table-head.eps}
    & \includegraphics[width=1.0cm]{FIG/state-XYZ-1111.eps}
    & \includegraphics[width=1.0cm]{FIG/state-XYZ-1010.eps}
    & \includegraphics[width=1.0cm]{FIG/state-XYZ-0110.eps}
    & \includegraphics[width=1.0cm]{FIG/state-XYZ-0011.eps}\\
    \hline
    \parbox{1.0cm}{\rule{0cm}{1.2cm}\includegraphics[width=1.0cm,height=1.0cm]{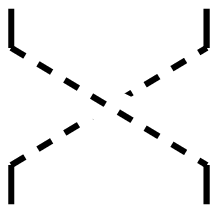}}
    &  0 & $\frac14(J_x-J_y)$ &  0 &  1\\
    \hline
    \parbox{1.0cm}{\rule{0cm}{1.2cm}\includegraphics[width=1.0cm,height=1.0cm]{FIG/horizontal-graph-anti.eps}}
    &  0 & $\frac14(J_x+J_y)$ &  1 &  0\\
    \hline
    \parbox{1.0cm}{\rule{0cm}{1.2cm}\includegraphics[width=1.0cm,height=1.0cm]{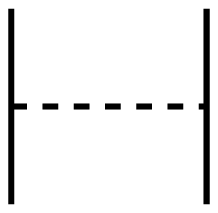}}
    &  0 & $\frac12(|J_z|-J_x)$ &  0 &  0\\
    \hline
  \end{tabular}
\end{table}

\begin{table}[htbp]
  \centering
  \caption{
    The same as Table \ref{tab:XYZ-graph-1} for 
    the case $0 \ge J_z, \quad  |J_z| \le J_x$.
}
  \tablabel{XYZ-graph-4}

  \begin{tabular}{|c|c|c||c|c|}
    \hline
    \rule{0cm}{1.2cm}
    \includegraphics[width=1.0cm,height=1.0cm]{FIG/table-head.eps}
    & \includegraphics[width=1.0cm]{FIG/state-XYZ-1111.eps}
    & \includegraphics[width=1.0cm]{FIG/state-XYZ-1010.eps}
    & \includegraphics[width=1.0cm]{FIG/state-XYZ-0110.eps}
    & \includegraphics[width=1.0cm]{FIG/state-XYZ-0011.eps}\\
    \hline
    \parbox{1.0cm}{\rule{0cm}{1.2cm}\includegraphics[width=1.0cm,height=1.0cm]{FIG/cross-graph.eps}}
    &  $\frac14(J_x-|J_z|)$ &  0 &  $\frac{J_x-|J_z|}{J_x+J_y}$ &  0\\
    \hline
    \parbox{1.0cm}{\rule{0cm}{1.2cm}\includegraphics[width=1.0cm,height=1.0cm]{FIG/cross-graph-anti.eps}}
    &  0 &  $\frac14(J_x-J_y)$ &  0 &  1\\
    \hline
    \parbox{1.0cm}{\rule{0cm}{1.2cm}\includegraphics[width=1.0cm,height=1.0cm]{FIG/horizontal-graph-anti.eps}}
    &  0 &  $\frac14(|J_z|+J_y)$ & $\frac{|J_z|+J_y}{J_x+J_y}$ &  0\\
    \hline
  \end{tabular}
\end{table}


\subsubsection{Quantum $XYZ$ spin model with large spins}

For the $XYZ$ model with spins larger than $s=1/2$, 
we can in general use the split-spin technique presented below.
The resulting algorithm are generally efficient 
when the corresponding algorithm for $s=1/2$ are efficient.
However, for models with the easy-plane anisotropy, such as the
$XY$ model, the coarse-grained algorithm described in
\msss{CoarseGrained} is recommended since there is no need for
working with multiple split-spins for each site.
On the other hand,
for models with the easy-axis anisotropy with no magnetic field,
the split-spin algorithm presented below, to our knowledge, is
the best choice among the algorithms discussed in the present article.

We consider $2s$, instead of one, vertical lines at each site,
each representing a split spin, or a Pauli spin that carries $s=1/2$.
Correspondingly, for each pair of nearest-neighbor sites,
we apply the same graph-assignment procedure as in the $s=1/2$ algorithm
to each one of $(2s)^2$ combinations of split spins.
In addition, in order to eliminate unphysical states, we stochastically
assign a graph that represents a permutation among the $2s$ split spins,
to the end points of the $2s$ vertical lines.
(One example is shown in \Fig{LoopAlgorithmS1} for $s=1$.)
All the permutation graphs that match the current state is chosen
with equal probability.
For example, when the $l$ split spins among $2s$ are up and the others
are down, there are $l!$ ways to connect up-spins and $\bar l!$ 
($\bar l \equiv 2s-l$) for down-spins.
Therefore, every matching graph is chosen with the probability
$1/(l!\, \overline{l}!)$.

Thus the split-spin algorithm for the $XYZ$ model with an arbitrary $s$
can be obtained by replacing Step 2 in \msss{LoopAlgorithmXYZ} by
\begin{description}
\item[Step 2:]
  For each pair of interacting split-spins, $(i,\mu)$ and $(j,\nu)$, 
  do the following.
  Decompose the interval $(0,\beta)$ into UIs.
  Then for each UI, place graph elements randomly 
  with the density given in
  Table \ref{tab:XYZ-graph-1}-\ref{tab:XYZ-graph-4}.
\end{description}
and inserting between Step 3 and Step 4 the following
\begin{description}
\item[(Insertion of the permutation graphs):]
  For each site, connect $2s$ end points of vertical lines
  at $\tau = \beta$ to those at $\tau = 0$, pairwise,
  so that an up-spin is connected to an up-spin and
  likewise for down-spins.
  (Choose one of $(l!)(\overline{l}!)$ ways of connection
  with equal probability.)
\end{description}

\subsubsection{Transverse Ising model\cite{RiegerK1999}}
Next we consider the transverse Ising model,
\begin{equation}
  \label{eq:hamiltonian-transverse}
  {\cal H} =  - J \sum_{(ij)}
  \sigma_i^z \sigma_j^z - B\sum_i \sigma_i^x
\end{equation}
with $J \ge 0$ and $B \ge 0$.
Since the transverse field breaks the conservation of the total
magnetization, the world-lines are discontinuous;
a world-line can terminate anywhere.
Accordingly, we need a new type of graphs
that cut segments at a single point. 
Therefore, we need two types of graph elements
in this case as shown in \Fig{graphs-transverse}.

\deffig{graphs-transverse}{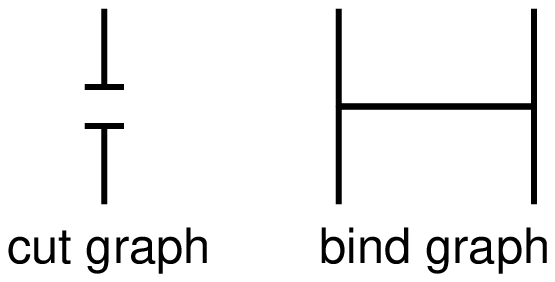}{0.2}
{The graphs for the transverse Ising model.}

Except for the difference in the types of the graph elements,
the algorithm is almost the same as the one for the $XYZ$ model
described above.
One cycle of the loop algorithm for the transverse-field Ising model 
is as follows:
\begin{liststep}
\item 
  Delete the graph.
\item 
  For each pair of the interacting sites $(ij)$, do the following.
  Decompose the interval $(0,\beta)$ into UIs.
  Then for every UI in which the two spins are parallel
  (i.e., the local spin state on $i$ is the same as that on $j$),
  place one of the graph elements randomly with the density $J/2$.
  \label{st:transverse-2}
\item 
  For each site $i$ (i.e., vertical line), do the following.
  Place ``cut'' graphs along the line with the density $B/2$.
  \label{st:transverse-2p}
\item 
  Place a ``cut'' graph on every kink.
\item 
  Draw vertical lines to connect legs of the graph elements.
\item 
  Identify the clusters.
\item 
  Flip every cluster independently with probability $1/2$.
\item 
  Do measurements.
  (For the estimators, see \msss{Estimators}.)
\end{liststep}

%
\subsubsection{Quantum $s=1$ bilinear-biquadratic model\cite{HaradaK2001}}
\ssslabel{BIQ}
The Hamiltonian for the model is
\begin{equation}
  {\cal H} 
  = -\sum_{(ij)} \left( 
    J_L \vect{S}_i\cdot\vect{S}_j + J_Q (\vect{S}_i\cdot\vect{S}_j)^2 \right).
\end{equation}
It is convenient to introduce the parameter $\theta$ by
\begin{equation}
  J_L = -J\cos\theta, \quad J_Q = -J\sin\theta \qquad (J >0).
\end{equation}
There is no negative-sign problem when the coupling constant $J_Q$ is 
positive ($-\pi \le \theta \le 0$). 
We consider this case in what follows.

Using the split-spin technique, the original spin is decomposed into two
$s=1/2$ spins.
In addition to the ordinary kinks,
we have {\it double kinks} at which two particles jump to the neighboring site.
(This is because of the biquadratic term.)
The biquadratic term also requires new types of graph elements
that have eight legs as shown in \Fig{graphs-BLBQ}
in addition to the ordinary four-legged graph elements.
(The third and the fourth graph elements in \Fig{graphs-BLBQ} are
eight-legged representation of the ordinary single graphs,
whereas the first and second ones are the new ones.)
We call the first {\it double-cross} and the second {\it double-horizontal}.

\deffig{graphs-BLBQ}{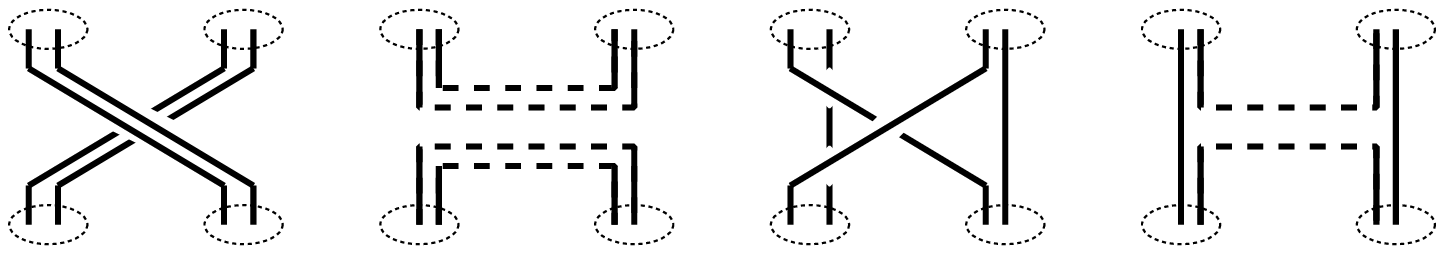}{0.5}
{The graph elements for quantum $s=1$ bilinear-biquadratic model.
  As before, the spins connected by a solid (dashed) line
  must be parallel (anti-parallel).}

From an algorithmic point of view,
the region $-\pi \le \theta \le 0$ is decomposed into three
sub-regions: 
$[-\pi, -3\pi/4]$, $[-3\pi/4,-\pi/2]$ and $[-\pi/2,0]$.  
In the first region, we need only single-cross and double-cross graphs.
In the third region, only single-horizontal and double-horizontal
graphs are required.
In the second region, no single graph is used.
In this region, only double-cross and double-horizontal graphs 
are sufficient for decomposing the Hamiltonian.
The procedure of a cycle can be summarized as follows.

\begin{liststep}
\item 
  Delete the graph.
\item 
  For each pair of the nearest-neighbor sites $(ij)$, do the following.
  First decompose the interval $(0,\beta)$ into UIs.
  Then, for each UI, and for each type of the graph elements, do the following.
  Place graph elements on the UI uniform-randomly with the 
  density given in Tables \ref{tab:BLBQ-graph-1}-\ref{tab:BLBQ-graph-3}.
  There are generally multiple ways of placing a graph element of
  a given type to a given position, i.e., there are multiple ways of wiring
  so that the spins connected by a solid (dashed) line
  are parallel (anti-parallel).
  Choose one of the consistent ways of wiring with equal probability.
\item 
  Place a graph element on every kink with the probability
  given in Tables \ref{tab:BLBQ-graph-1}-\ref{tab:BLBQ-graph-3}.
  Again choose one of the consistent ways of wiring with equal
  probability.
\item
  For each site, 
  connect two end points of vertical lines
  at $\tau = \beta$ to those at $\tau = 0$, pairwise,
  so that an up-spin is connected to an up-spin and
  likewise for down-spins. Choose one of such ways of
  wiring with equal probability.
\item 
  Draw vertical lines so that each line connects
  a leg of a graph element to a leg of another.
\item 
  Identify clusters.
\item 
  Flip every cluster independently with probability $1/2$.
\item 
  Do measurements.
\end{liststep}

\begin{table}[htbp]
  \caption{
    The density (a) and the probability (b) of the graph assignment
    for the bilinear-biquadratic model
    with $-\pi \le \theta \le -3\pi/4$. 
    \tablabel{BLBQ-graph-1}
  }
\ \\
\centering
(a)
  \begin{tabular}{|c|c|c|c|c|}
    \hline
    \rule{0cm}{8mm}
    \includegraphics[width=8mm,height=8mm]{FIG/table-head.eps}
    & \includegraphics[width=8mm]{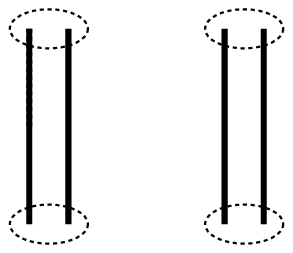}
    & \includegraphics[width=8mm]{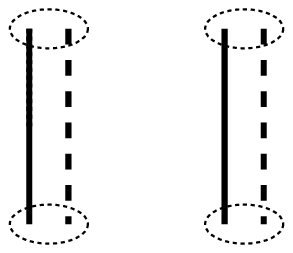}
    & \includegraphics[width=8mm]{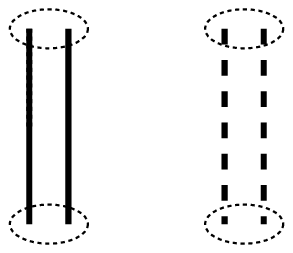}
    & \includegraphics[width=8mm]{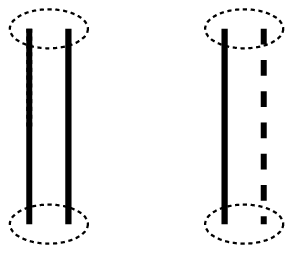}\\
    \hline
    \parbox{8mm}{\rule{0cm}{8mm}\includegraphics[width=8mm]{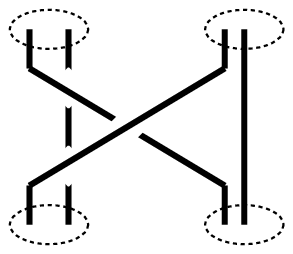}}
    & $2(J_L-J_Q)$ & $J_L-J_Q$ & $0$ & $J_L-J_Q$\\
    \hline
    \parbox{8mm}{\rule{0cm}{8mm}\includegraphics[width=8mm]{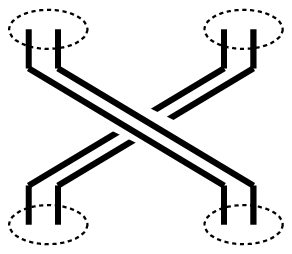}}
    & $J_Q$& $J_Q$ & $0$ & $0$\\
    \hline
  \end{tabular}
\ \\
\ \\
(b)
  \begin{tabular}{|c|c|c|c|}
    \hline
    \rule{0cm}{8mm}
    \includegraphics[width=8mm,height=8mm]{FIG/table-head.eps}
    & \includegraphics[width=8mm]{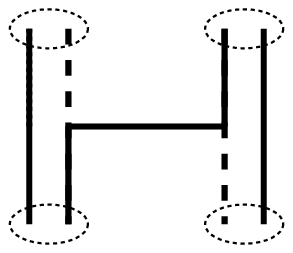}
    & \includegraphics[width=8mm]{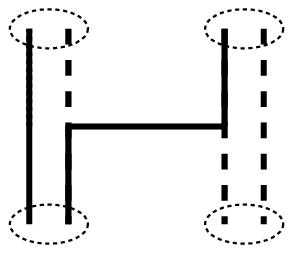}
    & \includegraphics[width=8mm]{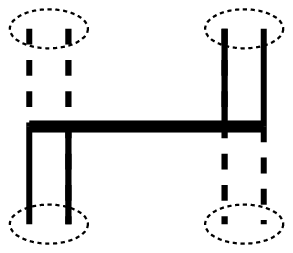}\\
    \hline
    \parbox{8mm}{\rule{0cm}{8mm}\includegraphics[width=8mm]{FIG/cross-BLBQ.eps}}
    & $1-\frac{J_Q}{J_L}$ & $1$ & $0$\\
    \hline
    \parbox{8mm}{\rule{0cm}{8mm}\includegraphics[width=8mm]{FIG/double-cross-BLBQ.eps}}
    & $\frac{J_Q}{J_L}$&$0$ & $1$\\
    \hline
  \end{tabular}
\end{table}

\begin{table}[htbp]
  \caption{
    The same as Table \ref{tab:BLBQ-graph-1}
    for the case $-3\pi/4 \le \theta \le -\pi/2$.
  }
  \tablabel{BLBQ-graph-2}

\centering
\ \\
(a)
  \begin{tabular}{|c|c|c|c|c|}
    \hline
    \rule{0cm}{8mm}
    \includegraphics[width=8mm,height=8mm]{FIG/table-head.eps}
    & \includegraphics[width=8mm]{FIG/state-BLBQ-1111.eps}
    & \includegraphics[width=8mm]{FIG/state-BLBQ-1010.eps}
    & \includegraphics[width=8mm]{FIG/state-BLBQ-1100.eps}
    & \includegraphics[width=8mm]{FIG/state-BLBQ-1110.eps} \\
    \hline
    \parbox{8mm}{\rule{0cm}{8mm}\includegraphics[width=8mm]{FIG/double-cross-BLBQ.eps}}
    & $J_L$ & $J_L$ & $0$ & $0$\\
    \hline
    \parbox{8mm}{\rule{0cm}{8mm}\includegraphics[width=8mm]{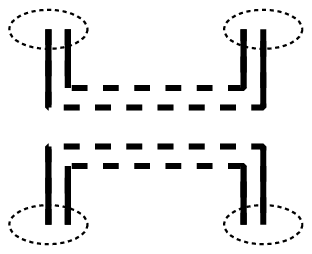}}
    & $0$ & $J_Q-J_L$ & $J_Q-J_L$ & $0$\\
    \hline
  \end{tabular}
\ \\
\ \\
(b)
  \begin{tabular}{|c|c|c|c|}
    \hline
    \rule{0cm}{8mm}
    \includegraphics[width=8mm,height=8mm]{FIG/table-head.eps}
    & \includegraphics[width=8mm]{FIG/state-BLBQ-kink-1.eps}
    & \includegraphics[width=8mm]{FIG/state-BLBQ-kink-1-anti.eps}
    & \includegraphics[width=8mm]{FIG/state-BLBQ-kink-2.eps}\\
    \hline
    \parbox{8mm}{\rule{0cm}{8mm}\includegraphics[width=8mm]{FIG/double-cross-BLBQ.eps}}
    & $1$ & $0$ & $\frac{J_L}{J_Q}$\\
    \hline
    \parbox{8mm}{\rule{0cm}{8mm}\includegraphics[width=8mm]{FIG/double-horizontal-BLBQ.eps}}
    & $0$ & $1$ & $1-\frac{J_L}{J_Q}$\\
    \hline
  \end{tabular}
\end{table}

\begin{table}[htbp]
  \caption{
    The same as Table \ref{tab:BLBQ-graph-1}
    for the case $-\pi/2 \le \theta \le 0$.
  }
  \tablabel{BLBQ-graph-3}
\centering
\ \\
(a)
  \begin{tabular}{|c|c|c|c|c|}
    \hline
    \rule{0cm}{8mm}
    \includegraphics[width=8mm,height=8mm]{FIG/table-head.eps}
    & \includegraphics[width=8mm]{FIG/state-BLBQ-1111.eps}
    & \includegraphics[width=8mm]{FIG/state-BLBQ-1010.eps}
    & \includegraphics[width=8mm]{FIG/state-BLBQ-1100.eps}
    & \includegraphics[width=8mm]{FIG/state-BLBQ-1110.eps} \\
    \hline
    \parbox{8mm}{\rule{0cm}{8mm}\includegraphics[width=8mm]{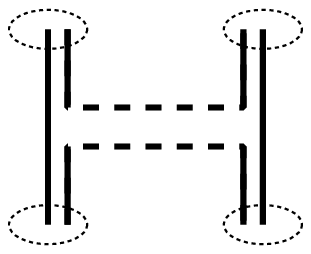}}
    & $0$ & $-J_L$ & $-2J_L$ & $-J_L$\\
    \hline
    \parbox{8mm}{\rule{0cm}{8mm}\includegraphics[width=8mm]{FIG/double-horizontal-BLBQ.eps}}
    & $0$ & $J_Q$ & $J_Q$ & $0$\\
    \hline
  \end{tabular}
\ \\
\ \\
(b)
  \begin{tabular}{|c|c|c|c|}
    \hline
    \rule{0cm}{8mm}
    \includegraphics[width=8mm,height=8mm]{FIG/table-head.eps}
    & \includegraphics[width=8mm]{FIG/state-BLBQ-kink-1.eps}
    & \includegraphics[width=8mm]{FIG/state-BLBQ-kink-1-anti.eps}
    & \includegraphics[width=8mm]{FIG/state-BLBQ-kink-2.eps}\\
    \hline
    \parbox{8mm}{\rule{0cm}{8mm}\includegraphics[width=8mm]{FIG/horizontal-BLBQ.eps}}
    & $1$ & $\frac{-J_L}{J_Q-J_L}$ & $0$\\
    \hline
    \parbox{8mm}{\rule{0cm}{8mm}\includegraphics[width=8mm]{FIG/double-horizontal-BLBQ.eps}}
    & $0$ & $\frac{J_Q}{J_Q-J_L}$ & $1$\\
    \hline
  \end{tabular}
\end{table}

We have described the loop algorithm with the split-spin 
representation, in which all the loops are binary loops.
In the region $-3\pi/4 \le \theta \le -\pi/2$, however,
we can apply the loop algorithm with non-binary loops
as described below.

In the algorithm with the non-binary loops, we do not need to split
spins into $2s$ Pauli spins.
We need two types of graph elements,
cross and horizontal, as we use for the $s=1/2$ $XY$ model.
However, each loop has three states, $0$, $1$ and $2$, rather than two.
Accordingly, the constraint imposed by the graph elements must be
generalized;
the local spin variables on the two points connected by a cross graph 
must be equal whereas those connected by the horizontal one must be 
complementary to each other ($0$ and $2$, or $1$ and $1$).
To be more specific, the procedure of a cycle is as follows:
\begin{liststep}
\item 
  Delete the graph.
\item 
  For each pair of the interacting sites $(ij)$, do the following.
  Decompose the interval $(0,\beta)$ into UIs.
  Place cross graphs with the density $\rho_{\rm C} \equiv J_L$ to the 
  UIs for which the local spin states on $i$ and $j$ are the same.
  Then, place horizontal ones with the density $\rho_{\rm H} \equiv J_Q - J_L$ 
  to the UIs for which the local spin states are complementary to each other.
  \label{st:biq-2}
\item 
  For each kink, place a graph (cross or horizontal) that matches the local state.
  The only local states that match both the graph elements are
$$
  \left(\begin{array}{cc} l' & m' \\ l & m \end{array} \right) = 
  \left(\begin{array}{cc} 2 & 0 \\ 0 & 2 \end{array} \right)\quad\mbox{and}\quad
  \left(\begin{array}{cc} 0 & 2 \\ 2 & 0 \end{array} \right).
$$
  For these kinks, we choose the cross graph with the probability
  $\frac{J_L}{J_Q}$ and the horizontal graph with $1-\frac{J_L}{J_Q}$.
  For kinks with the other local states, the matching graph element is unique.
\item 
  Draw vertical lines to connect legs of graph elements.
\item 
  Identify the clusters.
\item 
  Choose one of the three possible states (0,1 or 2) 
  for each loop with equal probability.
\item 
  Do measurements.
\end{liststep}

%
\subsubsection{Quantum SU($N$) model\cite{HaradaKT2003}}
An SU($N$)-invariant generalization of the
Heisenberg model can be formally written as
\begin{equation}
  \eqlabel{SUN}
  H = \sum_{(ij)} H_{ij}
    = - J \sum_{(ij)} \sum_{\alpha,\beta} J^{\alpha\beta}_i J^{\beta\alpha}_j.
\end{equation}
The symbols $J^{\alpha\beta}_i$ ($\alpha,\beta=1,2,\cdots,N$)
denote the generators of the SU($N$) algebra that satisfy
\begin{equation}
  [J^{\alpha\beta}_i, J^{\mu\nu}_j]
  =
  \delta_{ij}
  \left(
  \delta_{\alpha,\nu} J^{\mu\beta}_i - \delta_{\mu,\beta} J^{\alpha\nu}_j
  \right).
\end{equation}
We consider the model with the fundamental representation 
(and its dual representation)
where the local Hilbert space is $N$-dimensional.
The matrix elements of the pair Hamiltonian $H_{ij}$ can be
specifically written as
\begin{equation}
  \eqlabel{SUN-MATRIX}
  \langle l', m'| H_{ij} | l, m \rangle =
  \left\{
    \begin{array}[h]{rr}
    -J \   \delta_{l, m'} \  \delta_{l',m}  & (J > 0)
    \\
     J \   \delta_{l, \bar m}  \  \delta_{l',\bar m'} & (J < 0) 
  \end{array}\right.,
\end{equation}
where $l,m,l',m' = 0, 1, \cdots, N-1$ are the local `spin' variables
anologous to the ones used in the algorithms for the SU(2) spin models,
and $\bar l \equiv (N-1)-l$.
Then, it can be easily shown that the Hamiltonian is expressed
by a single type of graph elements;
a cross graph if $J$ is positive (ferromagnetic) or
a horizontal graph if $J$ is negative (antiferromagnetic).
(Two special cases have been described in \msss{BIQ}.
The $s=1$ bilinear-biquadratic models at $\theta = -\pi/2$ and $\theta = -3\pi/4$
are the same as the present SU(3) model with $J<0$ and $J>0$, respectively.)

The prescription for the SU($N$) model directly follows from 
this graphical expression:
\begin{liststep}
\item 
  Delete the graph.
\item 
  For each pair of the interacting sites $(ij)$, do the following.
  Decompose the interval $(0,\beta)$ into UIs.
  Place cross (horizontal) graphs with the density $|J|$ to the 
  UIs for which the local spin states on $i$ and $j$ are the same
  (complemantary to each other), if $J>0$ ($J<0$).
\item 
  For each kink, place a cross ($J>0$) or a horizontal ($J<0$) graph.
\item 
  Draw vertical lines to connect the legs of graph elements
  and form closed loops.
\item 
  Identify the loops.
\item 
  Choose one of the $N$ possible states for each loop 
  with equal probability.
\item 
  Do measurements.
\end{liststep}

\subsubsection{Estimators}
\ssslabel{Estimators}
In order to obtain the thermal average of a quantity $\hat Q$,
we compute the Monte Carlo averages of the corresponding estimator $E(S,G)$.
In other words,
$$
  \langle \hat Q \rangle_{\rm thermal}
  = \langle E(S,G) \rangle_{\rm MC}.
$$
While the correspondence between a quantity and an estimator 
is straightforward in many cases, 
it is not so obvious in some other cases.
In the following, we present a list of the estimators of frequently
computed quantities.
Many estimators depends on the world-line configuration, $S$, only,
whereas improved estimators depend only on the graph $G$.
For the derivation, see \mssc{EstimatorsAndEfficiency}.

\paragraph{Estimator of a diagonal operator:}
In this case, the estimator is simply
$$
  E_Q(S) \equiv Q(\psi(0)) \equiv \langle \psi(0) | \hat Q | \psi(0) \rangle,
$$
where $\hat Q$ is the diagonal operator and $\psi(0)$ is the
spin configuration at $\tau = 0$.
Because of the time-translational invariance, a better estimator is
\begin{equation}
  E_Q(S) \equiv \frac1{\beta} \int_0^{\beta} d\tau Q(\psi(\tau)).
  \eqlabel{EstimatorOfDiagonalOperator}
\end{equation}
For a conserved quantity such as $\hat Q \equiv M^z \equiv \sum_i S^z_i$
in the $XXZ$ quantum spin model, these two estimators are identical.
In this particular case, the estimator is simply
$$
  E_{M^z}(S) = \sum_i \psi_i(0).
$$
The time-dependent correlation function of two diagonal operators
$$
  \Gamma_{AB}(\tau,\tau') \equiv \langle \hat A(\tau) \hat B(\tau') \rangle
$$
can be computed with the estimator
$$
  E_{AB}(S) \equiv A(\psi(\tau)) B(\psi(\tau')).
$$
By integrating over the imaginary time, we obtain an estimator for the
generalized susceptibility 
$\chi_{AB} \equiv \int_0^{\beta} d\tau \Gamma_{AB}(\tau,0)$ as
\begin{equation}
  E_{\chi_{AB}}(S) \equiv \beta^{-1}
  \left( \int_0^{\beta} d\tau A(\psi(\tau)) \right)
  \left( \int_0^{\beta} d\tau B(\psi(\tau)) \right).
  \eqlabel{EstimatorForDiagonalSusceptibility}
\end{equation}

\paragraph{Energy and Specific Heat:}
The estimator for the total energy, $\langle \Ham \rangle$, is
\begin{equation}
  E_{\Ham}(S) =  E_{\diag{\Ham}}(S) - \frac1\beta n_{\rm kink}(S),
\end{equation}
where $n_{\rm kink}(S)$ is the total number of kinks in $S$, and
$E_{\diag{\Ham}}(S)$ is the estimator for the diagonal part of the Hamiltonian,
which is defined by \Eq{EstimatorOfDiagonalOperator} with 
$Q$ being $\diag{\Ham}$.
The specific heat is not measured with a single estimator.
Instead, it is computed using the following expression.
\begin{eqnarray}
  C & = & \beta^2 \left[
         \MCave{E_{\diag{\Ham}}^2} - {\MCave{E_{\diag{\Ham}}}}^2
         \right] \nonumber \\
    &   & \ 
      + \MCave{n_{\rm kink}^2}
      - \MCave{n_{\rm kink}}^2
      - \MCave{n_{\rm kink}}.
\end{eqnarray}

\paragraph{Improved operator:}
For the diagonal magnatic susceptibility,
$$
  \chi_{zz} \equiv \int_0^{\beta} d\tau
                   \langle M^z(\tau)M^z(0) \rangle_{\rm thermal}
$$
at zero field, the following graphical estimator is often useful.
For the cases where the clusters can be flipped with probability $1/2$,
i.e., if no field breaks the up-down symmetry, the estimator is
\begin{equation}
  \eqlabel{measure-improved-sus}
  E_{\chi_{zz}}(G) = \frac1{\beta} \sum_{c} M_c^2,
\end{equation}
where the summation is over all the clusters in $G$, and
$M_c$ is the cluster magnetization
$$
  M_c \equiv \left|\int_c dX \psi(X)\right|.
$$
The symbol $\int_c dX$ denotes the summation/integration over
the cluster $c$.

\paragraph{Non-diagonal susceptibility:}
The susceptibility of the spin components perpendicular to the
quantization axis, i.e.,
$$
  \chi_{xx} \equiv \int_0^{\beta} d\tau
                   \langle M^x(\tau)M^x(0) \rangle_{\rm thermal}
$$
can be measured for the $XXZ$ spin model with the estimator
\begin{equation}
  \eqlabel{measure-xx-cor}
  E_{\chi_{xx}}(G) = \frac1{4\beta} \sum_{c} V_c^2,
\end{equation}
where $V_c$ is the cluster volume
$$
  V_c \equiv \int_c dX 1.
$$

%
\subsection{Worm Algorithm\cite{ProkofevST1998}}
\ssclabel{WormAlgorithmRecipe}

The loop algorithm often becomes very inefficient (worse than the local
updating algorithm) when the Hamiltonian contains some terms 
that conflict with each other.
Such is the case in the anti-ferromagnetic $XXZ$ model in a
uniform magnetic field.
This difficulty can be removed by introducing discontinuity points
in the world-lines. 
However, the $XXZ$ models with an easy-axis anisotropy should not be
dealt with the worm algorithm if the magnetic field is not conflicting with
the exchange couplings, since the loop algorithm usually performs much better
in such cases especially near the transition point.

A cycle in the worm algorithm consists of a creation 
and an annihilation of a worm, and
a vertical move, a jump and an anti-jump of the head.
(See \Fig{WormMovements}).
A head or a tail of the worm is called positive (negative) if
the local state above it is greater (smaller) than 
the local state below by one.
If the head is positive the tail must be negative and vice versa.
When the positive one is created above the negative one,
the worm is called a ``lowering'' worm since the local state
between the two discontinuity points is lower than the original state by one
(\Fig{worm-possibility}).
A ``raising'' worm is the one in which the positive discontinuity point
is below the negative one.


In what follows, we describe the worm algorithm for the generic Hamiltonian
\begin{equation}
  \eqlabel{hamiltonian-boson-worm}
   {\cal H} = \sum_{(ij)}\left(U_{ij} + V_{ij}\right) - \eta \sum_i Q_i,
\end{equation}
where $V_{ij} \equiv \diag{{\Ham}_{ij}}$ is the diagonal part of the pair Hamiltonian,
$U_{ij} \equiv \Ham_{ij} - V_{ij}$ is the off-diagonal part, and
$Q_i$ is an operator defined on the site $i$ that has no diagonal part.
The last term is a source term which is included only for a technical purpose.
The constant $\eta$ is any finite real number of $O((N\beta)^{-1/2})$
where $N$ is the total number of sites in the whole system.
%

The procedure of one cycle, starting from a state with no worm,
is the following:
\begin{description}
\item[Step 1-1]
    [Creation (\Fig{WormMovements}(a))]
    Choose one of the two types (raising or lowering) 
    of the worm with equal probability.
\item[Step 1-2]
    Choose a point from the whole space-time.
    (In what follows, we consider a continuous part of a vertical line 
    delimited by kinks in which the site under consideration is involved.
    The head and the tail of the worm themselves are also regarded as kinks here.
    We call such a part, a single-site UI.)
    We denote the sincle-site UI in which the chosen point resides
    by $I \equiv \{(i,\tau) | \tau \in [t_1, t_2]\}$.
\item[Step 1-3]
    Choose two points $(i,\tau_1)$ and $(i,\tau_2)$ from $I$
    so that $\tau_1 < \tau_2$ with the probability density
    \begin{equation}
      p_{\rm c}(\tau_1,\tau_2) = \frac{\exp[-\overline{\Delta V}\times (\tau_2-\tau_1)]}{A}.
    \end{equation}
    These points are candidates for the positions at which a worm may be created.
    The constant $A$ is the normalization factor and
    $\overline{\Delta V}$ is the average excess axion per unit time defined as
    \begin{equation}
      \overline{\Delta V} \equiv \frac{1}{t_2-t_1} \int_{t_1}^{t_2}dt\,
      \sum_{j\in\delta(i)} \Delta V_{ij}^{\rm create}(t),
    \end{equation}
    where $\delta(i)$ denotes the set of the nearest neighbors of $i$,
    and $\Delta V_{ij}^{\rm create}(t)$ is the increase in
    $V_{ij}(t) \equiv \langle \psi(t) | V_{ij} | \psi(t) \rangle$ 
    that occurs if the two discontinuity points are created at the bottom and the
    top of $I$.
\item[Step 1-4]
    Accept the proposed creation of a worm
    at $\tau_1$ and $\tau_2$
    with the probability $\min(1,R_{\rm create})$ where $R_{\rm create}$ 
    is defined as
    \begin{eqnarray}
       & & 
       R_{\rm create} \equiv \frac{2N\beta\eta^2 A}{t_2-t_1} 
       \nonumber \\
       & & \quad
       \times \exp\left[ \int_{\tau_1}^{\tau_2}dt\, 
       \left( \overline{\Delta V} - \sum_{j\in\delta(i)} \Delta V_{ij}^{\rm create}(t)\right)
       \right] 
       \nonumber \\
       & & \quad
       \times \langle \psi'(\tau_1+0)|Q_i|\psi'(\tau_1-0) \rangle 
       \nonumber \\
       & & \quad
       \times \langle \psi'(\tau_2+0)|Q_i|\psi'(\tau_2-0) \rangle,
       \eqlabel{CreateProbability}
    \end{eqnarray}
    where $\psi'$ is the state after the creation was accepted.
    If the proposal is rejected, go to Step 6.
\item[Step 1-5]
   Choose one of the two discontinuities with equal probability,
   and make it the head.
\item[Step 2]
   [Vertical Move (\Fig{WormMovements}(b))]
   Move the head along the UI where it currently resides.
   The new time $\tau$ is chosen with the probability density
   \begin{equation}
     p_{\rm vertical} (\tau) \equiv 
     \frac{\exp[-\sum_{j \in \delta (i)} \Delta V_{ij}^{\rm move}(\tau)]}{B},
   \end{equation}
   where $B$ is the normalization factor and $\Delta V_{ij}^{\rm move}(\tau)$
   is the increase in the diagonal contribution $\int_{\rm UI} dt\, V_{ij}(t)$ that occurs if the vertical move to the time $\tau$ is accepted.
\item[Step 3] 
  [Jump (\Fig{WormMovements}(c))] 
  For each nearest-neighbor site $j$ of the current site $i$,
  consider two intervals; the one delimited by the head
  itself and the nearest kink above it, and the other 
  delimited by the head and the neaerst kink below it.
  (Here, we only consider kinks in which the sites $i$ and/or $j$ 
  are involved.)
  For each of the intervals, do Steps 3-1 and 3-2.
\item[Step 3-1] 
  Generate a time $\tau$, 
  which is the candidate for the temporal position of placing a kink,
  with the probability density
  \begin{equation}
    p_{\rm jump}(\tau) \equiv \frac{\exp(-\Delta V_{\rm jump}(\tau))}{C},
  \end{equation}
  where $C$ is the normalization factor and
  $\Delta V_{\rm jump}(\tau)$ is the increase in the diagonal contribution
  to the total weight that occurs if the kink is created 
  at the proposed time $\tau$. Specifically, 
  \begin{eqnarray*}
    \Delta V_{\rm jump}(\tau) & \equiv &
    \int_{t_1}^{t_2} dt\, \Big[
    \sum_{k\in\delta(i)} \Delta V_{ik}(t) \\
    & &
    + \sum_{k\in\delta(j)} \Delta V_{jk}(t)
    - \Delta V_{ij}(t) \Big],
  \end{eqnarray*}
  where $t_1$ and $t_2$ denote the starting and the ending time of the interval,
  respectively, and $\Delta V_{ij}(t)$ is the increase in the $V_{ij}(t)$
  after the jump was accepted.
\item[Step 3-2]
  Let the head jump from $i$ to $j$ at $\tau$, creating a kink there,
  with the probability $\min(1,R_{\rm jump})$. (Otherwise, do nothing.)
  Here, $R_{\rm jump}$ is defined as
  \begin{eqnarray}
    & &
    R_{\rm jump} 
    \equiv {C \langle \psi'(\tau+0) | U_{ij} | \psi'(\tau-0) \rangle} 
    \nonumber \\
    & & \times
    \frac{\langle \psi'(\tau_w+0) | Q_j | \psi'(\tau_w-0) \rangle}
    {\langle \psi(\tau_w+0) | Q_i | \psi(\tau_w-0) \rangle},
    \eqlabel{JumpProbability}
  \end{eqnarray}
  where $\psi'$ is the state after the jump was accepted and
  $\tau_w$ is the head's current temporal position.
\item[Step 4] 
  [Anti-Jump (\Fig{WormMovements}(c))] 
  Consider two directions, upward and downward.
  For each nearest-neighbor site $j$, and
  for each direction, do the following.
  If the kink,
  involving $i$ and/or $j$ and 
  the nearest to the head in the chosen direction,
  is not the one between $i$ and $j$, 
  do nothing and go to Step 5.
  Otherwise, we consider the second nearest kink among those which
  involves $i$ and/or $j$, and the interval delimited by it and
  the head itself.
  Let the head anti-jump, i.e.,
  let it jump from $j$ to $i$ so that the kink between 
  $i$ and $j$ is annihilated,
  with the probability $\min(1,R_{\rm anti-jump})$, where $R_{\rm anti-jump}$ 
  is the reciprocal of \Eq{JumpProbability} 
  with $\psi'$ being the current state, 
  $\psi$ the state that the anti-jump would result in, 
  and $\tau$ the temporal position of the nearest kink
  to be erased by the anti-jump.
\item[Step 5]
  [Annihilation (\Fig{WormMovements}(a))] 
  If the head and the tail are located on the same site and
  there are no kinks between the two, annihilate the worm
  with the probability $\min(1,R_{\rm annihilate})$ where 
  $R_{\rm annihilate}$ is the reciprocal of \Eq{CreateProbability}
  with $\psi'$ being the current state, $\psi$ the state that
  the annihilation would result in, and 
  $\tau_1$ and $\tau_2$ the temporal positions of the
  head and the tail. Go to Step 6.
  If annihilation is not chosen, go to Step 2.
\item[Step 6]
  Do measurements. (The end of the cycle.)
\end{description}



%
%
%
\subsection{Directed-Loop Algorithm}
\ssclabel{DirectedLoopAlgorithmRecipe}
%
The directed-loop algorithm can be thought of as a generalization of the
worm algorithm.
Therefore, remarks similar to the ones presented at the beginning of the 
last subsection apply to the directed-loop algorithm.
As for the simulations of $XXZ$ spin model, 
the directed algorithm should be used in the
cases with the isotropic couplings or the easy-plane couplings,
with or without a magnetic field.
For the easy-axis couplings, the loop algorithm should be used.

In the directed-loop algorithm proposed by Sylju\r{a}sen and Sandvik
\cite{SyljuasenS2002},
the spin configuration is updated through movements of the worm
as in the worm algorithm.
One ``sweep'' of the directed-loop algorithm consists of
an assignment of vertices and a number of cycles of worm update.
Vertices are represented as horizontal lines connecting 
nearest neighbor sites.
Unlike the worm algorithm, the head of the worm has a direction of motion
and it can only move in this direction.
(\Fig{DirectedLoopUpdate} shows a local configuration 
in which a worm and vertices are involved.)
It can change the direction of motion and its spatial location
only when it hits a vertex.
One cycle of the worm update, therefore, consists of 
the creation of a worm, the vertical movements in the direction of motion,
the scatterings at vertices, and the annihilation of the worm.

\subsubsection{Directed-Loop Algorithm for the $s=1/2$ $XXZ$ Model}
\ssslabel{DirectedLoopAlgorithmSpinOneHalf}
In the present subsection, we specialize in the description of
the $s=1/2$ $XXZ$ model,
$$
  \eqlabel{hamiltonian-XXZ-spin-one-half}
  {\cal H}_{ij} =  
  - J\big(\sigma_i^x \sigma_j^x + \sigma_i^y \sigma_j^y\big)
  - J^\prime \sigma_i^z \sigma_j^z 
  - \frac{h}{2}(\sigma_i^z +\sigma_j^z)
  - E_0
$$
with $J > 0$ and $\ h \ge 0$.
The whole parameter space is divided into six regions as shown in \Fig{xxz}.
The constant $E_0$ is chosen as $E_0 = (J-h)s^2+hs$ for the regions I and V,
$E_0 = -J's^2+hs$ for the regions II, III, and IV, and 
$E_0 = J's^2+h(s-2s^2)$ for the region VI.
(While $s=1/2$ in the present case,
the same expression can be used for the general $s$ case
discussed in the next section.)
Within each region, the scattering probability and the vertex density
are simple analytic functions of $J$, $J'$ and $h$.

\deffig{xxz}{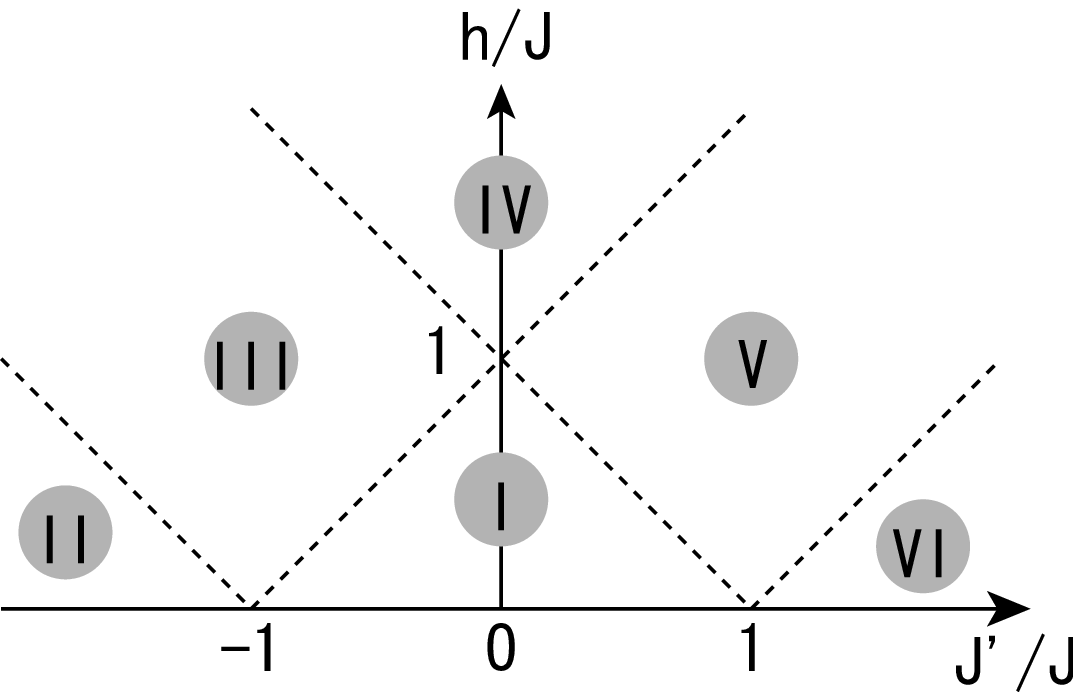}{0.4}{The six regions in the parameter space
of the $XXZ$ model.}

\begin{table}
\caption{
The local states around a vertex and their symbols used in
Tables \Tab{XXZ-spin-one-half}-\Tab{boson}.
\tablabel{LocalStates}
}
\begin{tabular}{|cc|cc|}
\hline
Symbol $\Sigma$ & State & Symbol $\Sigma$ & State \\
\hline
  $\left( \begin{array}{cc}l & m\end{array} \right)$ &
  \figbox{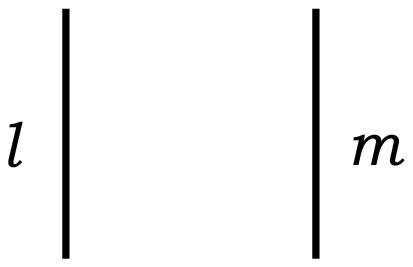}   &
                          &
\\
\hline
  \mattwo{l'}{m'}{l_+}{m} &
  \figbox{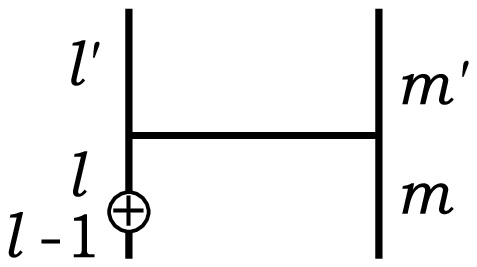}   &
  \mattwo{l'}{m'}{l_-}{m} &
  \figbox{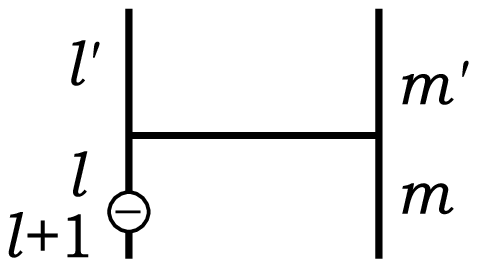} 
\\
\hline
  \mattwo{l'_+}{m'}{l}{m} & 
  \figbox{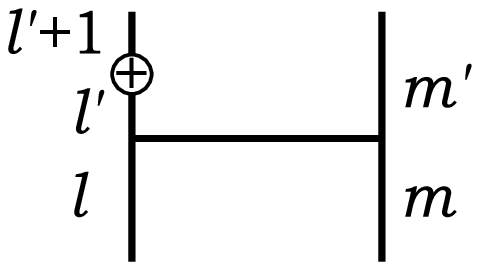}   &
  \mattwo{l'_-}{m'}{l}{m} & 
  \figbox{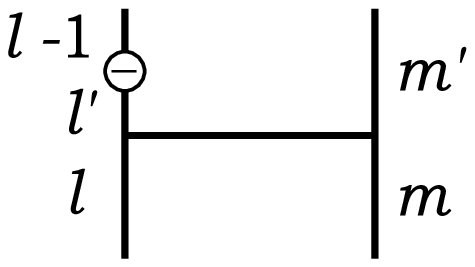}
\\
\hline
  \mattwo{l'}{m'_+}{l}{m} &
  \figbox{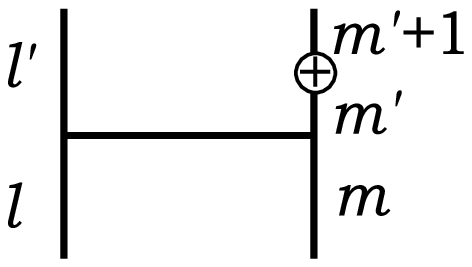}   &
  \mattwo{l'}{m'_-}{l}{m} &
  \figbox{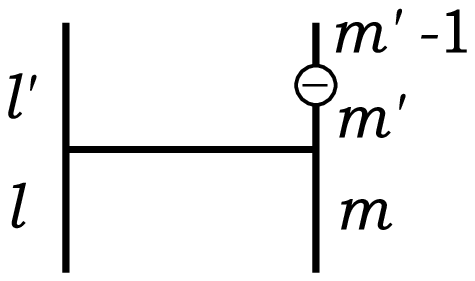} 
\\
\hline
  \mattwo{l'}{m'}{l}{m_+} &
  \figbox{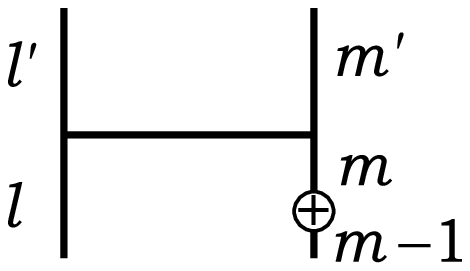}   &
  \mattwo{l'}{m'}{l}{m_-} &
  \figbox{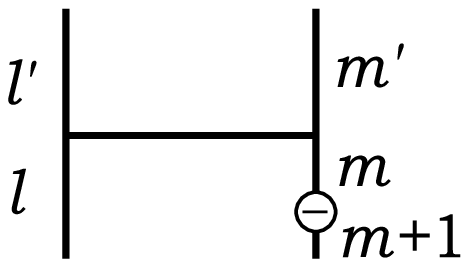} \\
\hline
\end{tabular}
\end{table}

When the head hits a vertex, 
one of four ways of scattering is chosen probabilistically:
turn-back, straight (vertical), diagonal, and horizontal,
as shown in \Fig{Scattering}. 
In what follows, the probability for choosing one out of these four
is denoted as $P(\downarrow|\Sigma)$, $P(\uparrow|\Sigma)$, 
$P(\nearrow|\Sigma)$, or $P(\rightarrow|\Sigma)$.
Here, $\Sigma$ is the local state at the vertex before
the head's arrival (\Tab{LocalStates}).

The whole space-time can be decomposed into a number of UI's.
(Here, a uniform interval (UI) is defined for a pair of 
nearest neighbor sites, rather than for a single site.)
Then, one Monte Carlo step in the directed-loop algorithm
consists of the following operations:
\begin{description}
\item[Step 1]
  Remove all the vertices with no kink on it.
  (As a result, only the vertices that delimit UIs remain.)
\item[Step 2]
  Place vertices in each UI uniform-randomly with 
  the density $\rho(\Sigma)$ given in \Tab{XXZ-spin-one-half}.
\item[Step 3]
  Repeat the cycle (Steps 3-1 through 3-4 or 3-4${}^\prime$) 
  $N_{\rm cycle}$ times.
\item[Step 3-1]
  Choose a point in the whole space-time uniform-randomly,
  and place the head and the tail both at the same point.
  (If the spin state at the point is 1 (up), 
  the initial type of the worm is loweing.
  Otherwise it is raising. (See \Fig{worm-possibility}.))
\item[Step 3-2]
  Choose one of the two discontinuities and make it the head.
  Choose the initial direction of motion, upward or downward,
  with equal probability. 
\item[Step 3-3]
  Let the head go until it hits a vertex or comes back to
  the original position where the tail stays.  
  If it hits a vertex, go to Step 3-4.
  If it hits the tail, go to Step 3-4${}^\prime$.
\item[Step 3-4]
  Choose the scattering direction $\Gamma$ with
  the probability $P(\Gamma|\Sigma)$ in 
  \Tab{XXZ-spin-one-half}.
  (The type of the head, positive or negative, is not changed by the
  scattering.)
  Then, go back to Step 3-3.
\item[Step 3-4${}^\prime$]
  Let the worm annihilate.
  (The end of one cycle.)
\item[Step 4]
  Do measurments. (The end of one Monte Carlo step.)
\end{description}
The number of the cycles $N_{\rm cycle}$ in one Monte Carlo step
is an arbitrary fixed number.
It is usually set so that every space-time point is visited
once on average during a Monte Carlo step.
(A sample program based on this procedure may be found at a web site.
\cite{WebSite})

\begin{fulltable}
\caption{The directed-loop algorithm for the quantum $s=1/2$
  $XXZ$ spin models. The density of vertices $\rho$, and  the scattering
  probabilities of the head $P(\Gamma|\Sigma)$ are shown. 
The latter are for presented only in the case where
the head is entering the vertex from below along the left line.
The scattering probabilities for the other cases can be obtained
by symmetry transformations.
  The probability of going through
  ($\Gamma = \uparrow$) is simply equal to $1-$ [the probabilities of the
  three proper scatterings].
  The symbol $h$ denotes the magnetic field per pair of spins,
  which is related to the magnetic field per original spin, $H$, by
  $h = H/d$ for the $d$-dimensional hyper cubic lattice.
}
\ \\[-5mm]
\tablabel{XXZ-spin-one-half}
\begin{tabular}{c|lcccccc}
\hline
\hline
$\Sigma$ &  & Region I & Region II & Region III & Region IV & Region V & Region VI \\
\hline
\hline
$\left( \begin{array}{cccc} l & m \end{array} \right)$ &
$\rho(\Sigma)= $ &
A & B & B & B & A & C \\
\hline
\hline
&
$P(\downarrow|\Sigma)=$ &
0 & 
0 &
0 &
0 &
0 &
$\frac{-J+J'-h}{2C}$ \\
$\left(\begin{array}{cccc} 0 & 0 \\ 0_- & 0 \end{array}\right)$ &
$P(\nearrow|\Sigma)=$ &
$\frac{J+J'-h}{4A}$ & 
0 &
0 &
0 &
$\frac{J+J'-h}{4A}$ &
$\frac{J}{2C}$ \\
  &
$P(\rightarrow|\Sigma)=$ &
0 &
0 &
0 &
0 &
0 &
0 \\
\hline
  &
$P(\downarrow|\Sigma)=$ &
0 & 
$\frac{-J-J'-h}{2B}$ &
0 &
0 &
0 &
0 \\
$\left(\begin{array}{cccc} 0 & 1 \\ 0_- & 1 \end{array}\right)$ &
$P(\nearrow|\Sigma)=$ &
0 & 
0 &
0 &
0 &
0 &
0 \\
  &
$P(\rightarrow|\Sigma)=$ &
$\frac{J-J'-h}{4A}$ &
$\frac{J}{2B}$ &
$\frac{J-J'-h}{4B}$ &
0 &
0 &
0 \\
\hline
  &
$P(\downarrow|\Sigma)=$ &
0 &
$\frac{-J-J'+h}{2B}$ &
$\frac{-J-J'+h}{2B}$ &
$\frac{-J-J'+h}{2B}$ &
0 &
0 \\
$\left(\begin{array}{cccc} 1 & 0 \\ 1_+ & 0 \end{array}\right)$ &
$P(\nearrow|\Sigma)=$ &
0 &
0 &
0 &
0 &
0 &
0 \\
  &
$P(\rightarrow|\Sigma)=$ &
$\frac{J-J'+h}{4A}$ &
$\frac{J}{2B}$ &
$\frac{J}{2B}$ &
$\frac{J}{2B}$ &
$\frac{J-J'+h}{4A}$ &
0 \\
\hline
  &
$P(\downarrow|\Sigma)=$ &
0 &
0 &
0 &
$\frac{-J+J'+h}{2B}$ &
$\frac{-J+J'+h}{2A}$ &
$\frac{-J+J'+h}{2C}$ \\
$\left(\begin{array}{cccc} 1 & 1 \\ 1_+ & 1 \end{array}\right)$ &
$P(\nearrow|\Sigma)=$ &
$\frac{J+J'+h}{4A}$ &
0 &
$\frac{J+J'+h}{4B}$ &
$\frac{J}{2B}$ &
$\frac{J}{2A}$ &
$\frac{J}{2C}$ \\
  &
$P(\rightarrow|\Sigma)=$ &
0 &
0 &
0 &
0 &
0 &
0 \\
\hline
  &
$P(\downarrow|\Sigma)=$ &
0 &
0 &
0 &
0 &
0 &
0 \\
$\left(\begin{array}{cccc} 1 & 0 \\ 0_- & 1 \end{array}\right)$ &
$P(\nearrow|\Sigma)=$ &
$\frac{J+J'+h}{2J}$ &
0 &
$\frac{J+J'+h}{2J}$ &
1 &
1 &
1 \\
  &
$P(\rightarrow|\Sigma)=$ &
$\frac{J-J'-h}{2J}$ &
1 &
$\frac{J-J'-h}{2J}$ &
0 &
0 &
0 \\
\hline
  &
$P(\downarrow|\Sigma)=$ &
0 &
0 &
0 &
0 &
0 &
0 \\
$\left(\begin{array}{cccc} 0 & 1 \\ 1_+ & 0 \end{array}\right)$ &
$P(\nearrow|\Sigma)=$ &
$\frac{J+J'-h}{2J}$ &
0 &
0 &
0 &
$\frac{J+J'-h}{2J}$ &
1 \\
  &
$P(\rightarrow|\Sigma)=$ &
$\frac{J-J'+h}{2J}$ &
1 &
1 &
1 &
$\frac{J-J'+h}{2J}$ &
0 \\
\hline
\hline
\multicolumn{8}{c}{
\begin{minipage}{130mm}
  \begin{equation*}
  A \equiv B +\frac14(J+J'-h), \qquad
  B \equiv \frac12(h-J')(l+m)+J'lm, \qquad
  C \equiv B+\frac12(J'-h)
  \end{equation*}
\end{minipage}
}\\
\hline
\hline
\end{tabular}

\caption{
The coarse-grained algorithm for the quantum $XXZ$ spin model
with arbitrary $s$.
The density of vertices, $\rho$, and 
the scattering probabilities of the head $P(\Gamma|\Sigma)$ are shown.
The symbol $h$ denotes the magnetic field per pair of Pauli spins,
which is related to the magnetic field per original spin, $H$, by
$h = H/(2ds)$ for the $d$-dimensional hyper cubic lattice.
($\bar l \equiv 2s - l$, $\bar m \equiv 2s - m$.)
}
\ \\[-8mm]
\tablabel{XXZ}
\begin{tabular}{c|lcccccc}
\hline
\hline
$\Sigma$ &  & Region I & Region II & Region III & Region IV & Region V & Region VI \\
\hline
\hline
$\left( \begin{array}{cccc} l & m \end{array} \right)$ &
$\rho(\Sigma)= $ &
A & B & B & B & A & C \\
\hline
\hline
  &
$P(\downarrow|\Sigma)=$ &
0 & 
$\frac{m (-J-J'-h)}{2B}$ &
0 &
0 &
0 &
$\frac{\bar m (-J+J'-h)}{2C}$ \\
$\left(\begin{array}{cccc} l & m \\ l_- & m \end{array}\right)$ &
$P(\nearrow|\Sigma)=$ &
$\frac{\bar m (J+J'-h)}{4A}$ & 
0 &
0 &
0 &
$\frac{\bar m (J+J'-h)}{4A}$ &
$\frac{\bar m J}{2C}$ \\
  &
$P(\rightarrow|\Sigma)=$ &
$\frac{m (J-J'-h)}{4A}$ &
$\frac{m J}{2B}$ &
$\frac{m (J-J'-h)}{4B}$ &
0 &
0 &
0 \\
\hline
  &
$P(\downarrow|\Sigma)=$ &
0 &
$\frac{\bar m (-J-J'+h)}{2B}$ &
$\frac{\bar m (-J-J'+h)}{2B}$ &
$\frac{
{\tiny \begin{array}{ll}
m (-J+J'+h) \\
\quad + \bar m (-J-J'+h)
\end{array}}
}{2B}$
&
$\frac{m (-J+J'+h)}{2A}$ &
$\frac{m (-J+J'+h)}{2C}$ \\
$\left(\begin{array}{cccc} l & m \\ l_+ & m \end{array}\right)$ &
$P(\nearrow|\Sigma)=$ &
$\frac{m (J+J'+h)}{4A}$ &
0 &
$\frac{m (J+J'+h)}{4B}$ &
$\frac{m J}{2B}$ &
$\frac{m J}{2A}$ &
$\frac{m J}{2C}$ \\
  &
$P(\rightarrow|\Sigma)=$ &
$\frac{\bar m (J-J'+h)}{4A}$ &
$\frac{\bar m J}{2B}$ &
$\frac{\bar m J}{2B}$ &
$\frac{\bar m J}{2B}$ &
$\frac{\bar m (J-J'+h)}{4A}$ &
0 \\
\hline
  &
$P(\downarrow|\Sigma)=$ &
0 &
0 &
0 &
0 &
0 &
0 \\
$\left(\begin{array}{cccc} l+1 & m \\ l_- & m+1 \end{array}\right)$ &
$P(\nearrow|\Sigma)=$ &
$\frac{J+J'+h}{\bar l\cdot 2J}$ &
0 &
$\frac{J+J'+h}{\bar l\cdot 2J}$ &
$\frac1{\bar l}$ &
$\frac1{\bar l}$ &
$\frac1{\bar l}$ \\
  &
$P(\rightarrow|\Sigma)=$ &
$\frac{J-J'-h}{\bar l\cdot 2J}$ &
$\frac1{\bar l}$ &
$\frac{J-J'-h}{\bar l\cdot 2J}$ &
0 &
0 &
0 \\
\hline
  &
$P(\downarrow|\Sigma)=$ &
0 &
0 &
0 &
0 &
0 &
0 \\
$\left(\begin{array}{cccc} l-1 & m \\ l_+ & m-1 \end{array}\right)$ &
$P(\nearrow|\Sigma)=$ &
$\frac{J+J'-h}{l\cdot 2J}$ &
0 &
0 &
0 &
$\frac{J+J'-h}{l\cdot 2J}$ &
$\frac1l$ \\
  &
$P(\rightarrow|\Sigma)=$ &
$\frac{J-J'+h}{l\cdot 2J}$ &
$\frac1l$ &
$\frac1l$ &
$\frac1l$ &
$\frac{J-J'+h}{l\cdot 2J}$ &
0 \\
\hline
$\left(\begin{array}{cc} l+1 & m \\ l_+ & m+1 \end{array} \right)$ &
  &
\multicolumn{6}{l}{
$
P(\downarrow|\Sigma)=
P(\nearrow|\Sigma)=
P(\rightarrow|\Sigma)= 0, \quad\mbox{and}\quad
P(\uparrow|\Sigma)=1
$
}
\\
\hline
$\left(\begin{array}{cc} l-1 & m \\ l_- & m-1 \end{array} \right)$ &
  &
\multicolumn{6}{l}{
$
P(\downarrow|\Sigma)=
P(\nearrow|\Sigma)=
P(\rightarrow|\Sigma)= 0, \quad\mbox{and}\quad
P(\uparrow|\Sigma)=1
$
}
\\
\hline
\hline
\multicolumn{8}{c}{
\begin{minipage}{130mm}
\begin{eqnarray*}
  A & \equiv & 
    \frac14 [lm (J+J'+3h) 
    + (l\bar m + \bar l m)(J-J'+h) + \bar l \bar m (J+J'-h)] \\
  B & \equiv &
    lm h + (l\bar m + \bar l m) \frac{-J'+h}2, \qquad
  C \equiv  \frac12 [lm(J'+h) + \bar l \bar m (J'-h)] 
\end{eqnarray*}
\end{minipage}
}\\
\hline
\hline
\end{tabular}

\end{fulltable}

\subsubsection{Directed-Loop Algorithm for $s>1/2$ Models 
(Coarse-Grained Algorithm)}
\ssslabel{CoarseGrained}

As stated in \mssc{DirectedLoopAlgorithm},
the head-scattering probability in the directed-loop algorithm
is not uniquely determined by the detailed balance condition.
In \mssc{CoarseGrainedAlgorithm}, we have explained how
we can use a solution to \Eq{TimeInversionSymmetry}
for $s=1/2$ to obtain a solution for larger spins.\cite{HaradaK2002}
Since the whole parameter space (i.e., $J$-$J'$-$H$ space) 
is devided in the six regions in the solution for $s=1/2$ (\Fig{xxz}),
the same devision applies to the larger spins.
It should be also noted that the turning-back 
probability in Region I is always zero for any $s$.
The directed-loop algorithm for large spins is
different from that for $s=1/2$ in the creation and the
annihilation of the worm.
Otherwise, the procedure is the same as the one
in the $s=1/2$ case described in 
\msss{DirectedLoopAlgorithmSpinOneHalf}.
Therefore, one Monte Carlo step of the coarse-grained algorithm 
for spin $s>1/2$ can be obtained by replacing 
the \Tab{XXZ-spin-one-half} by \Tab{XXZ},
and Step 3-1 and Step 3-4${}^{\prime}$ by the following operations:
\begin{description}
\item[Step 3-1]
  Choose a point in the whole space-time uniform-randomly,
  and create a worm there.
  Then, choose the initial type of the worm,
  raising or lowering, with the probability $\bar{l}/(2s)$ or $l/(2s)$, 
  respectively. 
  Here $l=0,1,\cdots,2s$ is the spin-state variable at
  the chosen point, and $\bar{l}\equiv 2s - l$.
\end{description}
and
\begin{description}
\item[Step 3-4${}^{\prime}$]
  Let the head annihilate with the tail, or 
  let it go through.
  The probability of the annihilation depends on the
  type of the worm and the local spin state $l$ 
  between the head and the tail just before the collision.
  If the worm is of the raising type,
  the annihilation probability is $l^{-1}$,
  while it is $(\bar{l})^{-1}$ otherwise.
  If the head goes through, go back to Step 3-3.
  If the worm is annihilated, go to Step 4.
\end{description}


\subsubsection{Soft-Core Boson Model with Repulsive Interactions}
\ssslabel{SoftCoreBosonModel}
It was pointed out that the algorithm for general $s$ presented above
can be used for bosonic models by taking the $s\to\infty$ limit
\cite{SmakovHK2003}.
Here, we consider the tight-binding soft-core boson model
(or the boson Hubbard model)
\begin{eqnarray}
  \eqlabel{hamiltonian-boson}
  {\cal H} &=& - \sum_{(ij)} \left[\frac{t}{2}(b_i^\dagger b_j + b_ib_j^\dagger)
  - V_1 n_i n_j \right]\nonumber\\
&&\qquad - \sum_i \left[\mu n_i-\frac{V_0}{2} n_i (n_i -1)\right],
\end{eqnarray}
where $b_i$ and $b_i^\dagger$ are an annihilation and a creation operator,
respectively, on the site $i$, and $n_i \equiv b_i^\dagger b_i$.
We here assume
that $V_0 \ge 0$ and $V_1 \ge 0$.
The model can be considered as a spin model with $s=\infty$.
In the present algorithm, 
the state is updated in much the same way as the worm algorithm.
A cycle consists of 
(i) the creation of a worm, 
(ii) the movements of the head, and 
(iii) the annihilation of the worm.
We do not use vertices explicitly.
Instead, ``scattering'' points are dynamically created with some 
density ahead of the head, and it is scattered at the
nearest scattering point.
One cycle of the coarse-grained algorithm 
for the soft-core boson model can be stated as follows:
\begin{description}
\item[Step 1]
  \label{st:bose-1}
  Choose a point uniform-randomly in the whole space-time, 
  and create there a worm. 
  (One is the head and the other is the tail.)
  The initial type of the worm is raising (\Fig{worm-possibility}).
\item[Step 2]
  Choose the initial direction of motion, upward or downward, 
  with equal probability. 
\item[Step 3]
  For each nearest-neighbor site $j$ to the current site $i$,
  consider the UI between $i$ and $j$ ahead of the head.
  For each one of the four scattering directions $\Gamma$,
  generate a time $\tau_{j,\Gamma}$ in the UI.
  Here $\tau_{j,\Gamma}$ is the closest to the current temporal
  position of the head among those generated with the density 
  given in \Tab{boson}.
  When this is done for all nearest neighbors $j$ and 
  all directions $\Gamma$, 
  choose from the $\tau_{j,\Gamma}$s
  the closest to the current temporal position.
  (Let it be $\tau_{j_0,\Gamma_0}$.)
\item[Step 4]
  Compare the nearest scattering point generated in Step 3,
  the nearest kink ahead of the head, and
  the tail
  if it resides on the same site as the head.
  If the nearest among these three
  is the scattering point, go to Step 5.
  If it is the kink, go to Step 5${}^{\prime}$.
  If it is the tail, go to Step 6.
\item[Step 5]
  Let the head scatter 
  as specified by $j_0$ and $\Gamma_0$.
  Go to Step 3.
\item[Step 5${}^{\prime}$]
  Choose the direction of the scattering with the probability
  $P(\Gamma | \Sigma)$ given in \Tab{boson},
  and let the head scatter.
  Go to Step 3.
\item[Step 6]
  If the head is negative and comes back to the tail
  from below, or if it is positive and comes back from above,
  let it pass the tail with the probability $1$ and go to Step 3.
  Otherwise, let it pass with the probability $1-1/n$
  where $n$ is the number of the particles ahead of the head
  just before the collision. Go to Step 3.
\item[Step 7]
  Let the worm annihilate.
\item[Step 8]
  Do measurements.
\end{description}

\begin{fulltable}
  \caption{
    The densities of scattering points $\rho(\Gamma|\Sigma)$
    and the scattering probability at kinks $P(\Gamma|\Sigma)$
    for a local state $\Sigma$.  
    The densities are shown only in the case where the head 
    is moving upward. (The densities in the other case can be
    obtained simply by changing the sign of the head.)
    The $\Gamma$ specifies the direction of the head after the scattering.
    The $z$ denotes the coordination number, e.g., $z=2d$
    for the $d$-dimensional hyper-cubic lattice.}
\ \\[-5mm]
  \tablabel{boson}

  \centering
  \begin{tabular}{|c|lccc|}
    \hline
    \hline
    $\Sigma$ & & $\mu/dt \le -1$ &  $-1 \le \mu/dt \le 1$ & $1 \le \mu/dt$
  \\
    \hline
    \hline
    & $\rho(\downarrow|\Sigma)=$
    & $-\frac{t}{2} + \frac{nV_0 - \mu}{z}+  m V_1$
    & $\frac{nV_0}{z}+m V_1$
    & $\max\left(0,\frac{t}{2}+\frac{nV_0-\mu}{z}+mV_1\right)$
  \\
    $\left( \begin{array}{cc}n_- & m\end{array} \right)$
    & $\rho(\nearrow|\Sigma)=$
    & $\frac{t}{2}$ & $\frac12\left(\frac{t}{2}-\frac{\mu}{z}\right)$ & 0
  \\
    & $\rho(\rightarrow|\Sigma)=$ & 0 & 0 & 0
  \\
    \hline
    & $\rho(\downarrow|\Sigma)=$
    & 0 & 0
    & $\max\left(0,-\frac{t}{2}-\frac{(n-1)V_0-\mu}{z}-mV_1\right)$
  \\
    $\left( \begin{array}{cc}n_+ & m\end{array} \right)$
    & $\rho(\nearrow|\Sigma)=$ & 0 & 0 & 0\\
    & $\rho(\rightarrow|\Sigma)=$ 
    & 0 & $\frac12\left(\frac{t}{2}+\frac{\mu}{z}\right)$ & $\frac{t}{2}$
  \\
    \hline
    \hline
    & $P(\downarrow|\Sigma)=$  & 0 & 0 & 0
  \\
    $\left( \begin{array}{cccc}
        n-1 & m \\ n_+ & m-1 \end{array} \right)$
     & $P(\nearrow|\Sigma)=$
     & $\frac{1}{n}$ & $\frac{1}{n}\left(\frac12-\frac{\mu}{zt}\right)$ & 0\\
     & $P(\rightarrow|\Sigma)=$ & 0 & $\frac{1}{n}\left(\frac12+\frac{\mu}{zt}\right)$
     & $\frac{1}{n}$\\
     & $P(\uparrow|\Sigma)=$ & $1-\frac1n$ & $1-\frac1n$ & $1-\frac1n$
  \\
     \hline
    $\left( \begin{array}{cccc}
        n-1 & m \\ n_- & m-1 \end{array} \right)$ & 
    \multicolumn{4}{c|}{
    $P(\downarrow|\Sigma) =    P(\nearrow|\Sigma) =    P(\rightarrow|\Sigma) = 0,$ and $P(\uparrow|\Sigma) = 1$
    }
  \\
    \hline
    $\left( \begin{array}{cccc}
        n+1 & m \\ n_+ & m+1 \end{array} \right)$ & 
    \multicolumn{4}{c|}{
    $P(\downarrow|\Sigma) =    P(\nearrow|\Sigma) =    P(\rightarrow|\Sigma) = 0,$ and $P(\uparrow|\Sigma) = 1$
    }
  \\
    \hline
    $\left( \begin{array}{cccc}
        n+1 & m \\ n_- & m+1 \end{array} \right)$ & 
    \multicolumn{4}{c|}{
    $P(\downarrow|\Sigma) =    P(\nearrow|\Sigma) =    P(\rightarrow|\Sigma) = 0,$ and $P(\uparrow|\Sigma) = 1$    }\\
    \hline
    \hline
  \end{tabular}
\end{fulltable}

\subsubsection{Observables}

In the directed-loop algorithm (and also in the algorithm with the
series-expansion representation), 
the energy and the specific heat can 
simply be computed by counting the number of the vertices,\cite{SyljuasenS2002}
$n_{\rm v}$, as
$$
  \langle \Ham \rangle = - \beta^{-1} \langle n_{\rm v} \rangle
$$
and
$$
  C = \langle {n_{\rm v}}^2 \rangle - \langle n_{\rm v} \rangle^2 - \langle n_{\rm v} \rangle.
$$

The off-diagonal Green's function $\Gamma(\vect{r},\tau)$ is,
as in the worm algorithm,\cite{ProkofevST1998}
proportional to the frequency of the occurrence of the configurations 
in which the head is separated from the tail
by the space-time vector $\vect{x} = (\vect{r},\tau)$.
Specifically, Green's function of the bosonic models,
for which the source term is proportional to $b+b^{\dagger}$,
can be expressed as
\begin{equation}
  \eqlabel{measure-Green}
  \langle b(\vect{x}) b^\dagger(\vect{y}) \rangle 
  = \langle n(\vect{x}-\vect{y}) \rangle
\end{equation}
where $N$ is the number of sites and 
$n(\vect{x})$ is the number of times the configurations
with the head and the tail separated from each other
by $\vect{x}$ appear during one cycle of the update 
(i.e., from the creation through the annihilation of the worm).
See Dorneich and Troyer\cite{DorneichT2001}
for the measurement of the time-dependent Green's function 
in the series-expansion representation with finite $L$.


\section{Summary and Future Problems}
We have reviewed recent developments in the Monte Carlo simulation
methods based on Markov processes in the space of the world-line 
configurations.
A few original results are also included; the non-binary algorithm
for the bilinear-biquadratic model (\mssc{NonBinaryLoops}, \msss{BIQ})
with the SU(2) symmetry,
the worm-scattering probability of the coarse-grained algorithm 
for the boson Hubbard model (\msss{SoftCoreBosonModel}),
and the continuous-time formulation of the extended ensemble method
(\mssc{ExtendedEnsembleMethods}).
We have focused on the three methods of updating the
world-line configurations; the loop algorithm, the worm
algorithm, and the directed-loop algorithm.
The detailed descriptions of the numerical procedures have been given.
The methods described solve, to a large extent, 
the problems in the local updating algorithm.
The solved (or eased) problems include 
(i) the slowing-down near the critical point or the zero temperature,
(ii) the discretization slowing-down (the Wiesler freezing), 
(iii) the systematic error due to the discretization, 
(iv) the non-ergodicity due to the artificially conserved quantities
(or the additional slowing-down due to the ad hoc global flips
introduced for ergodicity),
(v) the computation of off-diagonal quantities,
(vi) the freezing due to the external magnetic field competing against
    the exchange couplings, and
(vii) the negative-sign problem in a special case of spinless fermions.
On the other hand, a number of problems remain to be solved.
Particularly important among them are
(i) the general fermion sign problem,
(ii) the general frustration sign problem,
(iii) the models with interaction terms competing against each other,
    such as the spin model with a strong anisotropy (uniaxial, cubic,
    tetragonal, etc), and
(iv) the models for which the order parameter cannot be expressed
    as a simple sum of local operators.
The nature of the negative-sign problem may be quite different,
depending on its origin.
Whether it is due to the fermion sign or the frustration, however,
there is not even a clue to a general solution.
It is possible that a general solution does not even exist.
An example of the problems (iii) and (iv) can be found in the $s>1$ Heisenberg
models with the cubic anisotropy.
While such an anisotropy is quite common in real materials,
we are not aware of any efficient algorithms.
For example, the model
$$
  \Ham = -J\sum_{(ij)} \vect{S}_i\cdot\vect{S}_j
         -D\sum_i ((S^x_i)^4+(S^y_i)^4+(S^z_i)^4).
$$
with positive $J$ and $D$ does not cause the 
negative-sign problem when the ordinary $S^z$ representation basis is used.
In addition, it is easy to find a graphical decomposition such as
\Eq{GraphDecomposition} for the model\cite{HaradaKxxxx}.
However, the auto-correlation time of the simulation is extremely
long at low temperatures and the configuration is practically frozen.
A similar situation can be seen in another model with the cubic anisotropy
\begin{eqnarray*}
  \Ham & = & -J\sum_{(ij)} \vect{S}_i\cdot\vect{S}_j \\
       &   & -D\sum_{(ij)} ((S^x_i)^2(S^x_j)^2+(S^y_i)^2(S^y_j)^2
                           +(S^z_i)^2(S^z_j)^2).
\end{eqnarray*}
Again this model does not cause the 
negative sign problem when the $S^z$ representation basis is used.
However, when we use a representation basis in which the model's symmetry 
is manifest, i.e., when the three vectors 
$|\pm\rangle \equiv |\uparrow\rangle \pm |\downarrow\rangle$ 
and $|0\rangle$ are used
as the basis set, the model exhibits negative signs while
the computational auto-correlation time is small in this basis.
The Ising spin-glass problems may be considered as another example of the problem (iii).
It is well-known that the auto-correlation time of this model is
so long that an accurate numerical simulation is extremely difficult
near and below the critical temperature.
However, when the $S^x$ representation basis is used, 
it is easy to find a simulation method with a small auto-correlation time,
although the negative-sigh problem appears in the $S^x$ representation basis.
These facts appear to suggest that the difficulty of the problems (iii) and (iv) 
are closely related to (ii) in general.

\section*{Acknowledgment}
N.K.'s work was supported by the grant-in-aid (Program No.14540361)
from Monka-sho, Japan. K.H.'s work was supported by the 21st Century
COE Program ``Center of Excellence for Research and Education on
Complex Functional Mechanical Systems'' of Monka-sho, Japan.
The authors are grateful to N.~Hatano for his critical reading of the
draft and many helpful comments.
They also thank 
F.~Alet, 
H.-G.~Evertz,
N.~Prokov'ev,
A.~Sandvik, 
J.~\v{S}makov,
B.~Svistunov, 
and 
M.~Troyer
for stimulating conversations and/or useful comments.


\appendix
\section{Generation of Temporal Positions in Graph (or Vertex) Assignment}
\label{ap:poisson}
If a stochastic process consists of a time series of events with
a given density and if each event occurs independently, 
the process is a Poisson process. 
(Specifically in the present context, 
these events correspond to graph-elements, vertices or scattering points.)
In a Poisson process, a time interval between successive events 
obeys the exponential distribution.
When the density of events is $n$, the distribution of the intervals is
\begin{equation}
  \label{eq:exponential-dist}
  p(\Delta t) = \left\{\begin{array}{cl}
      n\exp(-n\Delta t) & (\Delta t > 0)\\
      0 & (\Delta t \le 0)
    \end{array}\right..
\end{equation}
We can generate a random variable $\Delta t$ that obeys this distribution 
from the uniform random variable $r \in (0,1]$ by using the transformation
\begin{equation}
  \eqlabel{uniform-exp}
  \Delta t = - \frac{\ln (r)}{n}.
\end{equation}
Therefore, placing events (objects) on a given time-window
(segment) with a given density $n$ can be done as follows:
\begin{liststep}
  \item
    Set the time variable $t$ to be the starting time of the window.
  \item
    \label{st:exp-2}
    Generate a uniform random number $r \in (0,1]$ and
    compute $\Delta t$ using \Eq{uniform-exp}.
  \item
    Increase the time $t$ by $\Delta t$, i.e., $t := t+\Delta t$.
  \item 
    If the new time $t$ is smaller than the ending time of the window,
    place an object at $t$, and go back to \Step{exp-2}.
    Otherwise, terminate the process.
\end{liststep}

\section{Cluster Identification}
\label{ap:identify}
In an actual computer program, 
the world-line configuration is represented as a linked-list 
data structure of objects, where each object corresponds to a segment. 
In order to identify loops and clusters, 
we define a variable, which will eventually be the cluster ID number,
for each object. 
(In the C$++$ language, for example, 
we add a member variable to the ``segment'' object.)
When a graph element is assigned and points are connected, 
the variables are updated as follows.

Let $c(s)$ be the variable of the object that is
specified (or is pointed to) by an index (or by a pointer) $s$.
In what follows, we identify an index with the object that is
specified by it.
The variables define a tree structure.
That is, $c(s)$ is a parent of $s$, and $c(s)=s$ if it
does not have a parent.
Every object initially has no parent.
When two segments, $s$ and $s'$, are connected by an edge 
in a graph element, the following operations are applied:
\begin{liststep}
\item 
  Find the root of each object. 
  Let $r$ and $r'$ be the roots of $s$ and $s'$, respectively.
  This can be done by repeating $r := c(r)$
  starting from $r := s$ until $r = c(r)$ holds.
  The same for $r'$.
\item 
  Let $R := \min(r,r')$.
  Then, let $c(a) := R$ for all $a$, where $a$ are
  the ancestors of $s$ and $s'$.\label{app:root}
\end{liststep}
After these procedures have been done for all connections, 
$c(s)$ is the unique identifier of the cluster, i.e.,
$c(s)=c(s')$ if and only if $s$ and $s'$ belong to the same cluster.

Choosing the root which has more children than the other in Step
\ref{app:root}, we can make this procedure more efficient. It can be
done by using the new variable $n(r)$ which holds the number of
children of a root $r$. Starting from all $n(s)=1$, we only need to
update the variable as $n(R):=n(r)+n(r')$ in Step \ref{app:root}.




\end{document}